\documentclass{aastex63}
\usepackage[caption=false]{subfig}
\usepackage{graphicx}

\usepackage[normalem]{ulem}

\usepackage{array}
\usepackage{amsmath}
\usepackage{multirow}
\usepackage{xspace}
\usepackage{pgf}
\usepackage{placeins}
\usepackage{changepage}

\DeclareMathOperator*{\argmin}{arg\,min}

\newcommand\MyBox[2]{
  \pgfmathparse{#2>0.35?1:0}  
    \ifnum\pgfmathresult=0\relax\color{white}\fi
  \pgfmathsetmacro\compA{0}      
  \pgfmathsetmacro\compB{#2*270}    
  \pgfmathsetmacro\compC{50}      
  \definecolor{cellcolor}{RGB}{\compA,\compB,\compC}
  \fcolorbox{cellcolor}{cellcolor}{\lower0.82cm
    \vbox to 1.0cm{\vfil
      \hbox to 1.0cm{\hfil\parbox{1.0cm}{\centering{\bfseries #2}\\(#1)}\hfil}
      \vfil}%
  }%
}

\received{July 1, 2020}
\revised{}
\accepted{}
\submitjournal{ApJ}

\shorttitle{CNN for Star Clusters}
\shortauthors{Perez et al.}
\graphicspath{{./}{figures/}}

\newcommand{\starnet}{\textsc{StarcNet}\xspace}

\newcommand{\revision}[1]{\textcolor{black}{#1}}
\newcommand{\newrevision}[1]{\textcolor{black}{#1}}

\begin{document}

\title{: Machine Learning for Star Cluster Identification\footnote{Based on observations obtained with the NASA/ESA Hubble Space Telescope, at the Space Telescope Science Institute, which is operated by the Association of Universities for Research in Astronomy, Inc., under NASA contract NAS 5-26555.}}

\correspondingauthor{Gustavo P\'erez}
\email{gperezsarabi@umass.edu}

\author[0000-0003-3880-8075]{Gustavo P\'erez}
\affiliation{Department of Computer Science\\
University of Massachusetts Amherst \\
40 Governors Drive \\
Amherst, MA 01003, USA}

\author[0000-0003-1427-2456]{Matteo Messa}
\affiliation{Department of Astronomy\\
University of Massachusetts Amherst \\
710 North Pleasant Street \\
Amherst, MA 01003, USA}

\author[0000-0002-5189-8004]{Daniela Calzetti}
\affiliation{Department of Astronomy\\
University of Massachusetts Amherst \\
710 North Pleasant Street \\
Amherst, MA 01003, USA}

\author[0000-0002-3869-9334]{Subhransu Maji}
\affiliation{Department of Computer Science\\
University of Massachusetts Amherst \\
40 Governors Drive \\
Amherst, MA 01003, USA}

\author[0000-0003-2797-9979]{Dooseok E. Jung}
\affiliation{Department of Astronomy\\
University of Massachusetts Amherst \\
710 North Pleasant Street \\
Amherst, MA 01003, USA}

\author[0000-0002-8192-8091]{Angela Adamo}
\affiliation{Department of Astronomy,\\ Oskar Klein Centre, Stockholm University,\\ AlbaNova University Centre,\\
SE-106 91 Stockholm, Sweden}

\author{Mattia Siressi}
\affiliation{Department of Astronomy,\\ Oskar Klein Centre, Stockholm University,\\ AlbaNova University Centre,\\
SE-106 91 Stockholm, Sweden}



\begin{abstract}
We present a machine learning (ML) pipeline to identify star clusters in the multi--color images of nearby galaxies, from observations obtained with the Hubble Space Telescope as part of the Treasury Project LEGUS (Legacy ExtraGalactic Ultraviolet Survey).
\starnet (STAR Cluster classification NETwork) is a multi--scale convolutional neural network (CNN) which achieves an accuracy of \revision{68.6\% (4 classes)/86.0\% (2 classes: cluster/non--cluster)} for star cluster classification in the images of the LEGUS galaxies, nearly matching human expert performance. 
We test the performance of \starnet by applying pre--trained CNN model to \revision{galaxies not included in the training set, finding accuracies similar to the reference one. 
We test the effect of \starnet predictions on the inferred  cluster properties by comparing multi--color luminosity functions and mass--age plots from catalogs produced by \starnet and by human--labeling; distributions in luminosity, color, and  physical characteristics of star clusters are} similar for the human and ML classified samples.
There are two  advantages to the ML approach: (1) reproducibility of the classifications: the ML algorithm's biases are fixed and can be measured for subsequent analysis; and (2) speed of classification: the algorithm requires minutes for tasks that humans require weeks to months to perform. By achieving  comparable accuracy to human classifiers, \starnet will enable extending classifications to a larger number of candidate samples than currently available, thus increasing significantly the statistics for cluster studies.

\end{abstract}



\section{Introduction} \label{sec:intro}
The birth and evolution of star clusters are seamlessly tied to the process of star formation. Most stars are formed in clustered structures \citep{lada2003}, 
but only a fraction of them are forming in gravitationally bound clusters, while the remaining stars will be quickly dispersed in the stellar field of the galaxy 
\citep[e.g.][]{bastian2008,Longmore2014}. 
The formation, the temporal and spatial evolution, and the physical and chemical properties of star clusters trace the dynamical evolution of galaxies and their merger history, provide insights into the origin and persistence of spiral arms, and constrain the mechanisms that govern and regulate star formation. 
In recent years, the Hubble Space Telescope (HST) has provided detailed observations for large samples of young star clusters (YSCs, age$<$300~Myr) in nearby galaxies, allowing detailed studies of their physical properties, which are fundamental for understanding their formation and evolution.
The distributions of clusters' luminosities and masses are tracers of the mechanisms of cluster formation  \citep[e.g.][]{whitmore1999,larsen2002,gieles2006b,gieles2006c,mora2009,whitmore2010,whitmore2014,johnson2017} and of the fraction of star formation that takes place in bound clusters \citep[e.g.][]{bastian2012,adamo2015,chandar2015,johnson2016,messa2018a}, while the age distributions reveal how rapidly clusters are disrupted \citep[e.g.][]{gieles2006a,gieles2009,bastian2012,silva-villa2014,chandar2014,chandar2016,adamo2017, messa2018b}.

Throughout the literature, observations show that both the initial cluster  mass  and  luminosity  functions (CMF  and  CLF, respectively) are well described by a power-law slope $\sim -$2, which traces the hierarchical star forming structures from which YSCs emerge \citep[][among many others]{whitmore1999,whitmore2010, whitmore2014, larsen2002, bik2003,gieles2006a, gieles2006b, mora2009, chandar2014, adamo2017, messa2018a}. 
However, recent measurements have uncovered a dearth in the number of very massive clusters ($\gtrsim 10^5$~M$_{\odot}$) in nearby spirals, suggesting that the formation of massive clusters may be disfavored in these environments   \citep{adamo2015,johnson2017,messa2018a}. This dearth, or truncation mass, may depend on the galactic environment in a similar manner to what  is observed for the fraction of star formation in clusters and the disruption strength of clusters. Variations in the truncation mass appear to  be present both between galaxies with different global properties \citep[e.g.][]{johnson2017} and between sub-regions of the same galaxy that trace different environments \citep{silva-villa2014,adamo2015,messa2018b}, but the issue is far from settled \citep[e.g.][]{chandar2017,mok2019,mok2020}.

The main impediment to reach a consensus on the role of environment on clusters' formation and evolution is the absence of large samples of uniformly--selected YSCs across a wide range of galaxies' properties. The HST Treasury Program Legacy ExtraGalactic UV Survey (LEGUS; GO-13364) attempted to fill this gap by observing 50 nearby (d$\lesssim 16$ Mpc) galaxies in 5 broad bands, from the near--UV to the I, with the goal of extracting YSCs catalogs with well-defined physical properties  \citep{legus}. 
While star clusters are easily detectable up to $\approx100$ Mpc, at distances where individual stars are no longer so, their identification is challenging. The clusters, especially those younger than a few Gyr, are projected against the uneven background of the host galaxy’s disk, and their colors are often similar to those of the surrounding stellar populations or of background galaxies. Confusion, artifacts, and isolated or chance superposition of stars are the main reasons for the failure of automatic approaches that rely on physical or geometrical parametrization (source concentration, colors, luminosity, symmetry, etc.). These contaminants are usually culled from automatic catalogs via visual inspection, a labor--intensive approach that requires humans to evaluate each source individually, with often inconsistent results among different classifiers. The need for human intervention explains the limited number of star cluster catalogs commonly available in the literature and the fact that the LEGUS collaboration has released catalogs for only 31 of their 50 galaxies, two-thirds of which for sparsely--populated dwarfs \citep{cook2019}. 

Crowd--sourcing approaches (e.g., ``citizen science"), while effective when the galaxies are located in the Local Group and the clusters are projected against a sparse stellar field (like in the case of the galaxy M31; \citealp{johnson2015}), become ineffective once the galaxies are beyond a few Mpc, and the star clusters are projected against an unresolved and uneven background. In these conditions, even human experts have difficulties yielding reproducible classifications: the same individual is usually able to reproduce their own classifications less than 90\% of the times, and different experts do not agree among themselves to better than about $70\%$--75\% of the times, across four identification classes \citep{adamo2017, grasha2019, wei2020}. Thus, training of citizen classifiers becomes time--consuming and low--yield; an experiment run by the LEGUS collaboration using a citizen-science platform yielded close--to--random classifications for galaxies at about 10 Mpc distance. 

 Machine learning, and computer vision in particular, offer tools that can  potentially be game changers for the field: they can be trained to reproduce at least the level of quality of expert classifiers; their classifications are self--consistent and reproducible; and they require a tiny fraction of the time, thus enabling multiple classification trials to be applied to the same catalog, as the training sets improve. Visual recognition is a core research activity of the computer science community that is finding increasing applications in astronomy, including classification of galaxies \citep{dominguezsanchez2018,khan2019,barchi2020}, galaxies' mergers \citep{ackermann2018,iprijanovi2020} and galaxy morphology \citep{Dieleman2015,Walmsley2018}, and will be the key tool for future petabyte-size catalogs (e.g., from the Vera C. Rubin Observatory). 
 ML algorithms have been recently applied also to the morphological classification of YSCs \citep{grasha2019, wei2020}. In \citet{grasha2019}, a bagged decision  tree algorithm was implemented on the galaxy pair NGC~5194+5195, after training on a small number of LEGUS cluster catalogs. \citet{wei2020} trained and tested deep learning algorithms using a larger sample of LEGUS catalogs than \citet{grasha2019}'s. We will compare our results with these earlier papers in  \S~\ref{sec:discussion}.

\begin{figure}[t]
\begin{center}
\includegraphics[width=1.0\linewidth]{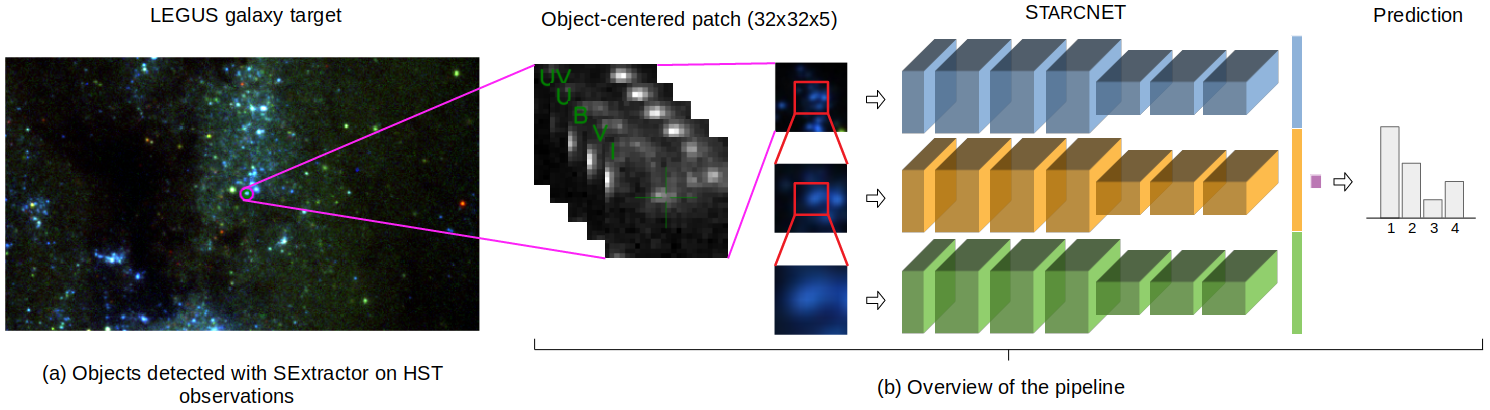}  
\caption{\textbf{The \starnet pipeline.} Graphic sketch of the machine learning pipeline used in this work to classify cluster candidates in the LEGUS images. (Left): The Hubble Space Telescope images as   processed by the LEGUS project through a custom pipeline to generate automatic catalogs of cluster candidates, which are part of the public LEGUS catalogs release \citep{legus, adamo2017}; we apply \starnet to the LEGUS catalogs and images. (Center--Left): The region surrounding each candidate is selected from the 5 band images at three magnifications, and is used as input to our multi-scale \starnet. (Center--Right and Right): Each of the three pathways of the CNN consists of 7 convolutional layers, which are later connected to produce a prediction for the candidate in one of four classes.  \label{fig:cnn}
}
\end{center}
\end{figure}

The goal of this paper is to design a deep network to classify star cluster candidates trained on the largest and most robust catalogs available and improve on previous approaches. 
We develop a three-pathway convolutional neural network (CNN) called \starnet\ to classify star clusters in the LEGUS five--band images (Figure \ref{fig:cnn}); our approach consists of applying a CNN to the region surrounding each cluster candidate at three increasing magnification levels, and then combining the resulting outputs to produce a classification using a 4--class morphological scheme \citep{adamo2017}. 
For training, we use the entire collection of identified LEGUS star clusters from all released catalogs \citep{adamo2017}, both to increase the number of examples and to use catalogs classified by multiple experts.  

This paper is organized as follows: in \S~\ref{sec:materials} we present the catalogs and images from the LEGUS project used in this work; in \S~\ref{sec:method} we describe the architecture of \starnet, while we test different configurations in \S~\ref{sec:experiments}. We discuss the performance of our approach in \S~\ref{sec:discussion}. In \S~\ref{sec:clusterscience} we study the average cluster properties in each of the four morphological classes, focusing on how the classification given by \starnet\  affects them. Finally, we briefly present the outlook for future developments in \S~\ref{sec:future_work} and summarize our results in \S~\ref{sec:conclusion}.

\section{Data and Catalogs} \label{sec:materials}
The LEGUS survey consists of 50 galaxies at distances between 3.5~Mpc and $\sim$16~Mpc, observed in 63 pointings with the Hubble Space Telescope in five broad bands, using either the WFC3/UVIS camera or archival ACS/WFC images when available. The five bands are: $NUV$ (WFC3-F275W filter), $U$ (WFC3-F336W filter), $B$ (either WFC3-F438W or ACS-F435W filter), $V$ (either WFC3-F555W, ACS-F555W or ACS-F606W
filter) and $I$ (either WFC3-F814W or ACS-F814W filter). A full description of the project, the sample selection, and the observing strategy is provided in \citet{legus}. Automatically--generated catalogs of star cluster candidates were produced using a six--step pipeline as described in \citet{adamo2017} and are publicly available at the Mikulski Archive for Space Telescopes (MAST) for 31 galaxies, in 37 separate pointings\footnote{\url{https://archive.stsci.edu/prepds/legus/}; the different pointings of
NGC5194$+$NGC5195 are combined into a single catalog, for a total of 34 separate star cluster catalogs. {DOI:10.17909/T9J01Z}}. 
The six steps are described in detail in \citet{adamo2017}; in brief, they consist of running SExtractor \citep{SExtractor} on the HST processed and aligned images\footnote{All LEGUS images are aligned and sampled to a common pixel scale, 0.04$^{\prime\prime}$/pix.}, applying basic
selection functions to remove as many stars and artefacts as possible, and generating aperture--corrected photometry. Subsequently, spectral energy distribution fits are performed on cluster candidates that are detected in at least four separate bands, to derive ages, masses, and extinction, and their uncertainties. Finally, visual classification is performed on candidates that are brighter than V=$-$6~mag \citep{adamo2017}, as the automatic selection still leaves about 50\%  contaminants in the catalogs (Table~\ref{table:dataset}). 

\begin{figure}
\begin{center}
\begin{tabular}{c}
\includegraphics[width=0.98 \linewidth]{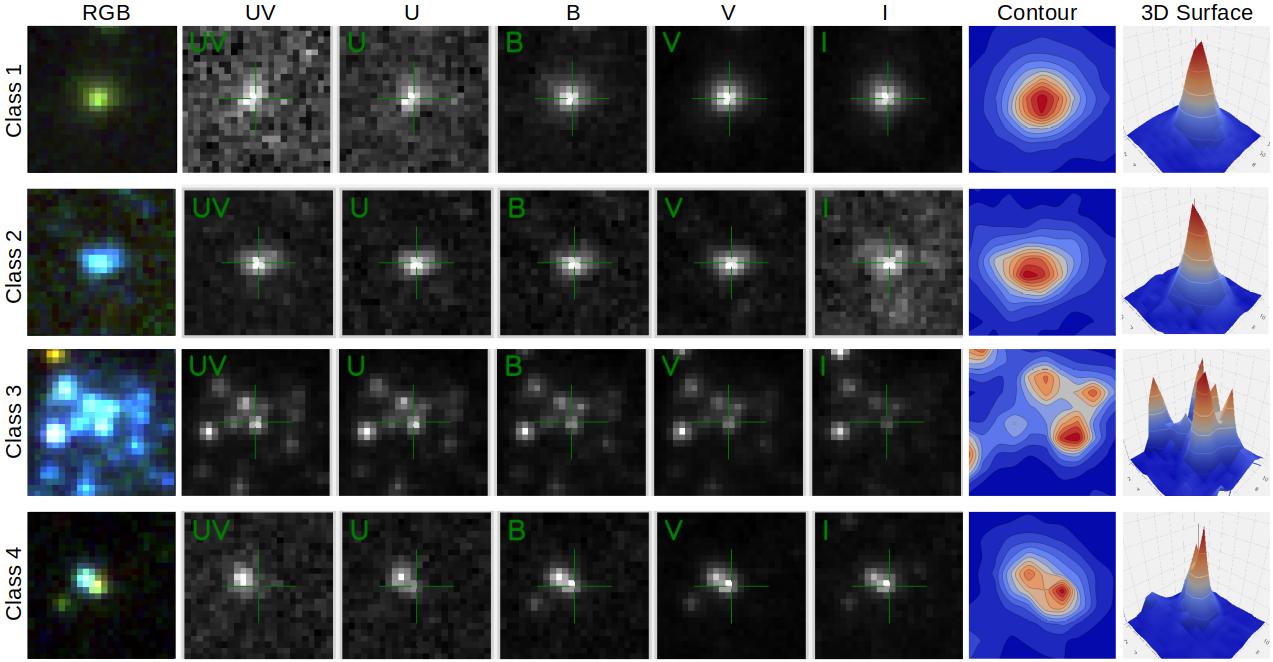}
\end{tabular}
\caption{\textbf{LEGUS classification scheme.} Examples of candidates from the LEGUS images of NGC 1566 classified as Class 1 (symmetric star cluster; top), Class 2 (elongated star cluster; middle-top), Class 3 (compact, multi--peak association; middle-bottom), and Class 4 (spurious object; bottom). The three--color image to the left is created using the   $NUV$ and $U$ bands for the blue channel, the $B$ band for green one, and the $V$ and $I$ bands for red one. The contour and 3D plots from the V--band are shown to the right of the figure.
\label{fig:examples}}
\end{center}
\end{figure}

The visual classification was performed by at least three human classifiers in the LEGUS team, with an additional one or two tie-breakers for ambiguous cases, using the following morphology--based classification scheme (Figure~\ref{fig:examples}; 
more details in \citealp{adamo2017}): 

\begin{enumerate}
    \item \textbf{Class 1:} symmetric, compact objects, with a light distribution more extended than the stellar one;
    \item \textbf{Class 2:} compact objects with slightly elongated density profiles; 
    \item \textbf{Class 3:} multiple peak systems on top of diffuse underlying emission, referred to as compact associations;
    \item \textbf{Class 4:} spurious detections (foreground/background sources, single bright stars, asterisms, artifacts); this class is highly inhomogeneous, as it contains everything that the classifiers deemed a `non--cluster'.
\end{enumerate}

The three (or more) independent classifications were then combined into a final \emph{class label}, 
defined as the mode of the classifications of each candidate. The catalogs with classified cluster candidates for the 31 LEGUS galaxies available from the MAST Archive include a total of around 15,000 candidates across the 4 classes.  
The 31 galaxies span the full range of distances of the LEGUS sample. A detailed presentation and discussion of the classification approach for the LEGUS sample will be given in a forthcoming paper (Hwi et al., in prep.) \revision{and we report in Appendix~\ref{sec:appC} a summary of the distribution of cluster candidates by  class for each galaxy, along with the galaxy distance.} These catalogs are the focus of the present work, and their cumulative statistics are listed in Table~\ref{table:dataset}.

We use the location of each source as listed in the catalogs to produce $32 \times 32$ pixels ($\sim$1.3$^{\prime\prime}\times$1.3$^{\prime\prime}$) cutouts from the images in the five bands, which  we use as inputs for our algorithm; we call these cutouts `input arrays' in the remaining of the paper. 
We split the dataset of the classified sources in a uniformly random fashion (as in Table \ref{table:dataset}) as follows: $80\%$ of the total classified sources are used for training and validation (\textit{trainval} set), and the remaining $20\%$ for testing. From the \textit{trainval} set, we use $90\%$ of the classified candidates for training, and the remaining $10\%$ for validation\footnote{The training set is the sample of data used to fit the model. The validation set is the sample of data used to provide an intermediate evaluation of a model fit on the training dataset while tuning model hyperparameters. The test set is the sample of data used to provide an unbiased evaluation of a final model fit on the training dataset.}.

\subsection{Accuracy of human classifications}
In order to evaluate the level of accuracy we can achieve with the ML predictions, we compare classifications from different individuals among themselves, as a metric for the highest possible agreement that can be reached between ML and humans. 
Individual classifications (as opposed to the mode or final classification) are available for a total of $\sim$6,000 sources across 13 galaxies\footnote{This number is smaller than the total $\sim15,000$ sources visually classified, as it is limited by the current availability of single-classifier files within the LEGUS collaboration.}. 
The fraction of sources with the same classification, weighted by the number of sources in each class, gives the agreement among the two classifiers. We take the mean of all the possible combinations of classifiers as the mean agreement. The resulting agreement among two separate classifiers is $57.3\%$, as shown by the confusion matrix in Figure~\ref{fig:cm_human} (left panel). When instead the labels given by each classifier are compared with the final classification, the mean agreement is $75.0\%$ (Figure~\ref{fig:cm_human}, right panel). The higher level of agreement between any individual classifier and the final classification is also  due to the degeneracy that the final classification includes the classifier's own classification. We can, therefore, expect that a well--constructed ML algorithm will yield accuracies above the 57.3\% of the individual--to--individual comparison, but not quite at the 75\% level of the individual--to--final comparison for our samples, because the ML--to--final classification comparison does not include the same degeneracy as the individual--to--final one. 
 
The agreement is not uniform across classes, as it is higher for classes 1 and 4, and is the lowest in class 3. Class~3 indeed remains the most challenging class to recognize, as the detection of diffuse  emission underlying multiple peaks depends on the depth of the image \citep{adamo2017}.
The agreement is not even uniform across different galaxies, as it goes from $\sim95\%$ in NGC 3738 (over 400 sources) to less than $50\%$ in NGC 628 ($\sim1800$ sources). The level of (dis-)agreement highlights the difficulty of performing morphological classifications of sources embedded in unresolved galaxies.
The variations in the level of disagreement may be contributed by the large number of people ($>10$) involved in the classification process, which introduces  personal biases. However, the overall agreement is in line with what found in similar comparisons by other authors \citep{grasha2019, wei2020}, indicating that: (1) the LEGUS classifications are as robust as any others found in the literature; and (2) we should expect the predictive accuracy of our ML algorithm to be no better than about 70\%--75\% when measured against human labels. In Appendix~\ref{sec:appB} we present the most frequent cases of mis-classification along with additional examples. 

\begin{figure}
\begin{center}
\begin{tabular}{c c c c c c}
\noindent
\renewcommand\arraystretch{1.0}
\begin{tabular}{ @{\hspace{-100pt}}c >{\hspace{-0.4em}\bfseries}c @{\hspace{-0.8em}}c @{\hspace{-1.2em}}c @{\hspace{-1.2em}}c @{\hspace{-1.2em}}c}
  \multirow{17.5}{*}{\rotatebox{90}{\parbox{2.3cm}{\bfseries\centering Classifier 1}}} & 
    & \multicolumn{4}{c}{\bfseries Classifier 2} \\
  & & \hspace{1.0em} \bfseries Class 1 & \hspace{1.0em} \bfseries Class 2 & \hspace{1.0em} \bfseries Class 3 & \hspace{1.0em} \bfseries Class 4 \\
  & \multirow{3}{*}{Class 1} & 
  \MyBox{796}{0.68} & \MyBox{215}{0.18} & \MyBox{61}{0.05} & \MyBox{103}{0.09} \\
  & \multirow{3}{*}{Class 2} & 
  \MyBox{213}{0.14} & \MyBox{713}{0.50} & \MyBox{281}{0.20} & \MyBox{223}{0.16} \\
  & \multirow{3}{*}{Class 3} & 
  \MyBox{44}{0.03} & \MyBox{260}{0.18} & \MyBox{700}{0.49} & \MyBox{437}{0.30} \\
  & \multirow{3}{*}{Class 4} & 
  \MyBox{123}{0.06} & \MyBox{242}{0.12} & \MyBox{398}{0.19} & \MyBox{1285}{0.63} \\
\end{tabular}
&  \hspace{1.8em}
\noindent
\renewcommand\arraystretch{1.0}
\begin{tabular}{ @{\hspace{-100pt}}c >{\hspace{-0.4em}\bfseries}c @{\hspace{-0.8em}}c @{\hspace{-1.2em}}c @{\hspace{-1.2em}}c @{\hspace{-1.2em}}c}
  \multirow{17.5}{*}{\rotatebox{90}{\parbox{3.1cm}{\bfseries\centering Classifier 1}}} & 
    & \multicolumn{4}{c}{ \bfseries Classifier (mode)} \\
  & & \hspace{1.0em} \bfseries Class 1 & \hspace{1.0em} \bfseries Class 2 & \hspace{1.0em} \bfseries Class 3 & \hspace{1.0em} \bfseries Class 4 \\
  & \multirow{3}{*}{Class 1} & 
  \MyBox{973}{0.81} & \MyBox{122}{0.10} & \MyBox{34}{0.03} & \MyBox{70}{0.06} \\
  & \multirow{3}{*}{Class 2} & 
  \MyBox{116}{0.09} & \MyBox{974}{0.72} & \MyBox{134}{0.10} & \MyBox{119}{0.09} \\
  & \multirow{3}{*}{Class 3} & 
  \MyBox{26}{0.02} & \MyBox{155}{0.11} & \MyBox{989}{0.71} & \MyBox{223}{0.16} \\
  & \multirow{3}{*}{Class 4} & 
  \MyBox{61}{0.03} & \MyBox{178}{0.08} & \MyBox{283}{0.13} & \MyBox{1635}{0.76} \\
\end{tabular}
&  \\
\noindent
\renewcommand\arraystretch{1.0}
\begin{tabular}{ @{\hspace{-70pt}}c >{\hspace{-0.4em}\bfseries}c @{\hspace{-0.8em}}c @{\hspace{-3.0em}}c}
  \multirow{10}{*}{\rotatebox{90}{\parbox{2.3cm}{\bfseries\centering Classifier 1}}} & 
    & \multicolumn{2}{c}{\bfseries \qquad Classifier 2\qquad  \ \ }  \\
  & & \hspace{1.0em} \bfseries C & \hspace{1.0em} \bfseries NC \\
  & \multirow{3}{*}{C} & 
  \MyBox{1937}{0.74} & \MyBox{668}{0.26}  \\
  & \multirow{3}{*}{NC} & 
  \MyBox{669}{0.19} & \MyBox{2820}{0.81}  \\
\end{tabular}
&  
\noindent
\renewcommand\arraystretch{1.0}
\begin{tabular}{ @{\hspace{-20pt}}c >{\hspace{-0.4em}\bfseries}c @{\hspace{-0.8em}}c @{\hspace{-3.0em}}c}
  \multirow{10}{*}{\rotatebox{90}{\parbox{2.3cm}{\bfseries\centering Classifier 1}}} & 
    & \multicolumn{2}{c}{\bfseries Classifier (mode) \quad \ \ }  \\
  & & \hspace{1.0em} \bfseries C & \hspace{1.0em} \bfseries NC \\
  & \multirow{3}{*}{C} & 
  \MyBox{2185}{0.86} & \MyBox{357}{0.14}  \\
  & \multirow{3}{*}{NC} & 
  \MyBox{420}{0.12} & \MyBox{3130}{0.88}  \\
\end{tabular}
\end{tabular}
\caption{\textbf{Consistency of human classification.} Mean confusion matrices for the comparison between independent human classifiers (left panels) and for the comparison of a human classifier with the final (mode) classification (right panels). The overall accuracies are \textbf{57.3\%} and \textbf{75.0\%} for the 4--class and \textbf{78.1\%} and \textbf{87.2\%} for the binary classifications, respectively.}
\label{fig:cm_human}
\end{center}
\end{figure}

\begin{table}
\centering
\caption{\textbf{Dataset statistics.} Classification statistics \revision{of the star clusters in training, validation, and test splits} of the LEGUS dataset. \revision{The distribution of star clusters across the 31 galaxies in the dataset is included in Appendix~\ref{sec:appC}.}}
\begin{adjustwidth}{0.5cm}{}
\begin{tabular}{l c c c c}
\hline\hline
  & Train set & Validation set & Test set & \textbf{Total dataset}\\

 \hline
Total class 1          & 1765 & 196 & 528 & \textbf{2489 (16.09\%)} \\
Total class 2          & 2225 & 247 & 612 &  \textbf{3084 (19.93\%)} \\
Total class 3          & 2192 & 244 & 617 & \textbf{3053 (19.73\%)} \\
Total class 4          & 4956 & 551 & 1338 & \textbf{6845 (44.24\%)} \\
\hline
 \textbf{Total} &  \textbf{11138 (72.00\%)} & \textbf{1238 (8.00\%)} & \textbf{3095 (20.00\%)} & \textbf{15471 (100.00\%)}  \\
 \hline
\label{table:dataset}
\end{tabular}
\end{adjustwidth}
\end{table}

\subsection{Binary classification}
\revision{For a number of applications, cluster samples do not need to have the detailed 4--class morphological classification developed by LEGUS, and a binary (cluster/non--cluster) classification suffice.
We will test the accuracy of \starnet for binary classifications on our samples by aggregating class 1 and 2 as the `clusters' class and class 3 and 4 as the `non-clusters' class (as e.g. in \citealt{bastian2012,adamo2015,hollyhead2016,messa2018a}).
The human agreement between classifiers, presented in the previous section, increases noticeably when a binary classification is adopted; the human--to--human agreement of the LEGUS sample increases from $57.3\%$ to $78.1\%$, while the human--to--mode agreement increases from $75.0\%$ to $87.2\%$ (bottom panels of Figure~\ref{fig:cm_human}).
There is no consensus in the literature about excluding class~3 sources from the `cluster' class; we discuss in \S~\ref{sec:binary_class} the accuracies resulting from considering different binary classifications. 
In the same section, we discuss the results of training \starnet\ directly on binary classification.}

\section{Method} \label{sec:method}

Our approach to star cluster classification is based on a \revision{deep} Convolutional Neural Network (CNN).
Over the last decade CNNs have emerged as the leading model in many visual recognition tasks such as categorization of images, 
semantic segmentation, and object detection.
The proposed network called \starnet is based on networks used for color image classification, but is modified to take into account the multiple channels, normalization and multi-scale spatial context of the input.
\starnet chains simple building blocks or layers to form a network, and uses gradient-based
optimization of all the parameters of these layers using backpropagation of a loss function 
which measures how far off the result produced by the model is from the expected result on training data (\S~ \ref{sec:background}).
Modern libraries for deep learning (e.g., Tensorflow \citep{tensorflow}, PyTorch \citep{pytorch}) allow a modular definition of the network architecture and support gradient based learning of parameters given a dataset of objects with \emph{class labels}. Below we provide an overview of the relevant building blocks.

\subsection{Background \label{sec:background}}
\paragraph{Convolutional Neural Network (CNN)} A CNN is a parameterized function $y=f(x; \mathbf{\theta})$, mapping inputs $x$ to outputs $y$ given parameters $\theta$.
The network consists of layers denoting sequential operations that transform the input to its output.
We will denote the inputs to layer $l$ as $\Phi^{(l)}$ and its output as $\Phi^{(l+1)}$. 
Thus a convolutional network with $n$ layers has $\Phi^{(1)}=x$ and with $\Phi^{(n+1)}=y$.
Common layers for convolutional network are:
\begin{itemize}

    \item Convolution layer: A convolutional layer consists of applying a set of filters over inputs with spatial dimensions. For images, the inputs to the layer $l$ are 3D arrays of size $\Phi^{(l)} \in \mathbb{R}^{h \times w \times c}$ where $h$ and $w$ denote the spatial dimensions (e.g., height and width) and $c$ denotes the number of channels. 
    For example, the first convolutional layer of our network takes an input represented as a 3D array of 32$\times$32 pixels $\times$ 5 bands. 
    A convolutional layer with $k$ filters is parameterized by filter weights $w^{(l)} \in \mathbb{R}^{m \times n \times c \times d} $ and bias $b^{(l)} \in \mathbb{R}^d$, producing an output $\Phi^{(l+1)} \in \mathbb{R}^{h' \times w' \times d}$
    \revision{where $m$ and $n$ are the filter spatial dimensions, $c$ is the number of channels (which corresponds to the number of channels $c$ of the input to the layer $l$), and $d$ is the number of filters}. 
    The output $\Phi^{(l+1)}$ of a convolution layer is a set of $d$ feature maps. A feature map can be thought of as a representation of the input or as a response of a single filter $d_i$ applied to the input $\Phi^{(l)}$.
    The output at a location $(x, y)$ for a filter $d$ is given by:
    \begin{equation}
        \Phi^{(l+1)}[x,y,z] = \sum_{i=1}^{m} \sum_{j=1}^n \sum_{k=1}^c\Phi^{(l)}[x\!-\!i,y\!-\!j,k]w[i,j,k,z] + b^{(l)}[z] \revision{\text{, for }z=\{1,2,...,d\}} .
    \end{equation}

    Convolution layers are used as feature extractors of the input array. We use three different sets of convolution layers to extract features of each input array at three different magnifications. The extracted features from the three pathways are then combined using a fully-connected layer.
    
    \item Pooling layer:
    Pooling replaces the value in a neighborhood with an overall statistic, such as the $\max$ or $avg$, resulting in reduction of the spatial dimension of the input and adding invariance to small deformations.
    The layer is parameterized by the neighborhood size over which the overall statistic is computed and stride denoting the offset between neighborhoods. 
    For example, pooling an input $\Phi^{(l)} \in \mathbb{R}^{h \times w \times d}$ with a stride $k$ leads to an output $\Phi^{(l+1)} \in \mathbb{R}^{h/k \times w/k \times d}$. 
    The parameters are set manually and are typically not learned.
    Our network uses a pooling layer after the fourth convolution layer on each of the pathways of our network as shown in Figure \ref{fig:cnn}.
    
    \item Non-linear activation layer: These layers are based on applying a non-linear transformation to the input.  
    Some commonly used non-linear activation layers are --- Recified linear unit \citep{relu2}: 
    $\texttt{ReLU}(x)=\max(0,x)$, Leaky ReLU \citep{leakyrelu}: 
    $\texttt{LReLU}(x)=\max(0,x)+\rho\min(0,x)$ where $\rho$ is a positive number, and sigmoid: $\sigma(x) = 1/(1+\exp(-x))$.
    Non-linearities allow the network to learn complex mappings between inputs and outputs. These are typically applied after each convolution and fully-connected layer in a network. 
    
    \item Fully-connected layer:  These layers are common for inputs with no spatial dimensions and connect all inputs $\Phi^{(l)} \in \mathbb{R}^m$ to outputs $\Phi^{(l+1)} \in \mathbb{R}^n$ via a weight matrix $w^{(l)} \in \mathbb{R}^{m \times n}$ and bias $b^{(l)} \in \mathbb{R}^n$: 
    \begin{equation}
        \Phi^{(l+1)} = \Phi^{(l)}w^{(l)}+ b^{(l)}.
    \end{equation}
    Fully-connected layers are usually used as the latter layers to learn the classifier part of the network. In our case is also used to combine the features of each of the pathways as shown in Figure \ref{fig:cnn}. 
    
    \item \revision{Dropout layer: With dropout \citep{dropout}, a vector $\Vec{r}^{(l)}$ of independent Bernoulli random variables (each of which has probability $p$ of being $1$) is multiplied element-wise with the outputs of the previous layer $\Phi^{(l)}$ to produce thinned outputs $\Phi_{*}^{(l)}  = \Phi^{(l)} \odot \Vec{r}^{(l)}$ before using them as inputs for the next layer. If dropout is used, the output of a layer is given by: }
    \revision{
    \begin{equation}
        \Phi^{(l+1)} = \Phi_{*}^{(l)}w^{(l)}+ b^{(l)}.
    \end{equation}
    }
    \revision{Randomly dropping activations during training has been shown to improve generalization.}
    
\end{itemize}
    
CNNs stack several blocks of convolution---pooling---non-linear layers.
The hierarchical nature of the network allows the emergence of simple features such as edges and blobs in the early layers and complex features such as a human face in higher layers.
Modern CNNs contain up to hundreds of these blocks, totalling millions of parameters, and are highly effective for visual recognition.

\paragraph{Dataset and training} 

A labeled dataset consists of pairs $\{(x_i, y_i)\}_{i=1}^{N}$. In our case each $x_i$ is a cluster candidate and $y_i$ is its \emph{class label} ($1$ through $4$).
All the parameters $\theta$ corresponding to the layers of the network, e.g. the filter and bias weights of the convolutional layers, must be learned.
This is done by minimizing a loss $\ell$ over the training set plus a regularization term $R(\theta)$ over the parameters:
\begin{equation}
   \hat{\theta} \leftarrow \argmin_\theta \left( \sum_{i=1}^N \ell(f(x_i; \theta), y_i) + R(\theta) \right).
\end{equation}
The loss measures the error between the network's prediction and \emph{class labels}, while the regularization encourages simpler models. 
A common regularization is the squared $\ell_2$ norm of the parameters, i.e., $R(\theta) = ||\theta||_2^2$.
In the classification setting the predictions $\hat{y} = f(x; \theta) $ denote the probability of $C$ target classes, and \emph{class labels} $y$ denote the one-hot encoding of the correct class (i.e., a vector with length equal to the number of categories in the data set with value $1$ in the class position and $0$ elsewhere), and we use the cross-entropy loss denoted by:
\begin{equation}
    \ell(\hat{y}, y) = -\sum_{k=1}^C y^{(k)} \log \hat{y}^{(k)}.
\end{equation}
Training on large datasets is computationally demanding as computing gradients require summation over the entire dataset (Equation 4).
A common practice is to perform stochastic gradient descent where a small batch of training examples ${\cal B}$ are selected at random at each iteration, and the gradient $\Delta \theta^{(t)}$ of the loss with respect to the current parameters $\theta^{(t)}$ are obtained by back propagating the  gradients of the loss:
\begin{equation}
    \Delta \theta^{(t)} =  \left( \sum_{(x,y) \in \cal B} \left( \frac{\partial{ \ell(f(x; \theta), y)}}{\partial \theta} \right) +  \gamma \frac{\partial{R}}{\partial \theta}  \right) \Biggr|_{\theta = \theta^{(t)}}.
\end{equation}
The parameters are updated by taking a step $\eta$ in the negative direction of the gradient:
\begin{equation}
    \theta^{(t+1)} =  \theta^{(t)} - \eta \Delta \theta^{(t)}.
\end{equation}
The overall training consists of initialization of the weights $\theta$ and performing gradient updates for a number of iterations till the loss on a validation set stops decreasing.
Initialization is either random or obtained by training the network on a different task (e.g., in the setting of transfer learning).
A number of modifications have been proposed to this basic scheme that include feature normalization schemes that allow larger step sizes (e.g., batch normalization \citep{bn}, layer normalization \citep{ln}), novel layer blocks (e.g., highway \citep{highway}, residual \citep{resnet}, squeeze-excite \citep{squeeze-excite}, bilinear \citep{bilinear}), and optimization techniques (e.g., Adam \citep{Adam}, AdaGrad \citep{adagrad}).
These have allowed training larger networks on bigger datasets, often leading to improved generalization. See the book of \cite{goodfellow} for a detailed background on CNNs.

\paragraph{Data augmentation} A common practice in training deep networks is to synthetically augment the training data by adding random transformations to the input which do not change the \emph{class label}. These include injecting noise to the pixel values, performing image scaling, rotations, and translations. Our images are centered at the star cluster so we would expect full rotational and mirror symmetry, but not to scaling and translations. The effect of data augmentation is described in \S~\ref{sec:experiments}.

\paragraph{Transfer learning} 

Training deep networks with millions of parameters on small datasets poses a risk of overfitting. 
Transfer learning is a strategy to alleviate this problem.
In this scheme the network is first trained on a large dataset of labeled objects and then fine-tuned by modifying a small number of the parameters on a target dataset where labels or human classifications are limited. 
The efficacy of the transfer depends on 
how close the source and target datasets are, as well as the fine-tuning strategy that is employed. 
A key reason for the popularity of deep networks is that CNNs trained on the ImageNet dataset \citep{imagenet}, which consists of millions of images of common objects (plants, flowers, animals, etc.) taken from the Internet, have been shown to transfer well to a wide variety of visual recognition tasks.
However, such networks are trained on three--band RGB images, which is challenging to transfer to astronomical classifications if the latter use a larger number of bands. In the case of star clusters from the LEGUS project, the input arrays consist of 5 bands. When training a CNN from scratch we can directly design the convolutional layers to have the appropriate number of channels, one per band. But when performing transfer training, the input or the lower layers of the network need to be adapted to enable transfer. One strategy is to manually combine the astronomical images and reduce them to the three--band RGB color images; a second strategy is to learn the transformation as part of the transfer learning strategy. We discuss these schemes in \S~\ref{sec:experiments}. 
We note that, in the case of the present work, the $\sim$15,000 classified sources available across the four classes are enough to train our deep network, as shown by the by the accuracy achieved by \starnet (in particular see  Table~\ref{table:transfer} in \S~\ref{sec:experiments}). 

\paragraph{Evaluation metric} We use the accuracy for the 4--class classification on the test set as the primary metric to compare models. 
We also visualize the confusion matrices that indicates the distribution of the errors made by the model, as well as recall and precision.
The confusion matrix of our best model is shown in Figure \ref{fig:cm}.

\subsection{\starnet architecture and training}
\paragraph{Network architecture} The best performing \starnet is a three-pathway architecture that simultaneously processes three input arrays of the same source \revision{at three different scales.}  \revision{To build the magnified input arrays we center crop and resize them back to $32 \times 32$ pixels using nearest--neighbor interpolation, so all the input arrays have the same pixel size despite containing visual information at three different magnifications}, as shown in Figure~\ref{fig:cnn}.
\starnet is a function $y=f(x_1,x_2,x_3;\theta)$ that receives as input three object-centered arrays $x_1,x_2,x_3 \in \mathbb{R}^{32 \times 32 \times 5}$ and outputs $y \in \mathbb{R}^{4}$, the probability distribution over the four categories. 
The three input arrays $x_1,x_2,$ and $x_3$ contain the photometric information of a single object at three different magnifications (1$\times$, 1.6$\times$, and 3.2$\times$). 
Each input array is passed through a single pathway or sub-network $y_i=f(x_i;\theta_i)$ for $i=1,2,3$.
Each pathway is composed of 7 modules, with each module consisting of a convolutional layer, group normalization \citep{gn}, and leaky ReLU activation layer. 
All convolutional layers contain 3$\times 3$ filters. After the fourth module we add a pooling layer to obtain a global representation of the input. 
The extracted features $y_i \in \mathbb{R}^{m}$ from each pathway are then concatenated into vector $y_c=[y_1~ y_2~ y_3]$ and passed though a fully-connected layer $y=f(y_c;\theta_{fc})$ to output a probability distribution $y \in \mathbb{R}^4$ over the target labels.

\paragraph{Training} The learnable parameters $\theta=\{\theta_1, \theta_2, \theta_3, \theta_{fc}\}$ corresponding to the three pathways and the combination layer are initialized with Xavier initialization~\citep{xavier} and trained using ADAM optimizer~\citep{Adam}. Xavier initializes the weights by drawing them independently from a Gaussian distribution ${\cal N}(0, \sigma^2)$, with $\sigma^2=1/k$, where $k$ is the dimension of the input. The entire network is trained for 15 epochs (one epoch is a full pass over the training set) using a learning rate of $\eta$=1E-04 and cross-entropy loss as described in Equation 4. \newrevision{We determine the number of training epochs by selecting when the best validation performance is achieved.}

\section{Experiments} \label{sec:experiments}
This section is dedicated to ablation studies of \starnet, by altering parameters one by one in the input arrays and in the network, to investigate their effect on the output accuracy of the classifications.

\subsection{Classification accuracy on LEGUS}

\begin{figure}
\begin{center}
\begin{tabular}{c c c c }
\noindent
\renewcommand\arraystretch{1.0}
\begin{tabular}{ @{\hspace{-100pt}}c >{\hspace{-0.4em}\bfseries}c @{\hspace{-0.8em}}c @{\hspace{-1.2em}}c @{\hspace{-1.2em}}c @{\hspace{-1.2em}}c}
  \multirow{17.0}{*}{\rotatebox{90}{\parbox{5.1cm}{\bfseries\centering Human Classification}}} & 
    & \multicolumn{4}{c}{\bfseries \starnet Prediction} \\
  & & \hspace{1.0em} \bfseries Class 1 & \hspace{1.0em} \bfseries Class 2 & \hspace{1.0em} \bfseries Class 3 & \hspace{1.0em} \bfseries Class 4 \\
  
  & \multirow{3}{*}{1} & 
  \MyBox{410}{0.78} & \MyBox{75}{0.14} & \MyBox{6}{0.01} & \MyBox{37}{0.07} \\
  & \multirow{3}{*}{2} & 
  \MyBox{107}{0.17} & \MyBox{335}{0.55} & \MyBox{73}{0.12} & \MyBox{97}{0.16} \\
  & \multirow{3}{*}{3} & 
  \MyBox{3}{0.01} & \MyBox{92}{0.15} & \MyBox{279}{0.45} & \MyBox{243}{0.39} \\
  & \multirow{3}{*}{4} & 
  \MyBox{45}{0.03} & \MyBox{80}{0.06} & \MyBox{115}{0.09} & \MyBox{1098}{0.82} \\

\end{tabular}
&  
\noindent
\renewcommand\arraystretch{1.0}
\begin{tabular}{ @{\hspace{-55pt}}c >{\hspace{-0.4em}\bfseries}c @{\hspace{-0.8em}}c @{\hspace{-3.0em}}c}
  \multirow{10}{*}{\rotatebox{90}{\parbox{2.3cm}{\bfseries\centering Human}}} & 
    & \multicolumn{2}{c}{\bfseries \starnet Prediction}  \\
  & & \hspace{1.0em} \bfseries C & \hspace{1.0em} \bfseries NC \\
  
  & \multirow{3}{*}{C} & 
  \MyBox{927}{0.81} & \MyBox{213}{0.19}  \\
  & \multirow{3}{*}{NC} & 
  \MyBox{220}{0.11} & \MyBox{1735}{0.89}  \\
  
\end{tabular}
&  
\renewcommand\arraystretch{1.0}
\begin{tabular}{ @{\hspace{-30pt}}c}
\includegraphics[width=0.33\linewidth]{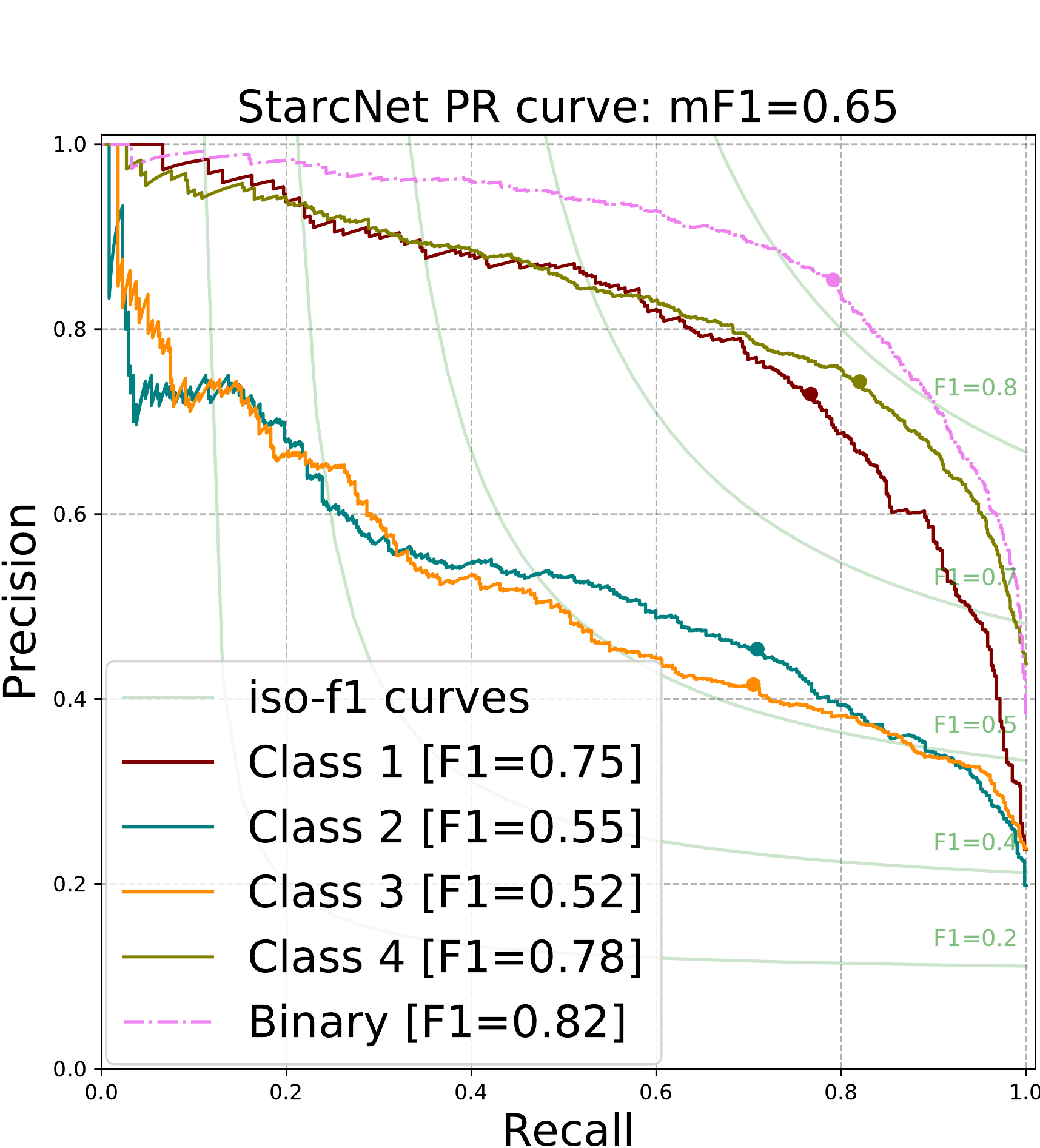}
\end{tabular}
\end{tabular}
\caption{\textbf{Performance of \starnet on the test set.} Confusion matrix normalized over the classes \revision{in \emph{test} set of the LEGUS dataset (20\% of the total sources or about 3000 objects)}. The rows show the distribution of the human--classified sources, while the columns are the predictions of \starnet. \revision{Parenthesis in the confusion matrix refers to the unnormalized values.} \revision{(Left)} Overall accuracy evaluated for 4 class classification using raw bands as input. \revision{(Middle) Results calculated with 2 classes (cluster/non-cluster classification). (Right) Precision-recall curves for each of the four classes, as well as for binary classification}. The overall accuracy is \textbf{68.6\%} \revision{with 4 classes and \textbf{86.0\%} with binary classification}.}
\label{fig:cm}
\end{center}
\end{figure}

As mentioned in \S~\ref{sec:materials}, the \emph{trainval} set (the set used for training and validation) contains 80\% of the total number of star  cluster candidates. 
We use 10 \% of the \emph{trainval} set as validation set to conduct hyper-parameter tuning and architecture choices. 
The \emph{test} set, containing the remaining 20\% of the total cluster candidates, is used for testing the pipeline. We carry out transfer learning experiments using state-of-the-art pre-trained models.  
We perform the evaluation using a confusion matrix normalized over the human classifications (rows) and the overall accuracy for comparison between models. \revision{In addition, we include precision-recall (PR) curves to show the trade-off between the precision and the recall for every possible cut-off in the prediction score for each class.} The performance of our model \starnet\ described in \S~3.2 is shown in Figure \ref{fig:cm}.
The overall accuracy of \starnet evaluated on the test set of the LEGUS dataset is 68.6\% \revision{when 4 classes are used and 86.0\% for binary classification.} \revision{In \S~4.2, we carry out hyper-parameter tuning and evaluate architecture choices experimentally using the validation set for calculating the accuracy, which yields different (slightly  lower) accuracy values from the test set.}

\subsection{Ablation studies}

\revision{Here we systematically evaluate the design choices for training \starnet. We define a modular architecture consisting of blocks with a convolution layer, a normalization layer, and an activation layer. We experimentally choose the filter size and number of convolution layers, the type of normalization, and activations. 
\newrevision{We experiment with different architecture depths, varying the number of blocks from 4 up to 12, with filters of sizes $3 \times 3$, $5 \times 5$, and $7 \times 7$ pixels. Furthermore, we experiment using batch and group normalization, with group sizes from 4 to 32, and ReLU and LeakyReLU activations.}
Our best initial model consists of 7 modules with convolution layers of 128 filters of size $3 \times 3$, group normalization \newrevision{(with a group size of 16)}, and LeakyReLU activations (with a pooling layer after the fourth module). Using our best initial model we perform tuning of the training hyper-parameters. 
\newrevision{We experiment with batch sizes from 16 to 256 samples and learning rate $\eta$ values from 1E-5 to 1E-2.}
We get the best performance when training with a batch size of 128 input arrays and a learning rate of $\eta$=1E-04. \newrevision{We prevent overfitting by controlling the complexity of our model, increasing our training set size with data augmentation, and adding regularization techniques like dropout layers~\citep{dropout}.}}

\revision{In addition, we experiment with different sizes and pre-processing of the input arrays, and data augmentation techniques. Lastly, we study the benefits of extracting features from different scales with multi-path architectures. We expand the description of our experiments in the remaining of this section and present a summary of the ablation studies in Table \ref{table:experiments}.}  \revision{The values in this Table should not be directly compared with the accuracy  quoted in  Figure~4; the accuracies in Table \ref{table:experiments} refer  to the validation set, while the accuracy  of  Figure~4  refers to the test set (Table~\ref{table:dataset}), which explains why the two accuracy values are slightly different.}

\paragraph{Size of input arrays} 
We test different input array sizes from $24\times 24\times 5$ (where the first two numbers refer to the number of pixels in the array, and subtend spatial scales  $\sim14-82$ pc, depending on galaxy distance) 
to $96\times 96\times 5$ ($\sim60-330$ pc). 
A smaller input makes for more efficient processing but might reduce the contextual information required to make an accurate prediction.
The best result is achieved by using input arrays of size $32\times 32 \times 5$ pixels ($\sim20 - 110$ pc ). Results using different input sizes are shown in Table \ref{table:experiments}a.

\paragraph{Input pre--processing} We consider various approaches to pre--processing of the input arrays (Table \ref{table:experiments}b), in order to test  whether any such approach improves the output accuracy. As mentioned in \S~3.1, transfer learning is generally applied to RGB inputs. We reduce the dimensionality of our input arrays from five to three, to create RGB arrays. In separate experiments, we  pre--process the input arrays by removing the galaxy's background, by inverting the gain, and by combining these two. In all cases, the unprocessed input arrays produce the highest output accuracy. 

\paragraph{Data augmentation} As shown in Table \ref{table:dataset}, the amount of objects per class is unbalanced. To balance the training dataset we apply horizontal and vertical (or both) reflections. After the training set is balanced, we apply data augmentation using scaling and rotations. The best performing model was obtained by data augmentation using scaling to a magnification of 1.07$\times$ with $50\%$ probability, and adding rotations of 90º and 270º (Table \ref{table:experiments}c). Resulting images after applying 180º rotations are the same as resulting images after applying reflections on both axes, therefore it is not used. 
The improvements provided  by scaling, resizing, and rotations are easily understood. Star clusters and compact associations do not have fixed projected sizes, and this characteristic is mimicked by small scaling and resizing transformations. Furthermore, the sources do not have fixed orientations, so adding rotations and reflections to the input arrays to increase their numbers help increase and diversify the input sample. Overall the benefits of data augmentation is significant, improving performance from 63.0\% to 67.9\%. 
After augmentation the training set consists of $\sim$115,000 sources, \revision{or about 10 times the original training  set (Table~\ref{table:dataset})}.

\paragraph{Architecture choices} Table~\ref{table:experiments}d shows the effect of varying the number of pathways in the network.
Compared to a single-path network the three-path network provides a 1.6\% improvement in accuracy  \revision{for 4--class classification, although the accuracy of the binary classification decreases slightly, by 0.8\%}. 

\begin{table}[!tb]
\centering
\caption{\textbf{\starnet ablation experiments on the validation set.} Ablation experiments showing the effect of different input sizes, preprocessing of the input, variations of CNN architectures, and data augmentation techniques. \revision{Results are presented using the overall accuracy on the \textit{validation} set for both the four-class and binary (cluster/non-cluster) classification}. \textbf{(a) }Quantitative results using different input image sizes, expressed in number of pixels. \textbf{(b) }Quantitative results using different pre-processing over 32$\times$32$\times$5 candidates. \textbf{(c) }Quantitative results using different techniques of data augmentation. Best results are obtained using data augmentation with scaling and rotations. Data augmentation with cropped arrays from the original input array reduces the performance of the model. Before performing the data augmentation we apply balancing on the training set using reflections. The multiplication factor over ``no data augmentation" is shown next to each data augmentation technique. \textbf{(d) }Quantitative results using different number of pathways for our model. 1-pathway CNN corresponds to a standard CNN.
}
     \subfloat[Input size]{\label{fig:sa}
        \begin{tabular}[t]{l c}
            \hline\hline
            Input size   &  Accuracy (\%)\revision{*}\\
            \hline
            24$\times$24$\times$5        & 61.8\revision{/83.7}            \\
            28$\times$28$\times$5        & 63.0\revision{/84.3}             \\
            32$\times$32$\times$5  & \textbf{67.9\revision{/85.5}}    \\
            48$\times$48$\times$5        & 65.1\revision{/83.8}             \\
            64$\times$64$\times$5        & 60.1\revision{/82.6}             \\
            96$\times$96$\times$5        & 57.3\revision{/80.1}             \\
            \hline
        \end{tabular}
        }\qquad
      \subfloat[Preprocessing]{\label{fig:sb}
        \begin{tabular}[t]{l c}
            \hline\hline
            Preprocessing     &  Accuracy (\%)\revision{*} \\
            \hline
            32$\times$32$\times$5 w/o preprocessing & \textbf{67.9\revision{/85.5}}  \\ 
            32$\times$32$\times$3 RGB image           &        63.3\revision{/83.1}            \\
            32$\times$32$\times$5 background removed   &        63.1\revision{/83.7}            \\
            32$\times$32$\times$5 gain inverse        &        64.4\revision{/84.2}            \\
            32$\times$32$\times$5 gain inverse + background removed &  63.1\revision{/83.3}  \\  
            \hline
        \end{tabular}
        }\qquad
      \subfloat[Data augmentation]{\label{fig:sd}
        \begin{tabular}[t]{l c c}
            \hline\hline
            Data augmentation   &  Accuracy (\%)\revision{*} \\
            \hline
            No data augmentation                     & 63.0\revision{/83.0}              \\
            Scaling only ($\times$2)                        & 64.1\revision{/84.3}              \\
            Rotations only ($\times$3)                      & 65.1\revision{/84.6}              \\
            Scaling and rotations ($\times$6)      & \textbf{67.9\revision{/85.5}}   \\
            Scaling, rotations, and cropping ($\times$30)   & 64.8\revision{/82.6}              \\ 
            \hline
        \end{tabular}
        }\qquad
     \subfloat[CNN architecture]{\label{fig:sc}
        \begin{tabular}[t]{l c}
            \hline\hline
            CNN architecture   &  Accuracy (\%)\revision{*}\\
            \hline
            1-pathway        & 66.3\revision{/86.3}            \\
            2-pathway       & 67.2\revision{/86.3}             \\
            3-pathway (\starnet) & \textbf{67.9\revision{/85.5}} \\  
            \hline
        \end{tabular}
        }\qquad
        \begin{flushleft}
        \revision{* Values correspond to 4-way (left) and binary classification (right) accuracy.} 
        \end{flushleft}
\label{table:experiments}
\end{table}

\paragraph{Transfer learning} 
To use transfer learning with a model trained on ImageNet we have to adapt our input (size and number of channels/bands) to what the models expect, which is typically a $224 \times 224 $ image with three channels (e.g., RGB), and replace the last layer of the CNN to predict the four categories of our application. A commonly used strategy is to only train the parameters of the last layer, or to   allow updating the entire network but with a small learning rate. 
To adapt the input we rescale our $32 \times 32$ input arrays of five bands to a size of $224 \times 224$ using bilinear interpolation. 
To adapt the five bands of the HST to the three RGB channels, we weight each image by the band's photometric zeropoint 
and combine them ($UV$ with $U$ to get blue channel, and $V$ with $I$ to get red channel). 
Table \ref{table:transfer} shows results using deep networks currently popular for image understanding. The number of parameters vary from 6.8--123 million and the models have been pretrained on the ImageNet dataset.
We obtained the best results using GoogleNet \citep{googlenet} as the network architecture 
and training over the entire network parameters. 
However, the performance is still below the proposed \starnet.

\begin{table}[!t]
\centering
\caption{\textbf{Transfer learning experiments on the validation set.} Results on the \textit{validation} set using ImageNet pre-trained networks. Left column indicate the network architecture used, middle column shows the number of parameters of each network, and right column the accuracy \revision{for the four class and binary  classification}. The last row shows the network developed in this work, on which no transfer learning was applied. \revision{\S~5.1 and \S~5.4 discuss confidence intervals for the accuracy.}}
\begin{tabular}{l c c}
\hline\hline
Network architecture       & \# Parameters      & Accuracy (\%)           \\
\hline
AlexNet \citep{Krizhevsky2012}          & 61M               & 63.0\revision{/84.7}              \\
VGG16 \citep{VGG}                       & 138M              & 65.3\revision{/84.2}              \\
VGG19\_BN \citep{VGG}                   & 144M              & 65.7\revision{/86.5}              \\
ResNet18    \citep{resnet}              & 12M               & 63.2\revision{/85.7}              \\
ResNet34    \citep{resnet}              & 22M               & 62.3\revision{/84.5}              \\
ResNet50    \citep{resnet}              & 26M               & 65.5\revision{/85.0}             \\
SqueezeNet1\_1 \citep{squeezenet}       & 1.2M              & 64.6\revision{/85.4}             \\
GoogleNet \citep{googlenet}             & 6.8M              & 66.3\revision{/85.4}     \\
ShuffleNet\_v2\_x1\_0 \citep{shufflenet}& 2.3M              & 63.1\revision{/84.1}              \\
DenseNet161 \citep{densenet}            & 29M               & 64.1\revision{/82.2}              \\
\starnet \textbf{(This work -- no transfer learning)}    & \textbf{15M}      & \textbf{67.9\revision{/85.5}}     \\
\hline
\label{table:transfer}
\end{tabular}
\end{table}

As shown in Table \ref{table:transfer}, the best result using transfer learning is worse than the best result that uses training from scratch (which is this work's approach). We speculate that this outcome is due to  the fact that by combining the five bands into three channels we lose the intrinsic information of each independent band.  \\

\section{Discussion}  \label{sec:discussion}
\subsection{\starnet\ accuracy}
We state in \S~\ref{sec:experiments} that we reach $68.6\%$ level accuracy when classifying LEGUS sources with our \starnet\ algorithm. In order to evaluate this performance we need to compare it to the agreement achieved by the \textit{human} classifiers. In fact, as the classification given by the LEGUS experts is used to train \starnet, their agreement act as an upper value achievable by our model. As reported in the confusion matrix in Figure~\ref{fig:cm_human} (right panel), LEGUS classification experts agree with the final classification, used to train and test the model, at an average value of $75\%$ for four classes and 87\% for two classes. As a reminder, this accuracy is higher than the true one, since the final classification includes the results from individual classifiers. For comparison, pairs of individual classifiers generally agree at the level of 57\% for four classes and 78\%  for two classes  (Figure~\ref{fig:cm_human}, left panel).  
It is therefore reasonable that the \starnet\ model cannot surpass the $\lesssim75\%$/87\% (4/2 classes)  level of accuracy, given that the training/testing sample is itself at that level. 
\revision{In order to estimate a confidence interval for the accuracy, we calculate the results of \starnet using bootstrapping (random sampling with replacement) on the training and validation sets ten times and average the result. We obtain an accuracy uncertainty of $\pm 0.8\%$ for 4 class classification and of $\pm 0.7\%$ for binary classification.} 
It is worth noticing that the accuracy mentioned is not uniform within classes or within galaxies, as we are going to discuss more in details in the following paragraphs.

\subsubsection{Accuracy by class}
\begin{figure}
    \centering
    \includegraphics[width=0.49\textwidth]{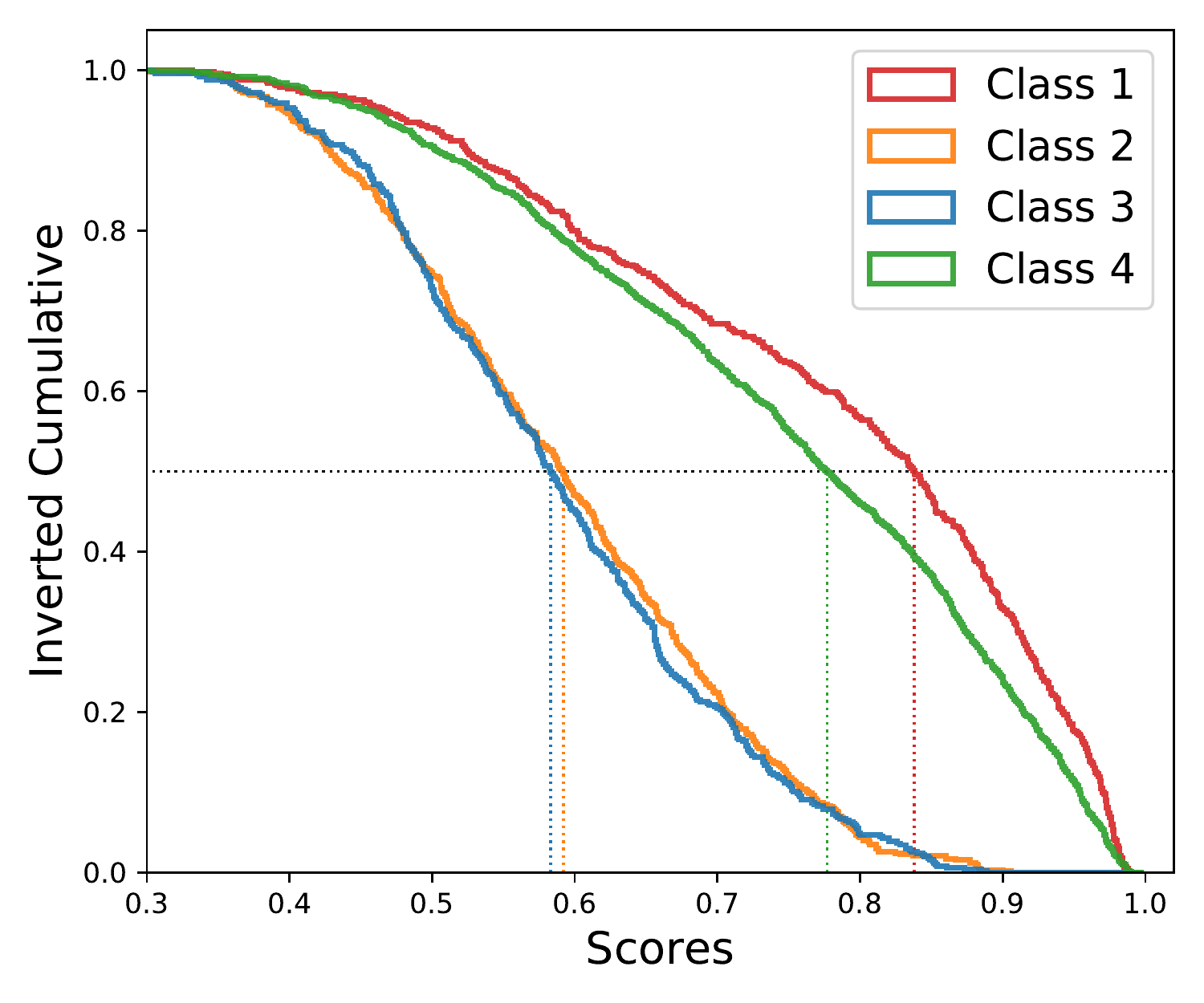}
    \caption{Cumulative distribution of the scores (ranging from 0 to 1) assigned by \starnet, divided by class. Only the highest score of each source is considered, which corresponds to the score of the predicted class. The vertical lines mark the median score for each of the classes.}
    \label{fig:scores}
\end{figure}

The larger difficulty encountered by human classifiers in recognizing class~2 and 3 sources (Figure~\ref{fig:cm_human}) is reflected in a lower accuracy given by the \starnet\ predictions in those two classes (Figure~\ref{fig:cm}). A high fraction of class 3 sources ($\sim40\%$) are misinterpreted as class 4 by \starnet, and a lower but considerable fraction ($\sim15\%$) as class 2. Similarly, almost half of class 2 sources are predicted by \starnet\ as class 1, 3 or 4, with a slight preference for class 1 ($\sim20\%$).
The difficulty in classifying class 2 and 3 sources can be also traced in the probability distributions given by \starnet\ to each classification. \starnet\ assigns a score (from 0 to 1) to each class whenever it runs a prediction (see Figure~\ref{fig:cnn}). We show in Figure~\ref{fig:scores} the scores assigned to each of the predictions in the test sample. Such scores can be interpreted as the ``confidence'' of \starnet\ for each classification, since when a source falls clearly in one of the classes it will receive a score close to 1, while when its classification is very uncertain it will have a score distributed across more than 1 class, and the final predicted class will have a score much lower than 1 (close to 0.25 if there is uniform uncertainty among all classes). Figure~\ref{fig:scores} indicates that \starnet\ is on average very ``confident'' when assigning a class 1, with $50\%$ of the sources predicted as class 1 receiving a score higher than $0.8$. On the other hand, for $50\%$ of the sources classified as either class 2 or 3 the score is lower than $0.6$, indicating a high degree of uncertainty. This behavior closely traces the accuracy retrieved by the confusion matrix in Figure~\ref{fig:cm}.
In addition to revealing the level of accuracy in each class, the scores can be used in a more practical way to select sub-samples of the catalogs, namely selecting on the base of the ``confidence'' in the classification. It must be noticed, however, that selecting only sources with a score above a certain limit, would bias the sample against class 2 and 3 sources.  

\subsubsection{Accuracy by galaxy distance}\label{sec:accuracy_distance}
The major difference among the galaxies of our sample, in terms of the possible effects on the training of \starnet, is their distance. The distance of a galaxy is inversely proportional to the angular size of star clusters, and therefore to the number of pixels subtended by each source.
In \S~\ref{sec:experiments} we tested different sizes in pixels for the input arrays used to train the model, finding that images of $32\times32$ pixels result in the best final accuracy. Our galaxies span the distance range $\sim$3--18 Mpc and 32 pixels (at the scale of 0.04$^{\prime\prime}$/pix) are equivalent to physical sizes $\sim20-110$ pc (Table~\ref{tab:legus_starcnet_class}). Clusters have on average effective radii of $2-3$ pc \citep{ryon2015,ryon2017} and are therefore fully contained in the $32\times32$ pix cutouts even in the closest galaxies. This remain true even for the magnified arrays, which subtend 6--7~pc in the closest  galaxies. In the case of distant galaxies it can be argued the cutouts contain a large fraction of the cluster surroundings, that may act as noise for the classification process. 

\begin{figure}
    \centering
    \includegraphics[width=0.95\textwidth]{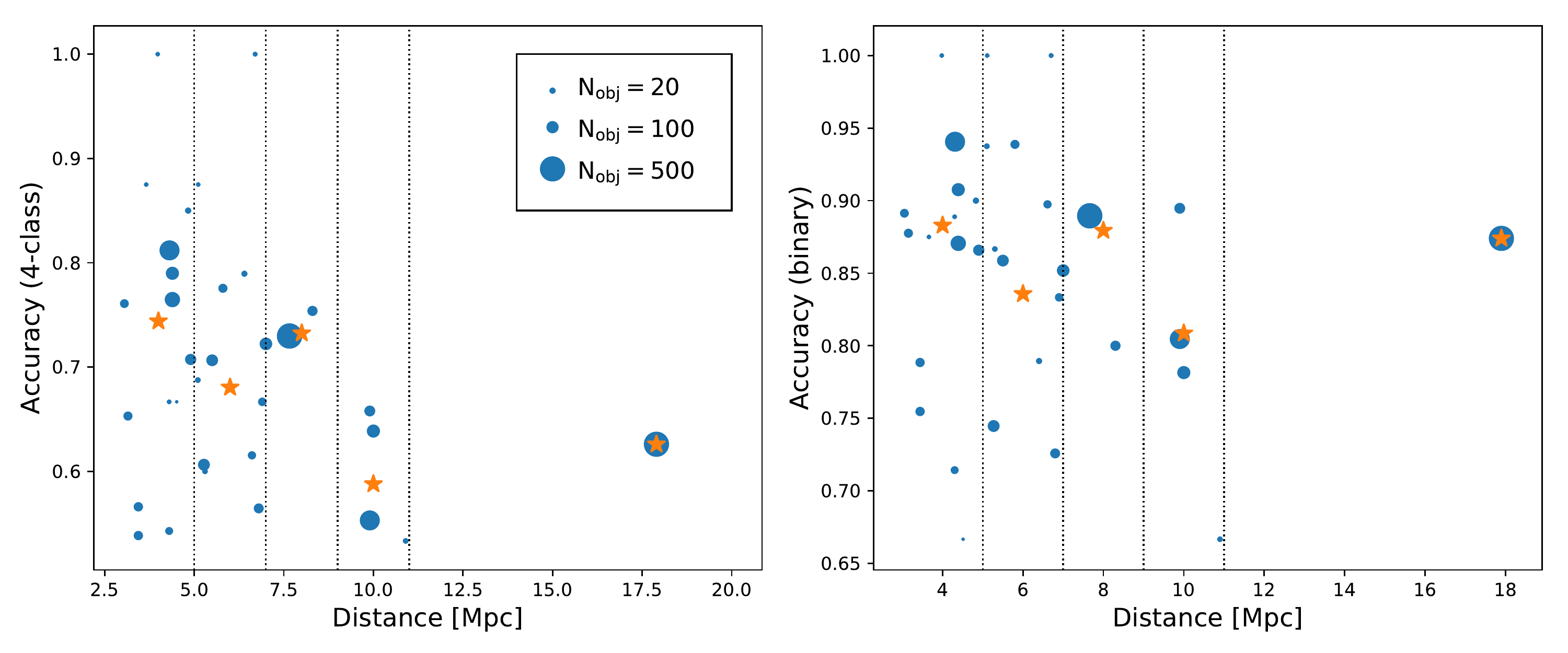}
    \caption{\revision{Accuracy by galaxy plotted against the galaxy distance (blue circles). The number of clusters is coded by the size of markers. Orange stars represent the weighted averages in distance bins (delimited by vertical dotted lines). Accuracies are calculated over the \textit{test} set. The left panel refer to 4 class classification, in the right panel to the binary classification.}}
    \label{fig:accuracy_distance}
\end{figure}

\revision{In order to check the impact of distance, \starnet\ accuracies for single galaxies are plotted against their distances in Figure~\ref{fig:accuracy_distance}.
We test for possible correlations using the Spearman's rank correlation test, but we do not find evidence for any (coefficient $\rho=-0.2$, $p_{value}=0.2$), not even considering a binary classification ($\rho=-0.2$, $p_{value}=0.3$). 
Accuracies cover the range from $\sim0.5$ to $1.0$ and  galaxies with few sources appear to  drive most of the scatter in accuracy  (Figure~\ref{fig:accuracy_distance}). We account for the sample size of each galaxy by calculating a weighted accuracy in 5 distance bins, using distance limits at 5, 7, 9 and 11 Mpc; in this way every bin contains $\sim500$ sources (the first bin contains $\sim1000$ sources). The distance-weighted accuracies are shown as orange stars in Figure~\ref{fig:accuracy_distance}. 
The Spearman's test suggests the presence of a distance--accuracy anti--correlation ($\rho=-0.8$), but  with low significance ($p_{value}=0.1$), which could be caused by having only 5 data points. In the case of a binary classification, the anti--correlation is weaker ($\rho=-0.5$), and still with low significance ($p_{value}=0.4$). 
}

One of the possible causes of the large scatter retrieved in the accuracies is the disagreement among classifiers. Even if the model were able, in principle, to label sources with $100\%$ accuracy, the measured accuracy would be lower because the human classifiers can only achieve $\lesssim 75\%$ level of internal accuracy over four classes. Thus, galaxies with lower overall \textit{human} accuracy could result in a lower recovered \starnet\ accuracy. We tested this possibility using the \revision{Spearman's test} to compare the \starnet\ accuracy with the human agreement; the result is a correlation coefficient $\rho=0.3$ with a $p_{value}=0.4$, \revision{indicating that the data cannot confirm the presence of a correlation. We conclude that we have no evidence that our results are impacted by this effect. Using the binary classification yields a similar result ($\rho=0.5$, $p_{value}=0.1$).}

\subsection{Accuracy for binary classifications (clusters vs non-clusters)}\label{sec:binary_class}
\revision{Throughout the paper we mentioned binary accuracies by merging together class 1 and 2 as `clusters' and class 3 and 4 as `non clusters'.}
To account for variations in the literature on what is considered a YSC, we also consider the case that classes 1, 2 and 3 are `clusters' and only class 4 are `non--clusters' (as, e.g., in \citealp{chandar2014}). Re--arranging the confusion matrix of the binary classification in Figure~\ref{fig:cm} results in an accuracy of $80.1\%$, \revision{lower than the accuracy  (86.0\%)  obtained  with  our default definition of binary classification}. Finally, some studies consider class 1 and 2 sources together as clusters, and keep class 3 and class 4 as separate classes (e.g. \citealt{adamo2017}). In this case \starnet\ would result in an accuracy of $74.4\%$. 

So  far, we have measured the accuracy for a binary classification starting from an algorithm trained on a 4-class labels. We can instead directly train the model using a binary classification. For this test we keep  the architecture of \starnet\ fixed and we use the same training and testing samples as described in the previous sections. The only difference is that we aggregate the class 1 and 2 sources in the `cluster' category and the class 3 and 4 sources in the `non-cluster' category prior to performing any training. Running our pipeline from start to finish under these conditions yields a final accuracy of $86.0\%$, identical to what obtained when training on 4 classes and only aggregating the final outputs (see Figure~\ref{fig:cm}). Repeating the experiment with classes 1, 2 and 3 as `clusters' and class 4 as `non-clusters' yields an  accuracy of $80.0\%$, again identical to training with 4 separate classes (see above). We conclude that training our model on four-- or two--class classification does not improve its final accuracy for binary classification, which remains  $\gtrsim$80\%.

\subsection{\revision{Classification heterogeneity among human classifiers}}
\label{sec:refining}

\revision{The LEGUS approach of obtaining classifications from several human  classifiers raises the issue of heterogeneity in the training catalogs. We thus  test  whether removing classifications with the largest disagreement among human  classifiers can lead to an improvement of the overall performance of \starnet. 
 As described in \S~\ref{sec:materials}, we train \starnet using the \emph{mode} of the various individual human classifications. In addition to the \emph{mode}, the LEGUS  catalogs report the \emph{mean} of the classifications for each object. For the present experiment, we leverage  both \emph{mode} and \emph{mean}, to train \starnet only on those objects where the $|mean - mode| \leq \epsilon$.  This mimics the situation of higher agreement among different human classifiers. We performed experiments varying the value of $\epsilon$ from $0.1$ to $3.0$ (the latter number is the maximum variance we expect over four classes). For every value of $\epsilon$ the classification accuracy of \starnet was worse than using all objects. The best result of this experiment was $65\%$ using a value of $\epsilon=2.0$.} \revision{This value is lower than the highest accuracy  we achieve when using {\em the entire training set}, irrespective of the (dis)agreement among human classifiers. The result in this section underscores that samples of classified clusters are still small in size, and reducing the number of training sets decreases the performance of automatic classification algorithms more drastically than including discrepant classifications.}

\subsection{Comparison with other algorithms in the literature}
\citet{grasha2019b} and \citet{grasha2019} presented an early attempt at developing a ML-driven cluster classification scheme based on a bagged decision tree algorithm, trained on a small sample of eight LEGUS  catalogs (all those available at the time) and applied to the galaxy NGC~5194. The results from the ML classifications were used in detailed analyses of the characteristics of the star cluster population in \citet{messa2018a,messa2018b} and \citet{grasha2019}. \revision{\citet{messa2018a} also compared the catalog of NGC~5194 with other existing catalogs, concluding  that the main  differences consisted in the quality of the data \citep{bastian2005} and in the definition of star cluster \citep{chandar2016}.} 

The major drawback of the ML classification of  star clusters described  in \citet{grasha2019b} and \citet{grasha2019} is the incapability of the algorithm to recognize class 3 sources, which were mostly labeled as class 4. For this reason, only class 1 and class 2 sources were included in all analyses using that catalog.
In NGC~5194 only \revision{47} class 3 sources were found by the bagged decision tree algorithm when classifying $\sim8400$ sources without human labeling ($0.6\%$). As a term of comparison, $\sim15\%$ of the sources with human labels in NGC~5194 were class 3 clusters.  
\starnet is an important improvement over that first attempt, in recognizing class 3 sources with better accuracy. A test run on NGC~5194 recovered \revision{$\sim10\times$} more class 3 sources relative to the earlier algorithm.

\begin{figure}[t]
\begin{center}
\begin{tabular}{ccc}
    \includegraphics[width=0.31\linewidth]{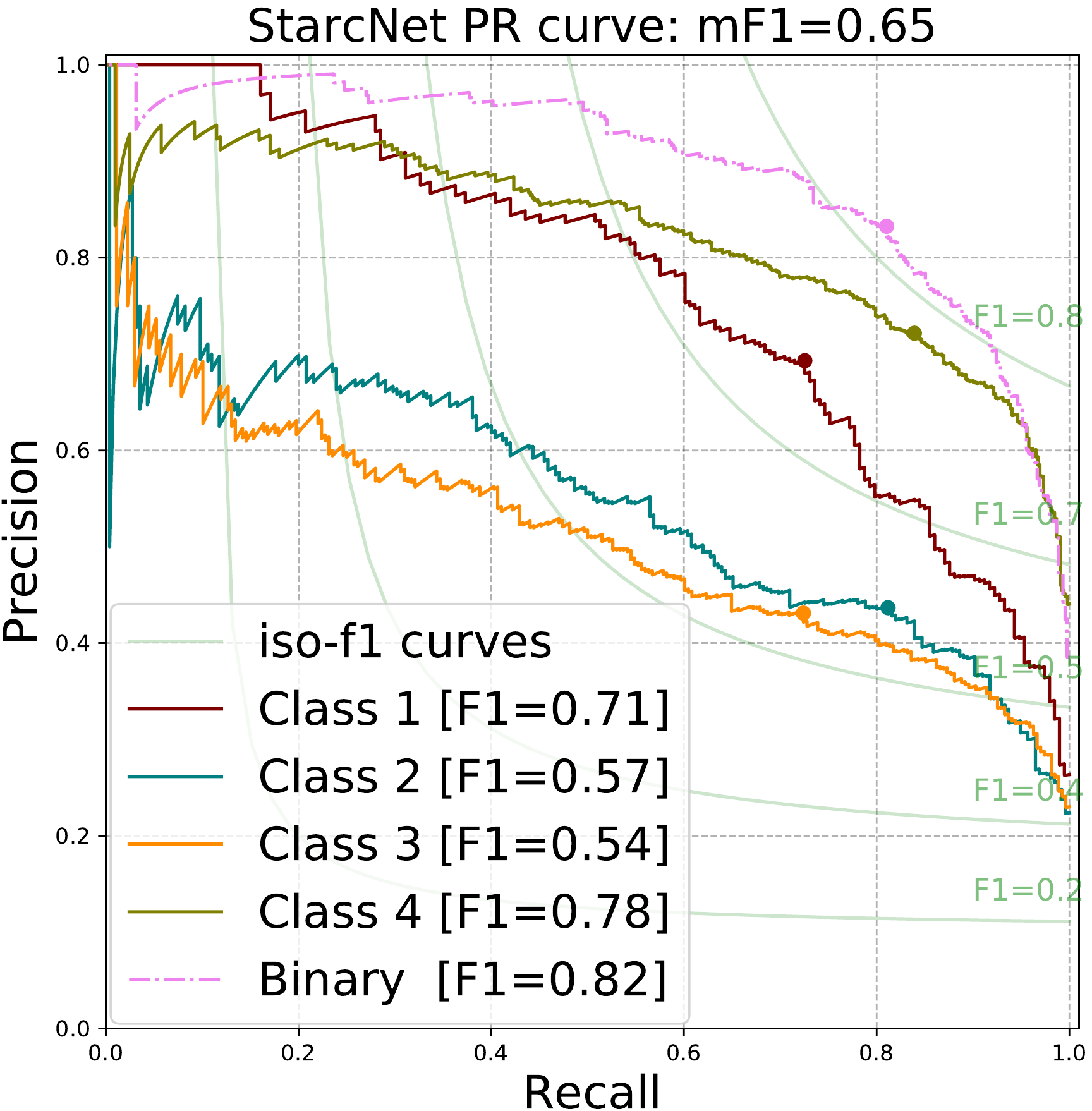}  &  
    \includegraphics[width=0.31\linewidth]{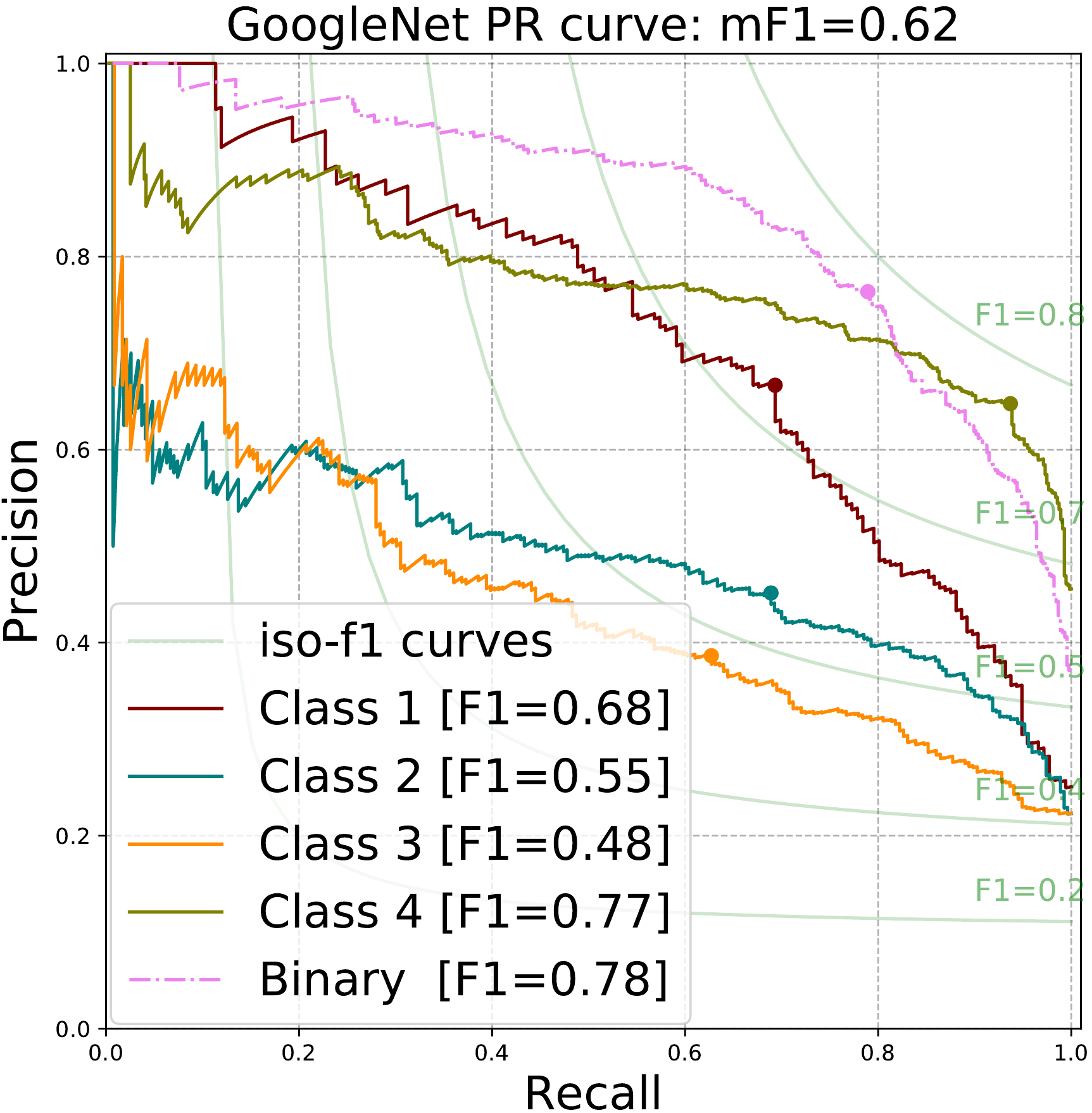} &
    \includegraphics[width=0.31\linewidth]{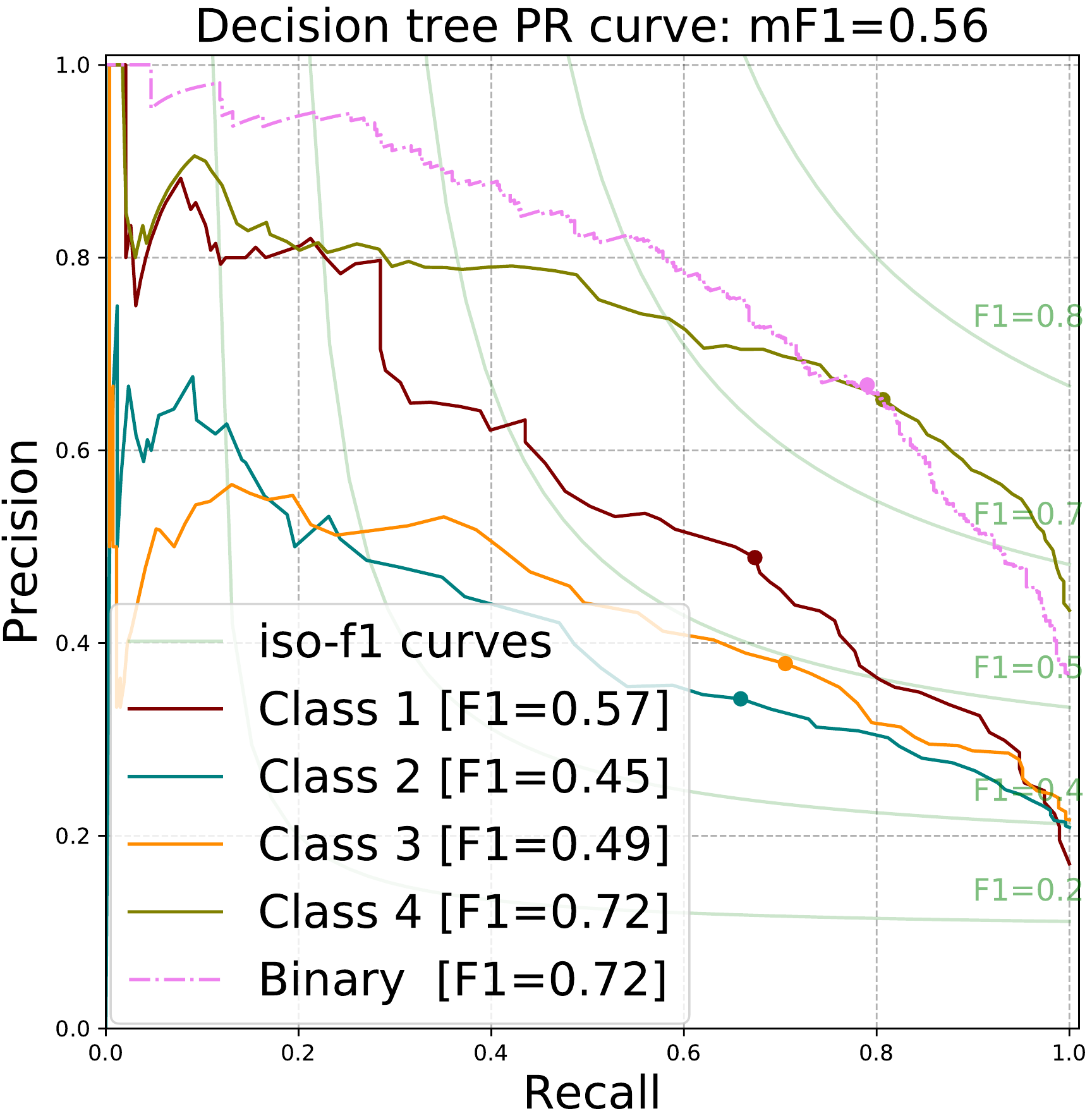}  \\
\end{tabular}
\caption{\label{fig:PRall}
\revision{Precision-recall curves for four class classification and binary (cluster/non-cluster) classification over the \textit{validation} set. Results using: (left) \starnet, (middle)  GoogleNet, and (right) the bagged decision tree approach of \cite{grasha2019}.}
}
\end{center}
\end{figure}

\revision{For a direct  and internally consistent  comparison between the two approaches, we trained a bagged decision tree model as described in \cite{grasha2019}, using our larger set of LEGUS catalogs. The model includes 100 trees, trained on object-centered patches of size $15 \times 15$ pixels corresponding to the HST filter bands F336W, F555W, and F814W, and data augmentation by rotations of 90, 180, and 270 degrees. We train the model using \texttt{ensemble.BaggingClassifier} of the \texttt{scikit-learn} library with a decision tree base estimator using the same train and validation set objects used to train \starnet. We choose the best split on each tree node using ``Gini impurity" as the function to measure its quality. The maximum depth of the tree is set to default (i.e. the nodes are expanded until all leaves are pure or until all leaves contain less than 2 samples).
As in \cite{grasha2019}, we train the model using a $1 \times 675$ vector with raw pixel values (i.e. after reshaping the $15 \times 15 \times 3$ patch). The accuracy of the model evaluated over our validation set is $57.1\%$ with four class classification (compared to $67.9\%$ achieved with \starnet) and $78.4\%$ with binary classification (compared to $85.5\%$ with \starnet). The PR curves for this model are shown in Figure \ref{fig:PRall}.}

\revision{Additionally, we train bagged decision tree models by varying the input size. First, use a vector of size $1\times3072$ ($32\times 32\times 3$) instead which corresponds to the input spatial size used for  \starnet. Next, we use a $1\times 5120$ ($32\times 32\times 5$) input which includes the remaining 2 bands not included in \cite{grasha2019} (F275W and F435W). Both cases lead to a reduction in accuracy ---  $56.4/81.3\%$ and $56.8/85.5\%$ respectively for four--class/binary classification. However, we get a small improvement ($58.2/78.6\%$) when training the model using an input vector of size $1\times 1125$ ($15\times 15\times 5$), which matches the spatial size as in \cite{grasha2019} but uses all the bands.} 

More recently, \citet{wei2020} applied deep transfer learning to the classification of star clusters. They utilized the same LEGUS 4-classes classification scheme and the training of their model was based on LEGUS images. It is therefore worthy a comparison between their results and the ones achieved by \starnet. 
The \citet{wei2020}'s experimental framework differs from ours in two important aspects. First, we have included every LEGUS galaxy with published cluster photometric catalogs, with the classifications provided by at least  three \emph{human} classifiers per source. \revision{This increases by 38\% our set of human--classified sources relative to \citet{wei2020}'s. Although larger datasets may introduce more variability in the classifications, the approach makes sense when training a general discriminator. \citet{wei2020} also used a second, smaller (10 galaxies, about 5,000 sources) dataset, obtained from classifications performed by a single human classifier; this approach provides more internally--consistent classifications throughout the entire sample but introduces the potentially systematic bias of the classifier. The second difference between the two works is that} \starnet\ is a custom model trained from randomly initialized weights in contrast to the transfer learning model used by \citet{wei2020}. Using a custom model gives us freedom regarding the architecture design which allows us to use directly the LEGUS  photometric information. 
\revision{We investigate this second aspect, by quantifying the difference in performance between the two architectures.} 

\revision{An exact comparison is not possible as the code and experimental setup to reproduce \citet{wei2020}'s  experiments is not publicly available.} In order to compare our approach to theirs, we run their algorithms using our catalogs, which provide a consistent platform for evaluating performance. Table \ref{table:transfer}  shows results using transfer learning with different  architectures. \citet{wei2020} tested both VGG19-BN and ResNet18; those algorithms applied to our more diverse catalogs yield accuracies of 65.7\% and 63.2\%, respectively, while \starnet\ yields an accuracy of 67.9\%. Thus, the accuracy achieved with \starnet\ \revision{on a 4--class classification} is higher than using transfer learning. \revision{For binary classification the accuracies are 86.5\%, 85.7\%, and 85.5\%, respectively, closer to each other than the 4--class case (Table \ref{table:transfer}).}

\revision{Our best result for the 4--class classification from transfer learning is with the  GoogleNet architecture, where we obtain an overall accuracy of 66.3\% (Table \ref{table:transfer}). We show PR curves with the results over the validation set of \starnet and GoogleNet in Figure \ref{fig:PRall}. 
As we can see, \starnet consistently achieves better precision for all recall values compared to transfer learning with GoogleNet for all the four classes and binary classification.}

\revision{We estimate confidence intervals for GoogleNet using bootstrapping on the training and validation sets ten times and average the result, as done previously for \starnet. We obtain an uncertainty on the accuracy of $\pm 0.7\%$ for 4 class classification and $\pm 0.6\%$ for binary classification (to be compared with $\pm 0.8\%$ for 4 class classification and $\pm 0.7\%$ for binary classification obtained for \starnet).}

\subsection{Applying \starnet\ to galaxies outside the training sample}
\revision{\subsubsection{Leave-one-out test}
We study how \starnet\ performs on galaxies that are not part of the training sample. As first
test, we re--train \starnet, with the only difference that we leave one of our galaxies completely out of the training sample. We then use this newly trained \starnet\ to make predictions on that same galaxy. In this way we are treating the selected galaxy as if it were a galaxy outside the LEGUS sample. We repeat this process, in turn, for all the galaxies in our sample. 
We show in Figure~\ref{fig:CM_ngc3344} (left panels) the confusion matrix obtained as the mean of the confusion matrices of all the galaxies estimated with the new training setup. Note that the matrices in this case were evaluated over the validation sample and the accuracy of the mean matrix ($66.6\%$ for four  classes; $84.2\%$ for binary classification) should be compared to the accuracy over the validation sample for the reference training, as reported in Table~\ref{table:experiments} ($67.9$\% for four  classes; $85.5$\% for binary classification). 
For $50\%$ of the galaxies, the accuracy does not change compared to the one obtained from the reference \starnet\ training. The accuracy decreases with the new training scheme for $30\%$ of the galaxies while it actually increases for the remaining $20\%$ of the sample. The change in accuracy given by the mean matrix  ($-1.3\%$ with respect to the reference accuracy) suggests that, on average, the exclusion of a galaxy from the training sample does not heavily affect \starnet\ predictions.}
 
\revision{We show in Figure~\ref{fig:CM_ngc3344} (central panels) the confusion matrix for one of the LEGUS galaxies, NGC3344, which can be considered an average galaxy of the sample, both in terms of number of clusters (557 labelled sources; 449 used for the \emph{trainval} set) and of distance (7 Mpc). 
The accuracy of NGC3344 using \starnet\ trained on the default set (i.e. without excluding any galaxy from the training set) is 62.1\% for four--class classification, below the \starnet\ overall one. Leaving this one galaxy out of the training sample and then running predictions on it does not change the resulting accuracy.
For other galaxies in the sample this test produce variations in accuracies up to $\pm20\%$. In the case of galaxies with few sources, variations are expected,  due to low--number statistics. For galaxies with a numerous cluster population, their exclusion implies losing a consistent fraction of the training sample and, as seen in \S~\ref{sec:refining}, a reduction in the training sample size leads to worse accuracies.}

\subsubsection{NGC 1512+1510}
As a further test, we consider a galaxy in the LEGUS sample that does not have manually classified sources, and therefore is 
not included in either the training, validation, or testing sets. We focus on the galaxy interacting pair NGC 1512+1510 made of a barred-spiral and a dwarf galaxy at a distance of $11.6$ Mpc \citep{legus}. We consider a single source catalog made by the merging of the catalogs for the two galaxies.

The total number of sources in the merged catalog is 906, all classified using \starnet. Independently, 300 sources in the catalog, drawn to cover randomly the entire range of positions and luminosities of the parent sample, were classified by three \textit{human} classifiers (co-authors of this paper). \revision{Only one of these three classifiers had taken part in the LEGUS classification previously; the other two were trained to classify sources according to the LEGUS scheme described in \S~\ref{sec:materials}.}
The \textit{human} classification was performed without any knowledge of the ML classification and, at the same time, each classifier worked independently of the other two. The three independent classifications were merged into the final \textit{human} classification using the same methodology of the LEGUS project \citep{adamo2017}. 
The overall agreement of the \textit{human} classifiers amongst themselves is $54.3\%$, $64.0\%$ and $65.0\%$, respectively 	(considering pairs of two classifiers). The agreement between each classifier and the final classification is $77.7\%$, $76.7\%$ and $87.3\%$, for the three humans, respectively.  The agreement percentages were calculated in the same way as in \S~\ref{sec:experiments}, i.e. weighting the number of sources in each class (e.g. see Figure~\ref{fig:cm}). As noted before, the higher agreement between individual classifiers and final class is due to the  final  class including the classifications of all classifiers. 
The great majority of sources ($59.3\%$) are in class 4, stressing again the necessity of cleaning automatic catalogs of spurious entries. The other sources are distributed among the remaining classes as shown in Table~\ref{tab:1512_labels}. 

\revision{When we compare the ML predictions with the human classification for the 300 sources in common, we find an overall agreement of $58.7\%$, lower than the overall accuracy of \starnet, but consistent with the agreement among human classifiers (see above) }
The confusion matrix in Figure~\ref{fig:CM_ngc3344} (right panels) reveals a lower accuracy for class 2 and 4 compared with the one found when testing \starnet. 
We stress again the fact that, like humans, \starnet\ struggles between class 3 and class 4 and between class 1 and class 2. These distinctions however become less relevant when a binary classification (cluster/non-cluster) is considered, improving the accuracy to a much higher $83.3\%$. 
\revision{We point out that, while \starnet\ accuracy for this galaxy is well below the average value, it is consistent with the accuracies of galaxies at similar distances, as seen in \S~\ref{sec:accuracy_distance}. 
We conclude that the tests done in this section suggest that the accuracy of \starnet\ when classifying galaxies outside the training sample is consistent with its reference accuracy.}

\begin{table}
    \centering
    \begin{tabular}{lrrrrrr}
    \hline\hline
    \     & \multicolumn{2}{c}{Human classification} & \multicolumn{2}{c}{\starnet\ prediction} & \multicolumn{2}{c}{\starnet\ prediction}  \\
    \hline
    Class 1     & $42$   & ($14.0\%$)  & $65$  & ($21.7\%$) & 188 & ($20.8\%$)\\
    Class 2     & $55$   & ($18.3\%$)  & $34$  & ($11.3\%$) & 101 & ($11.1\%$)\\
    Class 3     & $25$   & ($8.3\%$)   & $68$  & ($22.7\%$) & 201 & ($22.2\%$)\\
    Class 4     & $178$  & ($59.3\%$)  & $133$ & ($44.3\%$) & 416 & ($45.9\%$)\\
    Total       & $300$ & \   & $300$ & \ & $906$ & \  \\
    \hline
    \end{tabular}
    \caption{Classifications of the sources in NGC 1512+1510 given by the mode of the human classification and by the \starnet\ predictions. For the predictions we report the statistics both for the 300 sources with human classifications (center) and for the entire sample of 906 sources  (right).}
    \label{tab:1512_labels}
\end{table}

\begin{figure}
\begin{center}
\begin{tabular}{c c c}

\begin{tabular}{ @{\hspace{-67pt}}c >{\hspace{-0.4em}\bfseries}c @{\hspace{-0.8em}}c @{\hspace{-1.2em}}c @{\hspace{-1.2em}}c @{\hspace{-1.2em}}c}
  \multirow{17.5}{*}{\rotatebox{90}{\parbox{5.1cm}{\bfseries\centering Human Classification}}} 
  & & \multicolumn{4}{c}{\bfseries Mean} \\
  & & \multicolumn{4}{c}{\bfseries \starnet Prediction } \\
  & & \hspace{1.0em} \bfseries Class 1 & \hspace{1.0em} \bfseries Class 2 & \hspace{1.0em} \bfseries Class 3 & \hspace{1.0em} \bfseries Class 4 \\
  & \multirow{3}{*}{1} & 
  \MyBox{39}{0.71} & \MyBox{10}{0.18} & \MyBox{1}{0.02} & \MyBox{5}{0.09} \\
  & \multirow{3}{*}{2} & 
  \MyBox{15}{0.19} & \MyBox{39}{0.51} & \MyBox{13}{0.17} & \MyBox{10}{0.13} \\
  & \multirow{3}{*}{3} & 
  \MyBox{2}{0.03} & \MyBox{11}{0.17} & \MyBox{33}{0.51} & \MyBox{19}{0.29} \\
  & \multirow{3}{*}{4} & 
  \MyBox{7}{0.04} & \MyBox{10}{0.06} & \MyBox{22}{0.12} & \MyBox{138}{0.78} \\
\end{tabular}

\noindent
\renewcommand\arraystretch{1.0}
\begin{tabular}{ @{\hspace{-80pt}}c >{\hspace{-0.4em}\bfseries}c @{\hspace{-0.8em}}c @{\hspace{-1.2em}}c @{\hspace{-1.2em}}c @{\hspace{-1.2em}}c}
  & & \multicolumn{4}{c}{\bfseries NGC 3344} \\
  & & \multicolumn{4}{c}{\bfseries \starnet Prediction } \\
  & & \hspace{1.0em} \bfseries Class 1 & \hspace{1.0em} \bfseries Class 2 & \hspace{1.0em} \bfseries Class 3 & \hspace{1.0em} \bfseries Class 4 \\
  & \multirow{3}{*}{1} & 
  \MyBox{70}{0.76} & \MyBox{15}{0.16} & \MyBox{1}{0.01} & \MyBox{6}{0.07} \\
  & \multirow{3}{*}{2} & 
  \MyBox{8}{0.08} & \MyBox{47}{0.49} & \MyBox{33}{0.35} & \MyBox{7}{0.07} \\
  & \multirow{3}{*}{3} & 
  \MyBox{2}{0.02} & \MyBox{10}{0.08} & \MyBox{90}{0.74} & \MyBox{20}{0.16} \\
  & \multirow{3}{*}{4} & 
  \MyBox{1}{0.01} & \MyBox{4}{0.03} & \MyBox{63}{0.45} & \MyBox{72}{0.51} \\
\end{tabular}

\noindent
\renewcommand\arraystretch{1.0}
\begin{tabular}{ @{\hspace{-80pt}}c >{\hspace{-0.4em}\bfseries}c @{\hspace{-0.8em}}c @{\hspace{-1.2em}}c @{\hspace{-1.2em}}c @{\hspace{-1.2em}}c}
  & & \multicolumn{4}{c}{\bfseries NGC 1512+1510} \\
  &  & \multicolumn{4}{c}{\bfseries \starnet Prediction} \\
  & & \hspace{1.0em} \bfseries Class 1 & \hspace{1.0em} \bfseries Class 2 & \hspace{1.0em} \bfseries Class 3 & \hspace{1.0em} \bfseries Class 4 \\
  & \multirow{3}{*}{1} & 
  \MyBox{35}{0.83} & \MyBox{1}{0.02} & \MyBox{0}{0.00} & \MyBox{6}{0.14} \\
  & \multirow{3}{*}{2} & 
  \MyBox{22}{0.40} & \MyBox{15}{0.27} & \MyBox{3}{0.05} & \MyBox{15}{0.27} \\
  & \multirow{3}{*}{3} & 
  \MyBox{0}{0.00} & \MyBox{3}{0.12} & \MyBox{18}{0.72} & \MyBox{4}{0.16} \\
  & \multirow{3}{*}{4} & 
  \MyBox{8}{0.04} & \MyBox{15}{0.08} & \MyBox{47}{0.26} & \MyBox{108}{0.61} \\
\end{tabular}
 \\
\noindent
\renewcommand\arraystretch{1.0}
\begin{tabular}{ @{\hspace{-50pt}}c >{\hspace{-0.4em}\bfseries}c @{\hspace{-0.8em}}c @{\hspace{-3.0em}}c}
  \multirow{10}{*}{\rotatebox{90}{\parbox{2.3cm}{\bfseries\centering Human Classification}}} & 
    & \multicolumn{2}{c}{\bfseries \starnet Prediction}  \\
  & & \hspace{1.0em} \bfseries C & \hspace{1.0em} \bfseries NC \\
  & \multirow{3}{*}{C} & 
  \MyBox{103}{0.78} & \MyBox{29}{0.22}  \\
  & \multirow{3}{*}{NC} & 
  \MyBox{30}{0.12} & \MyBox{212}{0.88}  \\
\end{tabular}
  
\noindent
\renewcommand\arraystretch{1.0}
\begin{tabular}{ @{\hspace{-60pt}}c >{\hspace{-0.4em}\bfseries}c @{\hspace{-0.8em}}c @{\hspace{-3.0em}}c}
  \multirow{10}{*}{\rotatebox{90}{\parbox{2.3cm}{\bfseries\centering Human Classification}}} & 
    & \multicolumn{2}{c}{\bfseries \starnet Prediction}  \\
  & & \hspace{1.0em} \bfseries C & \hspace{1.0em} \bfseries NC \\
  & \multirow{3}{*}{C} & 
  \MyBox{140}{0.75} & \MyBox{47}{0.25}  \\
  & \multirow{3}{*}{NC} & 
  \MyBox{17}{0.06} & \MyBox{245}{0.94}  \\
\end{tabular}

\noindent
\renewcommand\arraystretch{1.0}
\begin{tabular}{ @{\hspace{-60pt}}c >{\hspace{-0.4em}\bfseries}c @{\hspace{-0.8em}}c @{\hspace{-3.0em}}c}
  \multirow{10}{*}{\rotatebox{90}{\parbox{2.3cm}{\bfseries\centering Human Classification}}} & 
    & \multicolumn{2}{c}{\bfseries \starnet Prediction}  \\
  & & \hspace{1.0em} \bfseries C & \hspace{1.0em} \bfseries NC \\
  & \multirow{3}{*}{C} & 
  \MyBox{73}{0.75} & \MyBox{24}{0.25}  \\
  & \multirow{3}{*}{NC} & 
  \MyBox{26}{0.13} & \MyBox{177}{0.87}  \\
\end{tabular}
\end{tabular}

\caption{\revision{\textbf{\starnet performance on leave-one-out-experiments.} Confusion matrices for leave-one-out experiments on the \emph{trainval} set. (Left) Mean confusion matrix of all LEGUS galaxies in the leave-one-out experiments. The overall agreement is \textbf{66.6\%} for four  classes (top-left) and   \textbf{84.2\%} for binary classification (bottom-left). (Middle) Confusion matrix for the human classification and the ML predictions of 449 sources in NGC 3344. The overall agreement is \textbf{62.1\%} for four classes (top-middle) and  \textbf{85.7\%} for binary classification (bottom-middle). (Right) Confusion matrix for the human classification and the ML predictions of 300 sources in NGC 1512+1510. The overall agreement is \textbf{58.7\%} for four  classes (top-right) and \textbf{83.3\%} for binary classification (bottom-right).}}
\label{fig:CM_ngc3344}
\end{center}
\end{figure}

\section{Tests on cluster properties}  \label{sec:clusterscience}
Previous LEGUS studies suggest that the morphological classification of star clusters is linked to physical differences among classes. Clusters show younger ages and smaller masses with increasing class number (from 1 to 3, \citealt{grasha2015,grasha2017a,adamo2017,messa2018a}), on average. In addition, class 3 sources show a stronger degree of clustering at small spatial scales, compared to class 1 and 2, suggesting that they are still distributed according to the hierarchical structure typical of young ($<40-50$ Myr) star forming regions \citep{grasha2015,grasha2017a}. Conversely, class 1 sources are distributed more uniformly than class 2 or 3, which indicates the clusters are old enough to have had enough time to disperse from their natal area, i.e. are several tens to a few hundreds of Myr in age.
In this section we test whether the classification performed with our network \starnet\ maintains the same observed trends in the cluster population properties as the classifications performed by humans. 

\subsection{Overall distributions of cluster properties}\label{sec:CS1}
\begin{figure*}
\centering
\subfloat{\includegraphics[width=0.61\textwidth]{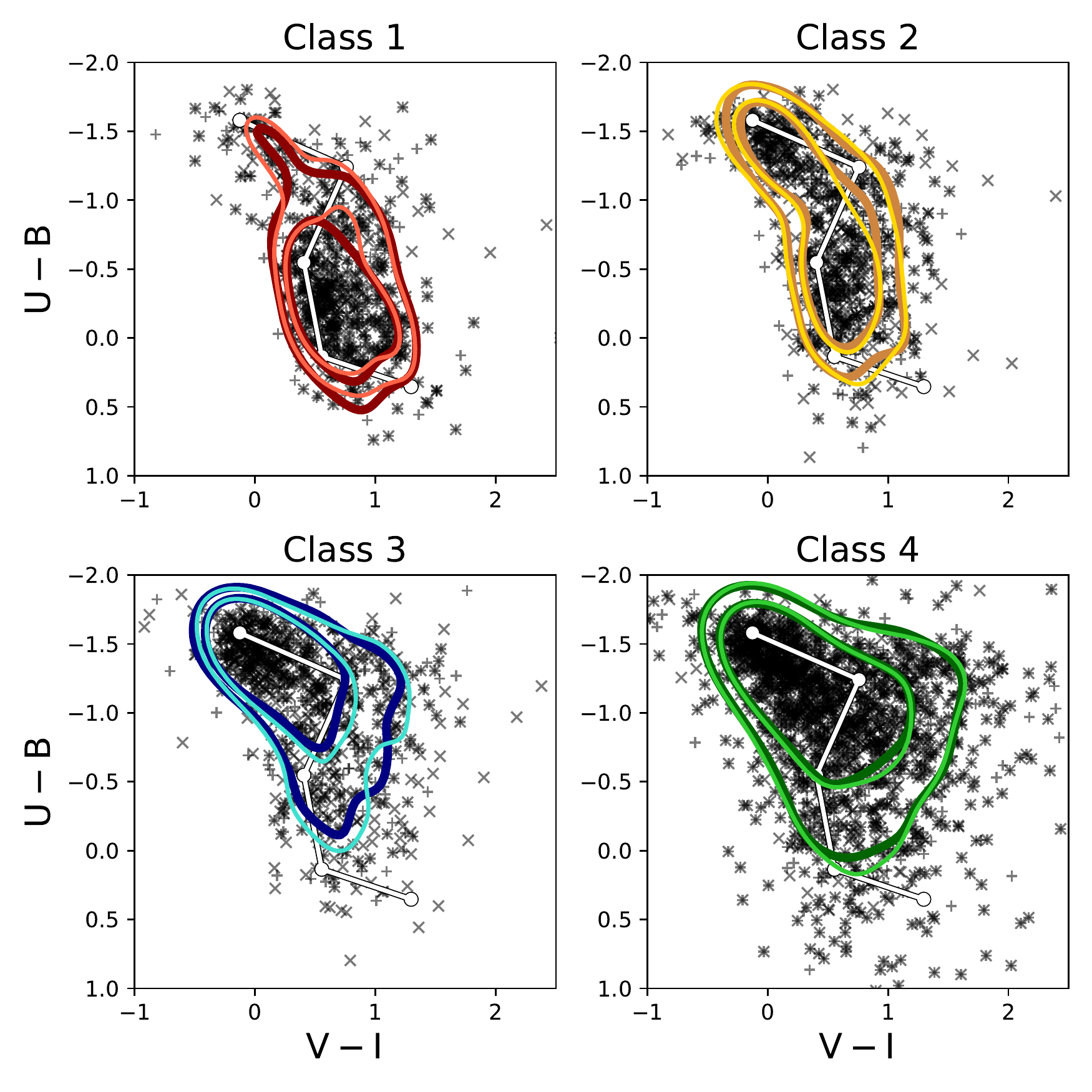}}
\subfloat{\includegraphics[width=0.37\textwidth]{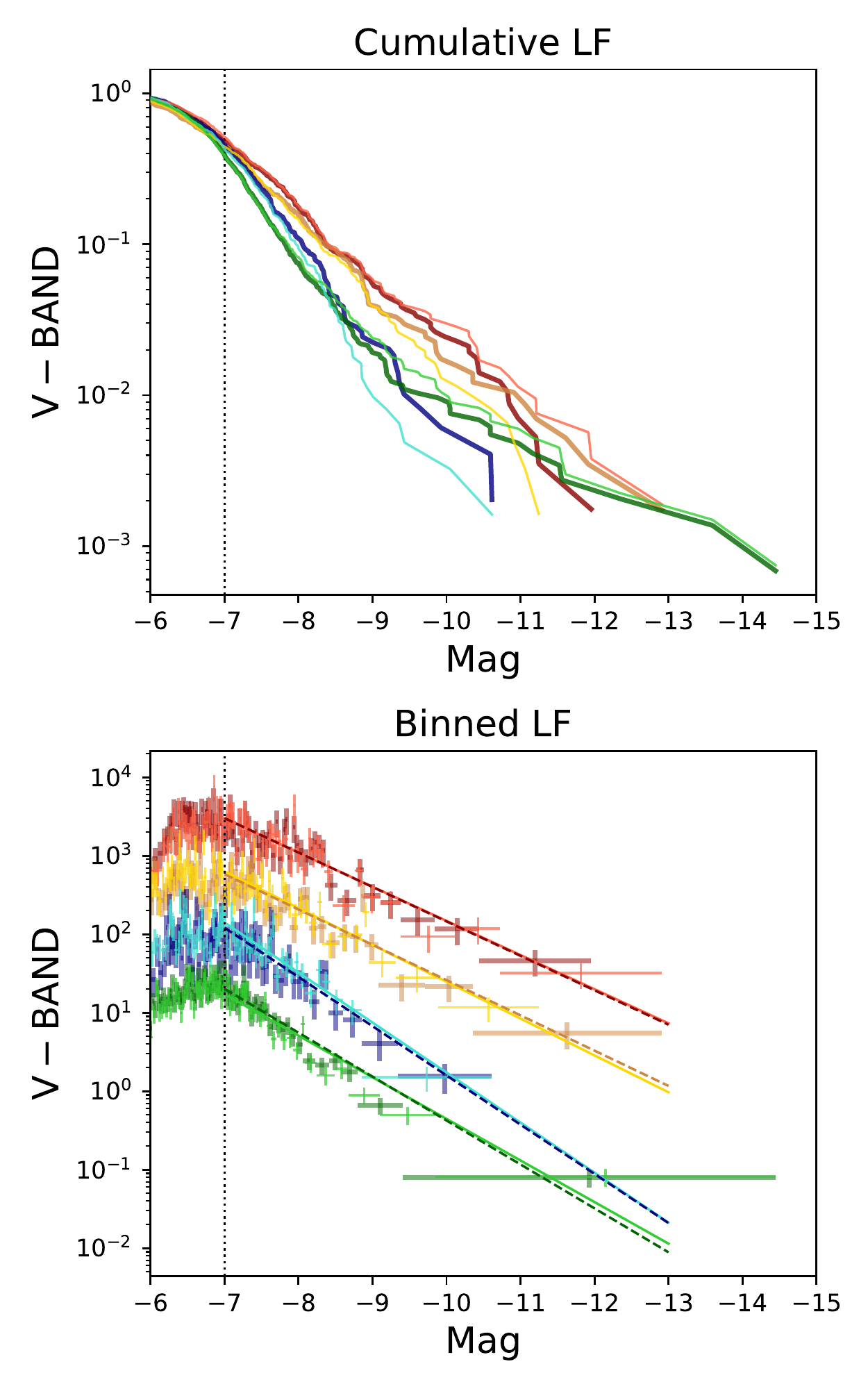}}\\
\caption{\revision{Properties of the \textbf{test} set divided by class. 
Class 1, 2, 3 and 4 are color-coded as red, yellow, blue and green respectively (light colors for the human-classified classes and dark colors for the predictions given by \starnet). 
The left panels shows color-color diagrams with contours enclosing $50\%$ and $75\%$ of the sources. Stellar evolutionary tracks for Padova-AGB models with solar metallicity indicate the evolution of color with age, from 1 Myr to 9 Gyr (white broken lines and circles). Crosses symbols ($\times$) are used for the clusters with human labeling, plus symbols ($+$) are used for clusters with \starnet\ predictions.
The right panels show the luminosity function both in the cumulative (top) and in the binned (bottom) form, with the best fits over-plotted as dashed (\starnet) and solid (human) lines; the fit was performed using only the bins brighter than $-7$ mag (black dotted vertical lines).} }
\label{fig:test1}
\end{figure*}

\begin{figure*}
\centering
\subfloat{\includegraphics[width=0.95\textwidth]{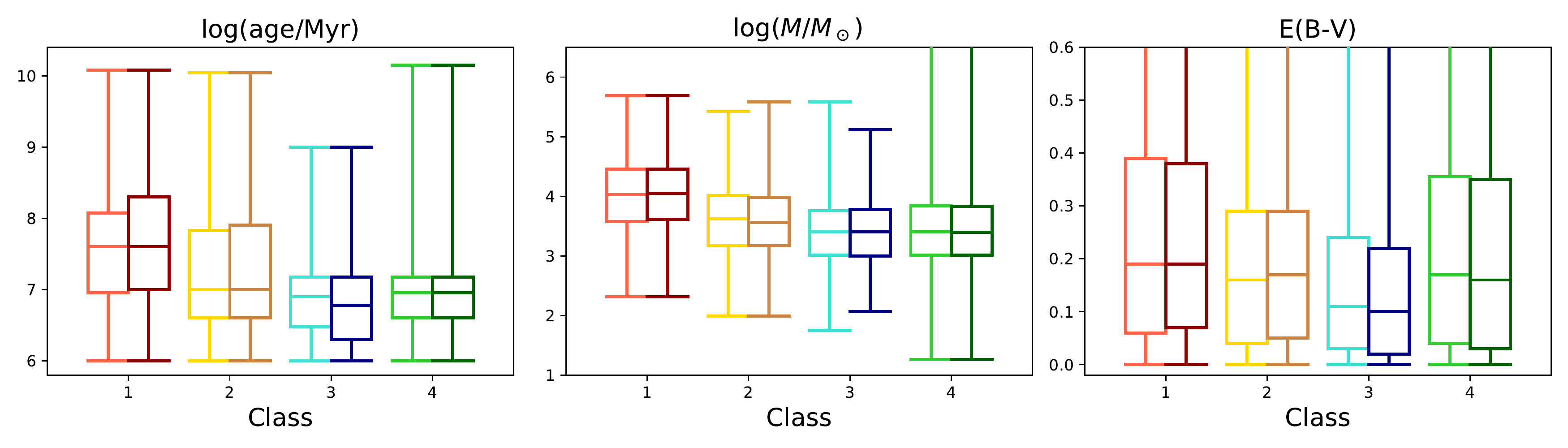}}\\
\subfloat{\includegraphics[width=0.62\textwidth]{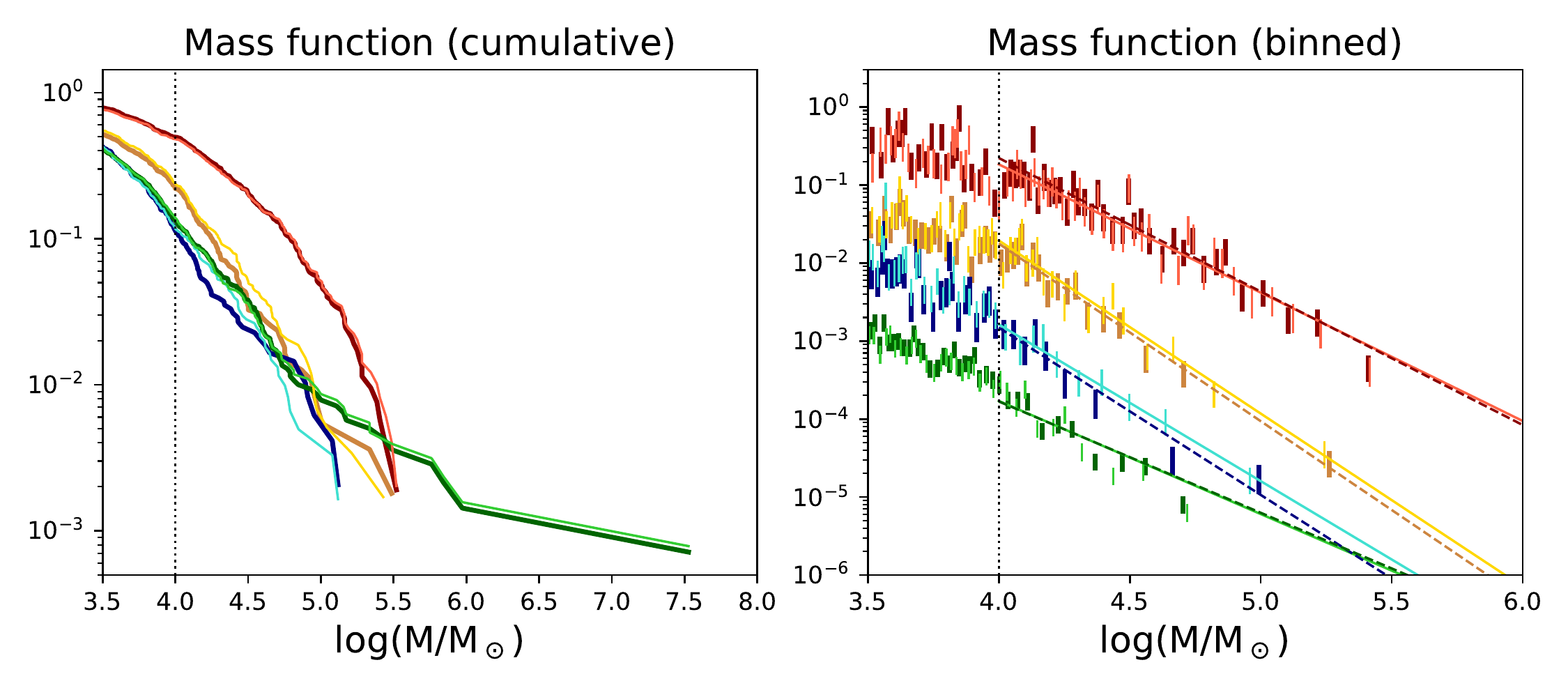}}
\subfloat{\includegraphics[width=0.35\textwidth]{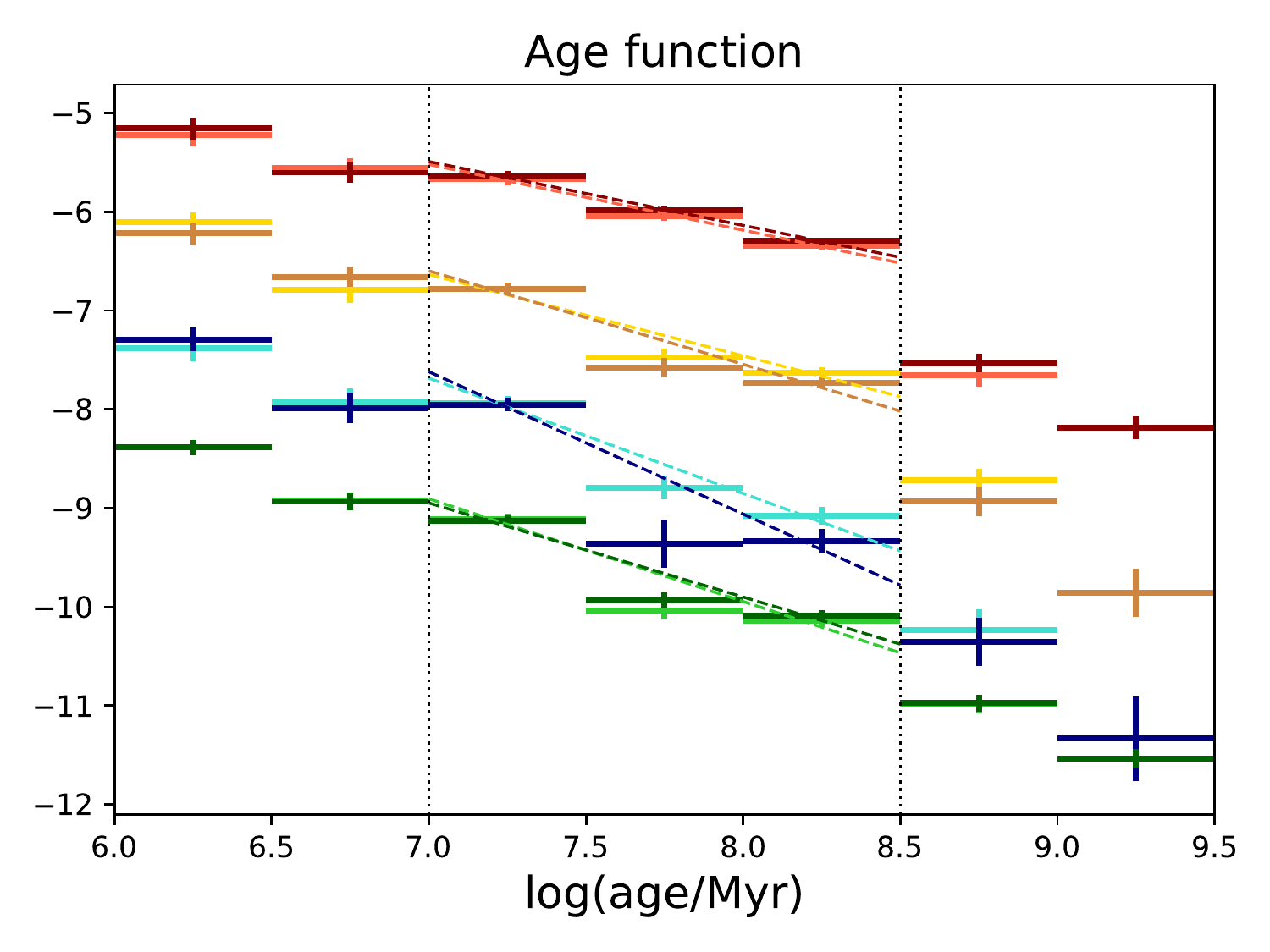}}
\caption{\revision{Physical properties of the \textbf{test} set divided by class. The same colors as Figure~\ref{fig:test1} are used. In the top row, the distribution of ages, masses and extinctions are shown as box-and-whisker plots. The distributions of the extinctions have a maximum value at $\rm E(B-V)=1.5$ mag in all classes. The bottom row shows the mass functions in the cumulative form (left panel), and in the binned form (central panel) and the age functions (right panel). Results of the best--fit slopes are over-plotted as dashed (for \starnet--predicted) and solid (for human--labelled) lines. The mass function was fitted down to masses of $\rm 10^4\ M_\odot$ in order to avoid incompleteness. For the same reason, and in order to avoid contaminants, the age functions were fitted in the age range between 10 and $\sim300$ Myr.}}
\label{fig:test2}
\end{figure*}

\revision{We consider the test sample, consisting of a little over 3,000 classified sources across 31 galaxies, the same used to test the performance of the \starnet\ classification, resulting in an overall accuracy of $68.6\%$ and the confusion matrix of Figure~\ref{fig:cm}. We report in Table~\ref{tab:test_labels} the distribution of the source classifications in this set. 
We compare the distributions of the main properties of cluster in this set, dividing them by class according to their human label and to the \starnet\ prediction, as shown in Figure~\ref{fig:test1} and Figure~\ref{fig:test2}.}

\revision{The photometric cluster properties are summarized in Figure~\ref{fig:test1}, which shows the  $V-$band luminosity functions (LF, defined as the number of clusters per luminosity bin, and usually modeled as a power law, $\rm{LF}\equiv dN/dL\propto L^{-\alpha}$) and the color--color diagrams. 
Classes labeled by humans and predicted by \starnet\ follow the same trends; in details:
\begin{itemize}
    \item the overall colors of the clusters move from red (bottom--right of the color--color diagrams) to blue (top--left) going from class 1 to class 3. Class 4 sources spans the entire range of colors. Colors are related to the cluster age, as highlighted by the stellar tracks on the color--color diagrams of Figure~\ref{fig:test1}, and therefore the evolution of colors suggests an evolution of cluster ages with class;
    \item the luminosity functions steepen going from class 1 to class 3, meaning that there are, on average, brighter sources in class 1 than in class 3. The luminosity function of Class 4 sources is steep at the low-luminosity end but then becomes the shallowest at the bright end, exhibiting double power-law shape. As a reminder, class~4 sources are not star cluster candidates, but the `rejects' (non-clusters). Luminosity functions are plotted using their two most popular parametrisations. The binned luminosity functions are fitted using a power-law (down to $-7$ Mag) and the two slopes found for each class agree with each other within $1\sigma$.
\end{itemize}}

\revision{The main physical cluster properties are summarized in the top row panels of  Figure~\ref{fig:test2}, which show the distribution of ages, masses and extinctions as a function of cluster class. Mass and age functions, with equivalent definition as the luminosity function, are shown in the bottom row of Figure~\ref{fig:test2}.
Again, the distributions obtained with human and \starnet\ classifications show consistent trends:
\begin{itemize}
\item Class 1 clusters are older and more massive than class 2, which in turn are older and more massive than clusters in class 3. Class 4 distributions have median values similar to the ones in class 3 but with larger scatter. The same trends are retrieved using the mass and age functions. The mass function steepens going from class 1 to class 3. In the case of class 4 sources  the shape of the mass function is similar to the luminosity function's one described above. Also age functions become steeper going from class 1 to class 3, with class 4 having a slope similar to class 2. In the case of both mass and age functions, the slopes obtained by fitting the human and \starnet\ classifications agree with  each other within $1\sigma$ in each class;
\item Extinction distributions move to lower values going from class 1 to class 3, in contrast to what found by for the median values by \citet{grasha2015}. The difference in median $E(B-V)$ values is however less than $0.1$ mag. Class 4 sources have a similar distribution as those in class 1. 
\end{itemize}
We conclude that the overall trends of cluster photometric and physical properties are not affected by considering \starnet\ predictions.}

\begin{table}
    \centering
    \begin{tabular}{lrrrr}
    \hline\hline
    \     & \multicolumn{2}{c}{Human label} & \multicolumn{2}{c}{\starnet prediction}   \\
    \hline
    Class 1     & $528$   & ($17.1\%$)  & $569$  & ($18.4\%$) \\
    Class 2     & $612$   & ($19.8\%$)  & $575$  & ($18.6\%$) \\
    Class 3     & $617$   & ($19.9\%$)  & $493$  & ($15.9\%$) \\
    Class 4     & $1338$  & ($43.2\%$)  & $1458$ & ($47.1\%$) \\
    Total       & $3095$  & \           & $3095$ & \          \\
    \hline
    \end{tabular}
    \caption{Classifications of the sources in the test samples, given by the mode of the human classification and by the \starnet\ predictions.}
    \label{tab:test_labels}
\end{table}


\subsection{Mis--classified clusters}\label{sec:CS2}
\begin{figure*}
\centering
\subfloat{\includegraphics[width=0.90\textwidth]{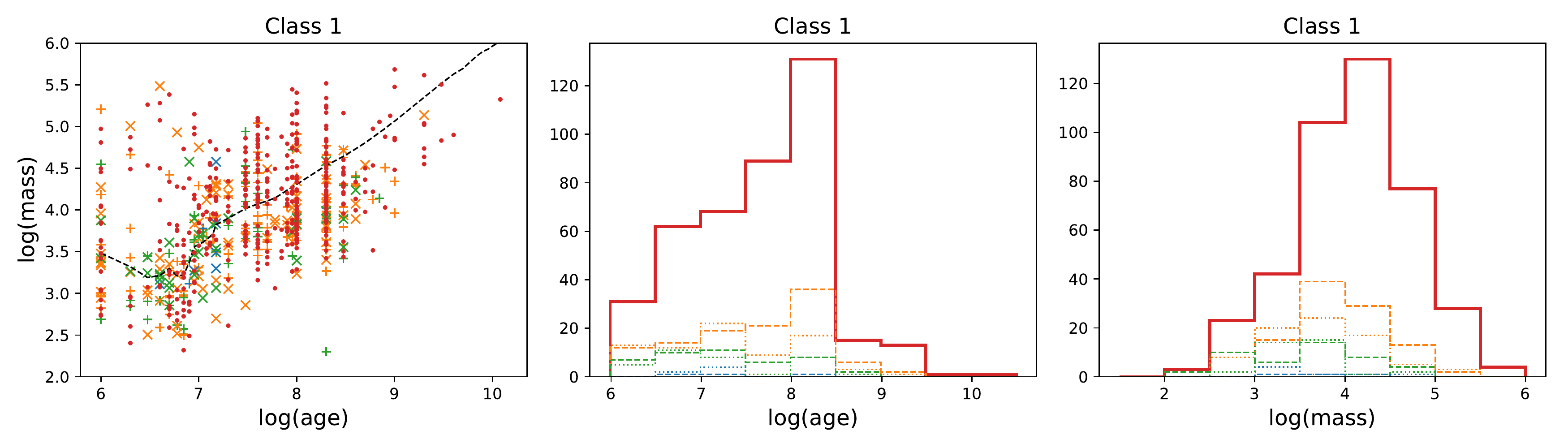}}\\
\subfloat{\includegraphics[width=0.90\textwidth]{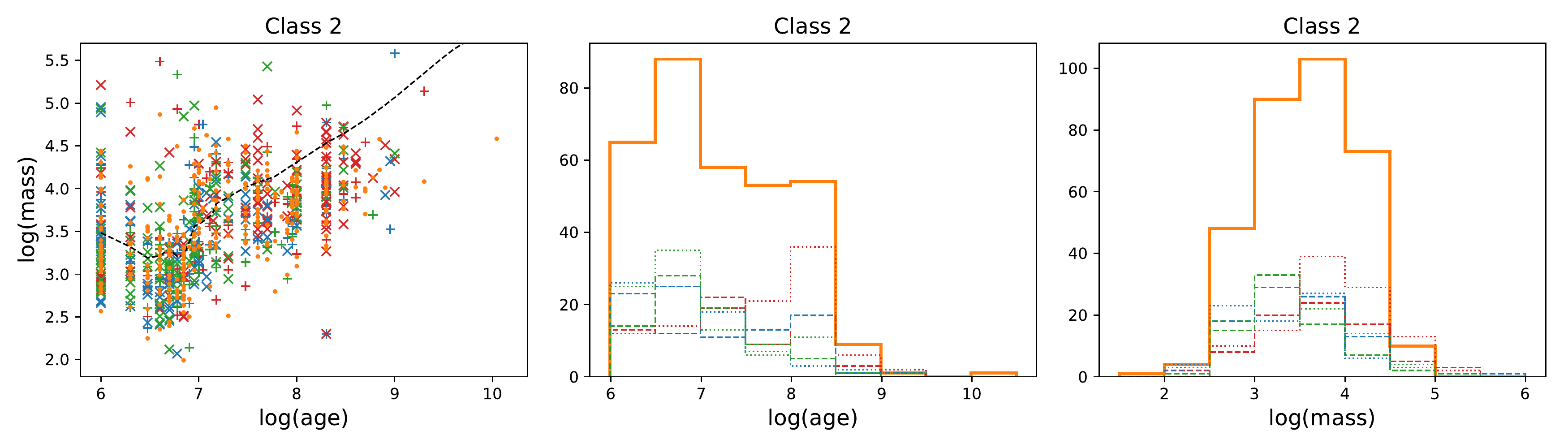}}\\
\subfloat{\includegraphics[width=0.90\textwidth]{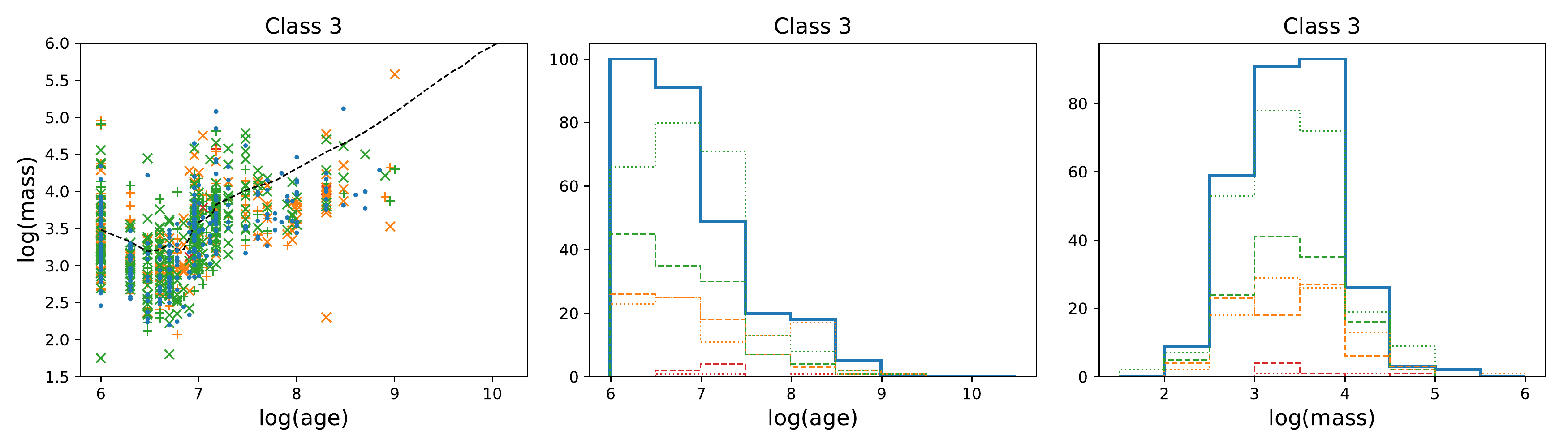}}\\
\subfloat{\includegraphics[width=0.90\textwidth]{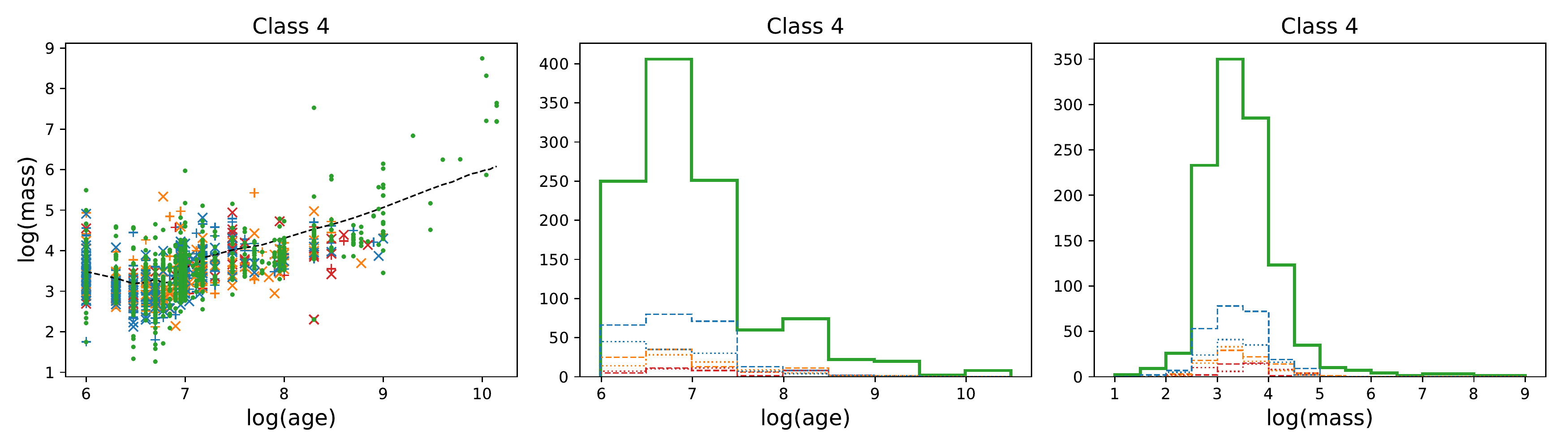}}\\
\caption{(Left column): age--mass plots showing sources with correct predictions (filled circles), false positives ($+$) and false negatives ($\times$). We use red color for class 1, orange for class 2, blue for class 3 and green for class 4. The dashed black lines indicate the limit of $\rm Mag_V=-8$. (Second and third columns): distribution of correct classifications (solid lines) FPs (dashed) and FNs (dotted) in the age and mass spaces.}
\label{fig:FPFN}
\end{figure*}

\revision{
Having discussed the overall distribution of properties, we focus in detail on how the populations of false positives (FPs) and false negatives (FNs) from \starnet\ predictions are distributed. We report, in the left column of Figure~\ref{fig:FPFN}, the age--mass distribution of clusters in each class, using different markers for the clusters correctly predicted, the FPs, and the FNs. We also report histograms of the distributions of FPs and FNs both in the age and mass spaces. Figure~\ref{fig:FPFN} suggests the following:
\begin{description}
\item[Class 1] We find a relatively high fraction of mis--classifications at the lowest masses; when we consider the high-mass end only, most of the mis--classifications are with class 2 clusters (both FPs and FNs). We do not observe a clear trend with age;
\item[Class 2] the most massive clusters in this class are actually FPs, from class 1 but also from the other classes, across all ages; 
\item[Class 3] we see many FPs (from class 2 and 4) also in class 3, especially at very young ages. From the previous analysis we already know that class 3 is the one with the lowest accuracy;
\item[Class 4] this class includes a number of very massive sources ($\rm M>10^5\ M_\odot$) with old ages ($\rm \ge 10^9$ yr) that are correctly classified. Visual inspection shows that they are mostly foreground stars and background galaxies.
\end{description}
}

\revision{In order to quantify the observed trends we focus on the high--mass clusters and calculate the fraction of mis--classified ones. 
We report in Figure~\ref{fig:cm_highmass} (left panels) the confusion matrix for clusters with $\rm \log_{10}(M)\ge4.5$; the deviation of the mass function from a power-law has been debated widely in the literature (see e.g. \citealp{bastian2012,adamo2015,chandar2016,mok2019}) and is observed above these masses in nearby galaxies (e.g. \citealp{messa2018a}). There are 77 class 1 clusters correctly classified, with the addition of 7 false-positives, 5 of which are actually class 2; 7 class 1 clusters have been predicted as class 2. We deduce that most of the confusion for massive clusters in class 1 is with class 2, and therefore not crucial in studies that consider those two classes together as `clusters'. This is supported by the confusion matrix for binary classification in Figure~\ref{fig:cm_highmass} (bottom panels). At the opposite end, 39 class 4 sources have been correctly classified, with only 8 inclusions from FPs (3 from class 3). 6 class 4 sources have been assigned to other classes (2 of them to class 3). Again, we conclude that the mis--classification in this class is not elevated. 
As previously noticed in Figure~\ref{fig:FPFN}, mis--classification at high-masses heavily affect sources of class 2 and 3. At the same time, the number of high-mass sources in these two classes is much smaller than in class 1 or 4. Therefore we conclude that, overall, mis--classification will have only a small impact on the study of the high--end of the mass function, as also suggested by the high accuracy of the binary classification in this case ($90.8\%$, bottom left panel of Figure~\ref{fig:cm_highmass}).}

\revision{Similarly, we can focus on the brightest sources only; the confusion matrix of sources with $\rm Mag_V<-8$ is shown in Figure~\ref{fig:cm_highmass} (right panels). This limit is 2 mag brighter than the one used as completeness in LEGUS catalogs and is indicated as the black dashed lines in Figure~\ref{fig:FPFN}, left panels.
Similar trends to the ones observed for the high--mass sources are recovered. The confusion matrix is not very different, in terms of accuracy, from the one found for the entire sample (Figure~\ref{fig:cm}). We conclude that \starnet\ performs similarly for bright sources as for the rest of the sample.
} 

\begin{figure}
\begin{center}
\begin{tabular}{c c c c c c}
\noindent
\renewcommand\arraystretch{1.0}
\begin{tabular}{ @{\hspace{-100pt}}c >{\hspace{-0.4em}\bfseries}c @{\hspace{-0.8em}}c @{\hspace{-1.2em}}c @{\hspace{-1.2em}}c @{\hspace{-1.2em}}c}
  \multirow{17.5}{*}{\rotatebox{90}{\parbox{5.1cm}{\bfseries\centering Human Classification}}} 
  & & \multicolumn{4}{c}{\bfseries High mass} \\
  & & \multicolumn{4}{c}{\bfseries \starnet Prediction } \\
  & & \hspace{1.0em} \bfseries Class 1 & \hspace{1.0em} \bfseries Class 2 & \hspace{1.0em} \bfseries Class 3 & \hspace{1.0em} \bfseries Class 4 \\
  & \multirow{3}{*}{Class 1} & 
  \MyBox{77}{0.92} & \MyBox{7}{0.08} & \MyBox{0}{0.00} & \MyBox{0}{0.00} \\
  & \multirow{3}{*}{Class 2} & 
  \MyBox{5}{0.33} & \MyBox{3}{0.20} & \MyBox{2}{0.13} & \MyBox{5}{0.33} \\
  & \multirow{3}{*}{Class 3} & 
  \MyBox{0}{0.00} & \MyBox{3}{0.33} & \MyBox{3}{0.33} & \MyBox{3}{0.33} \\
  & \multirow{3}{*}{Class 4} & 
  \MyBox{2}{0.04} & \MyBox{2}{0.04} & \MyBox{2}{0.04} & \MyBox{39}{0.87} \\
\end{tabular}
&  
\noindent
\renewcommand\arraystretch{1.0}
\begin{tabular}{ @{\hspace{-60pt}}c >{\hspace{-0.4em}\bfseries}c @{\hspace{-0.8em}}c @{\hspace{-1.2em}}c @{\hspace{-1.2em}}c @{\hspace{-1.2em}}c}
  \multirow{17.5}{*}{\rotatebox{90}{\parbox{5.1cm}{\bfseries\centering Human Classification}}} 
  & & \multicolumn{4}{c}{\bfseries High luminosity} \\
  & & \multicolumn{4}{c}{\bfseries \starnet Prediction } \\
  & & \hspace{1.0em} \bfseries Class 1 & \hspace{1.0em} \bfseries Class 2 & \hspace{1.0em} \bfseries Class 3 & \hspace{1.0em} \bfseries Class 4 \\
  & \multirow{3}{*}{Class 1} & 
  \MyBox{79}{0.84} & \MyBox{12}{0.13} & \MyBox{1}{0.01} & \MyBox{2}{0.02} \\
  & \multirow{3}{*}{Class 2} & 
  \MyBox{107}{0.17} & \MyBox{335}{0.58} & \MyBox{13}{0.15} & \MyBox{9}{0.10} \\
  & \multirow{3}{*}{Class 3} & 
  \MyBox{0}{0.00} & \MyBox{13}{0.23} & \MyBox{29}{0.51} & \MyBox{15}{0.26} \\
  & \multirow{3}{*}{Class 4} & 
  \MyBox{5}{0.05} & \MyBox{11}{0.10} & \MyBox{10}{0.09} & \MyBox{83}{0.76} \\
\end{tabular}
&  \\
\noindent
\renewcommand\arraystretch{1.0}
\begin{tabular}{ @{\hspace{-70pt}}c >{\hspace{-0.4em}\bfseries}c @{\hspace{-0.8em}}c @{\hspace{-3.0em}}c}
  \multirow{10}{*}{\rotatebox{90}{\parbox{2.3cm}{\bfseries\centering Human Classification}}} & 
    & \multicolumn{2}{c}{\bfseries \starnet Prediction}  \\
  & & \hspace{1.0em} \bfseries C & \hspace{1.0em} \bfseries NC \\
  & \multirow{3}{*}{C} & 
  \MyBox{92}{0.93} & \MyBox{0.07}{0.25}  \\
  & \multirow{3}{*}{NC} & 
  \MyBox{7}{0.13} & \MyBox{47}{0.87}  \\
\end{tabular}
&  
\noindent
\renewcommand\arraystretch{1.0}
\begin{tabular}{ @{\hspace{-20pt}}c >{\hspace{-0.4em}\bfseries}c @{\hspace{-0.8em}}c @{\hspace{-3.0em}}c}
  \multirow{10}{*}{\rotatebox{90}{\parbox{2.3cm}{\bfseries\centering Human Classification}}} & 
    & \multicolumn{2}{c}{\bfseries \starnet Prediction}  \\
  & & \hspace{1.0em} \bfseries C & \hspace{1.0em} \bfseries NC \\
  & \multirow{3}{*}{C} & 
  \MyBox{158}{0.86} & \MyBox{25}{0.14}  \\
  & \multirow{3}{*}{NC} & 
  \MyBox{29}{0.17} & \MyBox{137}{0.83}  \\
\end{tabular}
\end{tabular}
\caption{\revision{Confusion matrix for sources with high masses ($\rm \log_{10}(M/M_\odot)>4.5$, left panels), and with high luminosities ($\rm Mag_{AB}<-8$ mag, right panels). The accuracies, in these cases, are \textbf{79.7\%} and \textbf{69.6\%} for the 4-class and \textbf{90.8\%} and \textbf{84.5\%} for the binary classifications, respectively.}}
\label{fig:cm_highmass}
\end{center}
\end{figure}

\section{Future improvements}  \label{sec:future_work}
Throughout this work, we have shown that \starnet\ can be trained to reproduce the same level of accuracy as the human classifications. Thus, future improvements to the overall accuracy need to include larger human--classified catalogs, possibly with higher accuracy. While this goal can be achieved by using the classifications of a single expert \citep{wei2020}, it is not a desirable path especially if samples larger than several 10's of thousands classifications (i.e., larger than the LEGUS sample) need to be collected.  The creation of visualization and classification tools with easy access (e.g. browser-based visualization tools) that facilitate and speed--up human classifications may prove important for progress in this area. Larger and more accurate catalogs would increase the discriminating power of the algorithm  and reduce the confusion created by ambiguous human classifications.

 Faint star clusters are an unexplored region of the luminosity parameter space. The LEGUS collaboration  classified clusters brighter than V$=$-6~mag, leaving the bulk of faint clusters untouched. However, faint (low mass) star  clusters are important discriminants for evolution models. Absence of classified faint clusters makes it difficult to train \starnet\ on these sources. Faint clusters are difficult to classify also for human classifiers. A potential way around this problem is to generate artificial clusters by dimming the existing, classified (bright) clusters in the HST images, and train \starnet\ onto these artificial sources.  
 \revision{Additional applications of artificially generated clusters include the exploration of wavelength  regimes, such as the JWST one, which are outside those analyzed in  this paper.}

Finally, we point out that, although \starnet\ was trained and tested on the cluster catalogs of LEGUS, it can be used to classify sources in other nearby galaxies, at least within the distance range covered by LEGUS ($\lesssim$20~Mpc). The HST archive, for example, is already a large repository of multi--band images of nearby galaxies, \revision{from which cluster catalogs to be inspected by \starnet\ can be easily created with automatic extraction tools, like the one used by LEGUS (see \citealp{adamo2017})}. 
\newrevision{The current version of \starnet is publicly available at \href{https://github.com/gperezs/StarcNet}{\texttt{github.com/gperezs/StarcNet} (DOI:\texttt{10.5281/zenodo.4279715})}}.

\section{Summary and conclusions}  \label{sec:conclusion}
We developed \starnet, a multi--scale CNN, with the goal of morphologically classifying stellar clusters in nearby galaxies. 
\starnet\ aims at speeding up by orders of magnitude the process of visual cluster classification, which currently is the single--most important limitation to securing large catalogs for studies of these sources. Availability of reliable and fast ways to  classify  star clusters will become even more critical with the advent of extremely large surveys, such as those that will be produced by the Vera Rubin Observatory and the Nancy Roman Space Telescope.

\starnet\ is a three-pathway CNN that processes each  input source at three different magnifications. Each pathway consists of a set of modules containing a convolutional layer, a group normalization layer, and a Leaky ReLU activation with a single pooling layer after the fourth module. Each of the three pathways' extracted features are combined into a fully-connected layer to output a probability distribution of the corresponding source class.

The classification adopted consist of 4 classes, where class 1 and 2 are for spherical and elongated, but compact, clusters, respectively, class 3 includes multi--peaked systems with diffuse nebular emission that may be compact stellar associations, and class 4 is for spurious detections, i.e. all sources that can be defined as non--clusters.

More than $15,000$ sources, visually classified by at least 3 experts from LEGUS HST Treasury Project are used to train and test \starnet. 
We test different architectures, e.g. by changing the number of pathways in the network, and different inputs, e.g. by changing the size of input arrays. The final version of \starnet\ reaches an overall accuracy of $\sim69\%$, nearly matching the agreement among human classifiers. The accuracy is not uniform across classes, as a better performance is achieved for classes 1 and 4; this in-homogeneity traces the difficulty of the human classifiers in confidently identifying clusters in class 2 and 3. 

Since many cluster studies in the literature rely on a simpler classification scheme than the one adopted by LEGUS, i.e. they simply separate clusters (what LEGUS classified as class 1 and 2) from non--clusters (class 3 and 4), we re--measure the accuracy of our algorithm using this binary classification. The \starnet\ accuracy reaches $86\%$ when merging the four classes into a binary classification. Training \starnet\ directly on the binary classification does not bring improvement to the final accuracy, which remains  around $86\%$.

\revision{A low--significance anti--correlation between \starnet\ accuracy and galaxy distance is found. However, we do not find a correlation between the \starnet\ accuracy and the level of agreement among human classifiers, when different galaxies are considered individually. We test the performance of \starnet on galaxies not included in the training set, first by removing, in turn, one galaxy from the training sample and making predictions on its sources, and second by considering the galaxy pair NGC1510+1512, one of the LEGUS galaxies without a human--classified cluster  catalog. These tests highlight that the \starnet accuracy on new galaxies is comparable to the reference one.}

We analyze the outputs from our pipeline, paying particular attention to  whether the classifications given by \starnet\ affect the average physical properties of the sources within the four classes. We consider color--color diagrams and luminosity functions, as well as age, mass and extinction distributions, mass functions and age functions. We find that the ML classification does not introduce changes in the recovered statistical properties and median distributions. 

\starnet\ proves to be a successful improvement over an early LEGUS attempt to develop a ML--driven cluster classification algorithm  \citep{grasha2019b,grasha2019}. This early  attempt,  tested on the cluster population of NGC~5194, performed  poorly in recognizing class 3 sources; conversely,  \starnet\ has a \revision{$10\times$} higher recovery rate for class 3 clusters in NGC~5194 than the earlier classification code. Recently, \citet{wei2020} applied deep transfer learning to the classification of LEGUS star clusters, reaching an overall accuracy and per-class distribution very similar to the \starnet\ ones. A direct comparison of the two approaches is not straightforward, due to the \revision{absence of a publicly released code and experimental setup by \citet{wei2020}}. However, a comparison between our algorithm  and those of \citet{wei2020} by using our own catalogs shows that the algorithms presented by \citet{wei2020} \revision{reach accuracy $2.2\%-4.7\%$ lower than \starnet for four  class classification and $0.2\%-1.0\%$ higher for binary classification}.

Future developments of \starnet\ will include  applications to the faint (low mass) sources found in the HST images. With training on appropriate sets,  \starnet\ can be readily applied to a range of cases, from HST images of nearby galaxies to ground based images of Local Group galaxies.

\acknowledgments
This paper is based upon work supported by the National Science Foundation under Grant No. 1815267. The authors thank Dr. Hwihyun Kim for providing individual human classifications for a subset of the LEGUS catalogs.

%

\vspace{5mm}
\facilities{HST(ACS and WFC3)}


\software{Astropy\footnote{https://www.astropy.org/} \citep{2013A&A...558A..33A},  
          SciPy\footnote{https://www.scipy.org/} \citep{scipy},
          scikit-image\footnote{https://scikit-image.org/} \citep{scikit-image},
          scikit-learn\footnote{https://scikit-learn.org/stable/} \citep{scikit-learn},
          PyTorch\footnote{https://pytorch.org/} \citep{pytorch}
          }



\appendix

\section{Mis--Classifications}\label{sec:appB}
We explore here several of the reasons for classification disagreements between different human classifiers, which the confuson matrix of Figure~\ref{fig:cm_human} summarizes, by showing that  
the largest disagreement is usually found for class 2 and 3 sources. 
The type of classification requested by the LEGUS project's approach, i.e. the division of cluster candidates into 3 morphological classes plus a fourth class for non-clusters, requires high accuracy and is subject to judgement calls in many instances.   
Therefore, even a single highly--trained classifier cannot always be $100\%$ sure of their own choices, and is not able to always repeat their own classifications. This uncertainty has been quantified at $\le$80\% repeatibility across four classes \citet{wei2020}. 
This intrinsic difficulty, which is driven by subjective evaluations of the morphology of a source, explains the failure of crowd--sourcing approaches for cluster identification/classification in cases where the galaxy's background is semi--resolved or unresolved. 
We summarize below several of 
the most common causes of  mis-classification. 

\begin{figure*}
\centering
\subfloat[]{\includegraphics[width=0.30\textwidth]{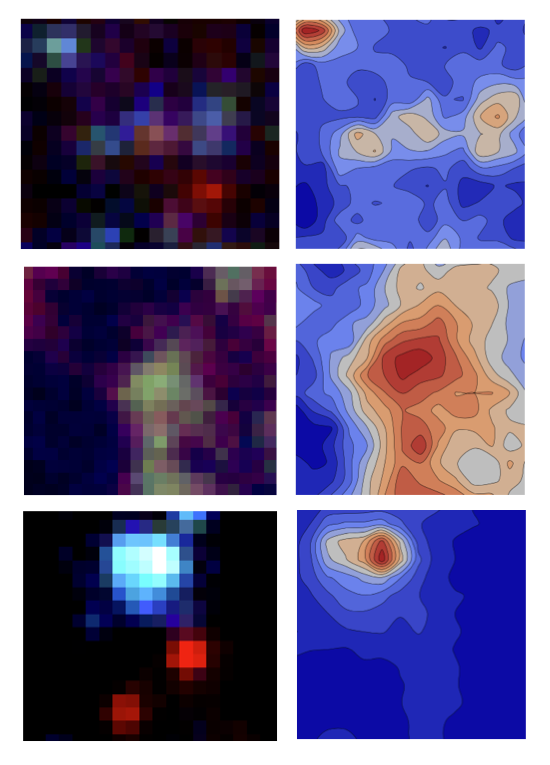}}\hfill
\subfloat[]{\includegraphics[width=0.61\textwidth]{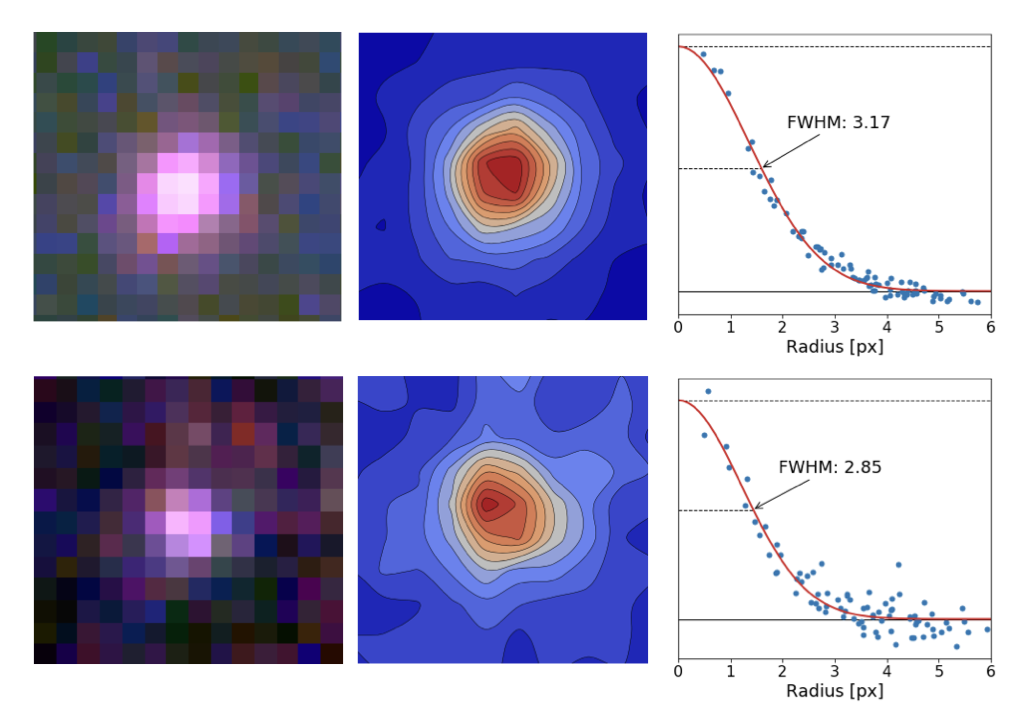}}\hfill
\subfloat[]{\includegraphics[width=0.45\textwidth]{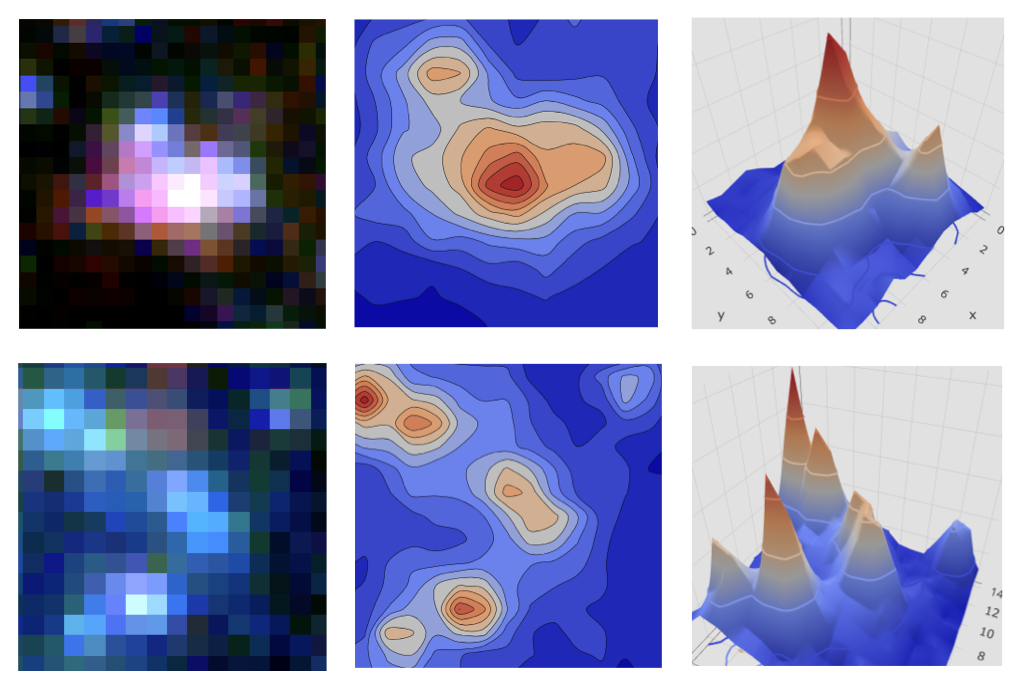}}\hfill
\subfloat[]{\includegraphics[width=0.31\textwidth]{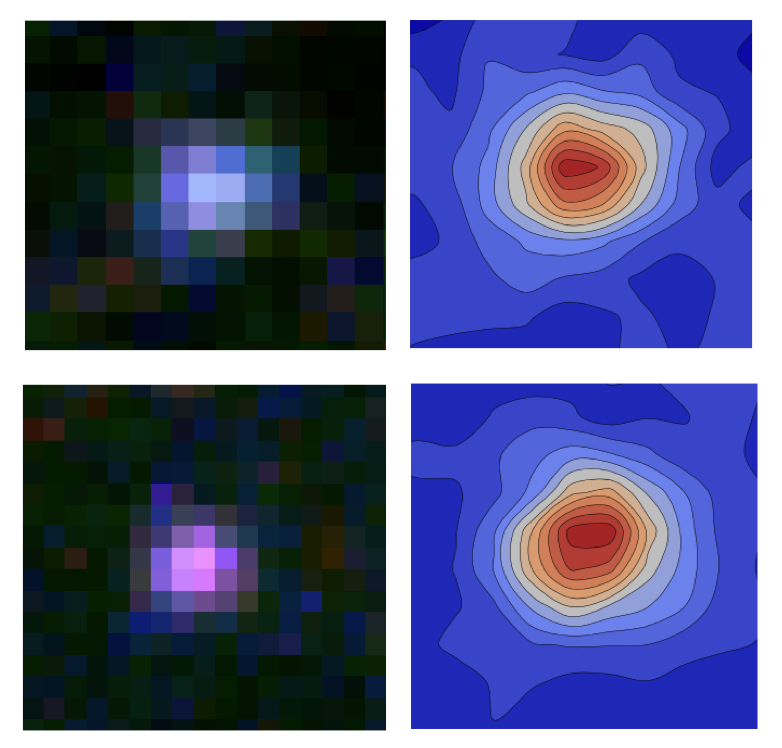}}\hfill
\subfloat[]{\includegraphics[width=0.15\textwidth]{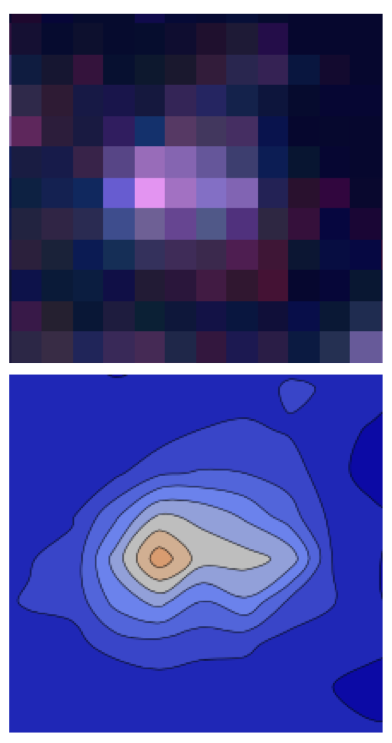}}\hfill
\caption{Examples of possible confusion and mis-classifications among classes: (a) sources that are hard to classify because they are faint, lie in a region of high background or next to a much brighter source; (b) isolated sources that can have similar light profiles, barely larger than the stellar PSF, and are here classified one as class 1 (upper source) and the other as a star, class 4 (bottom source); (c) class 3 sources (upper) that can be difficult to  discriminate from a serendipitous  collection of stars, i.e. class 4 (bottom); (d) slightly asymmetric class 1 sources (upper) that can be confused with class 2 ones with moderate elongation (bottom); (e) chance overlap of two sources with similar colors (classified as class 4) that can resemble a class 2 source. All the light profiles, 2D contours and 3D plots are referred to the $V$ band. All examples are from the LEGUS cluster catalog of the galaxy NGC 1566.} 
\label{fig:misclass}
\end{figure*}

\paragraph{Decreased contrast} A secure identification of a source, of any class, is facilitated if the contrast between the source and the background is high. The contrast can be low under several circumstances, including that the local galaxy background is high, a neighboring source is bright, or  the source to identify is intrinsically faint.
A few examples are given in Figure~\ref{fig:misclass}(a).
\paragraph{Compact sources} Although most  sources consistent with a stellar PSF are removed at the stage of catalog construction, sources with PSF barely larger than the stellar one are retained. These sources can be either classified as class 1 or 2 (cluster) or as class 4 (non-cluster) depending on the judgment of the classifier. 
An example of this situation is given in Figure\ref{fig:misclass}(b). However, as implied by the confusion matrices in Figure~\ref{fig:cm_human}, this  type of mis--classification is not frequent, occurring about $3.7\%$ of times for class 1 to 4 confusion, and increasing to $11.3\%$ if both class 1 and 2 mis--classifications are included.
\paragraph{Diffuse light in class 3} The main discriminant between the multi--peaked class 3 sources and random groupings of stars (asterisms, class 4) is the presence of diffuse emission between the peaks of the class 3 source. However, the diffuse emission can be faint, leading to potential confusion and mis--classifications.
An example is given in Figure~\ref{fig:misclass}(c). The confusion between class 3 and 4 is at the level of $13.7\%$ 
\paragraph{Separating class 1 from class 2  sources} The main difference between class 1 and class 2 sources is that the latter display either elongated or asymmetric light profiles. However, there is no strict value of the ellipticity that separates the two classes. 
In some cases this leads to confusion between 
a class 2 cluster candidate and a slightly elongated (or slightly asymmetric) class 1, which can be further complicated by an uneven background causing distortions in the light profiles (an example is given in Figure~\ref{fig:misclass}(d). This problem applies to approximately $7\%$ of the sources.  
\paragraph{Overlaps of source pairs} Asterisms can appear as a single distorted object when two sources align almost perfectly along the line of sight and have similar colors; in this case the two objects appears like a single elongated one (i.e. a class 2 sources). An example is shown in Figure~\ref{fig:misclass}(e). This case is different from the one shown in Figure~\ref{fig:examples}, where the two sources can be discriminated by their color difference, leading to the correct  class 4 classification.

\section{Summary of LEGUS galaxies with cluster classifications}\label{sec:appC}
\revision{We report in Table~\ref{tab:legus_starcnet_class} the number of cluster  candidates in each of the four classes for the 31 LEGUS galaxies with available catalogs, along with the galaxy distance and the physical scale subtended by one pixel (0$^{\prime\prime}$.04/pixel). }

\begin{table}
\centering
\caption{\revision{LEGUS star cluster catalog used for training, validation, and testing of \starnet.} }
\begin{tabular}{lrrrrrcc}
\hline\hline
Galaxy & Class 1 & Class 2 & Class 3 & Class 4 & Total & Dist.\tablenotemark{a} & Resolution \\
 & & & & & (by galaxy) & (Mpc) & (pc/pixel) \\
\hline
IC 4247 & 1 & 4 & 3 & 37 & 45 & 5.11 & 0.99 \\
IC 559 & 9 & 12 & 4 & 18 & 43 & 5.3 & 1.03 \\
NGC 1313 & 127 & 286 & 332 & 759 & 1504 & 4.39 & 0.85 \\
NGC 1433 & 51 & 61 & 56 & 138 & 306 & 8.3 & 1.61 \\
NGC 1566 & 477 & 404 & 692 & 878 & 2451 & 17.9 & 3.44 \\
NGC 1705 & 16 & 13 & 13 & 54 & 96 & 5.1 & 0.99 \\
NGC 3344 & 119 & 118 & 159 & 161 & 557 & 7 & 1.36 \\
NGC 3351 & 118 & 80 & 94 & 326 & 618 & 10 & 1.95 \\
NGC 3738 & 49 & 93 & 86 & 214 & 442 & 4.9 & 0.95 \\
NGC 4242 & 117 & 60 & 14 & 42 & 233 & 5.8 & 1.13 \\
NGC 4395 & 39 & 76 & 61 & 50 & 226 & 4.3 & 0.84 \\
NGC 4449 & 143 & 282 & 182 & 874 & 1481 & 4.31 & 0.84 \\
NGC 45 & 45 & 52 & 20 & 43 & 160 & 6.61 & 1.29 \\
NGC 4656 & 93 & 91 & 78 & 169 & 431 & 5.5 & 1.07 \\
NGC 5194+5195 & 363 & 502 & 365 & 1,261 & 2491 & 7.66 & 1.49 \\
NGC 5238 & 4 & 4 & 1 & 9 & 18 & 4.51 & 0.88 \\
NGC 5253 & 20 & 37 & 23 & 154 & 234 & 3.15 & 0.61 \\
NGC 5474 & 48 & 95 & 34 & 144 & 321 & 6.8 & 1.32 \\
NGC 5477 & 5 & 9 & 9 & 49 & 72 & 6.4 & 1.25 \\
NGC 628 & 426 & 437 & 413 & 664 & 1940 & 9.9 & 1.93 \\
NGC 6503 & 71 & 96 & 131 & 172 & 470 & 5.27 & 1.03 \\
NGC 7793 & 83 & 160 & 169 & 140 & 552 & 3.44 & 0.67 \\
UGC 1249 & 13 & 35 & 40 & 133 & 221 & 6.9 & 1.34 \\
UGC 4305 & 16 & 29 & 40 & 147 & 232 & 3.05 & 0.59 \\
UGC 4459 & 2 & 5 & 3 & 20 & 30 & 3.66 & 0.71 \\
UGC 5139 & 2 & 7 & 7 & 23 & 39 & 3.98 & 0.77 \\
UGC 685 & 7 & 4 & 3 & 6 & 20 & 4.83 & 0.94 \\
UGC 695 & 4 & 7 & 6 & 94 & 111 & 10.9 & 2.1 \\
UGC 7408 & 19 & 16 & 11 & 32 & 78 & 6.7 & 1.30 \\
UGCA 281 & 2 & 9 & 4 & 34 & 49 & 5.9 & 1.15 \\
Total (by class) & 2,489 & 3,084 & 3,053 & 6,845 & 15471 & & \\
\hline
\label{tab:legus_starcnet_class}
\end{tabular}
\tablenotetext{a}{The distances of the LEGUS galaxies from \citet{legus}. The  distance of NGC~1566 is from \citet{sabbi2018}.}
\end{table}


\FloatBarrier
\bibliography{main}{}

\begin{thebibliography}{}
\expandafter\ifx\csname natexlab\endcsname\relax\def\natexlab#1{#1}\fi
\providecommand{\url}[1]{\href{#1}{#1}}
\providecommand{\dodoi}[1]{doi:~\href{http://doi.org/#1}{\nolinkurl{#1}}}
\providecommand{\doeprint}[1]{\href{http://ascl.net/#1}{\nolinkurl{http://ascl.net/#1}}}
\providecommand{\doarXiv}[1]{\href{https://arxiv.org/abs/#1}{\nolinkurl{https://arxiv.org/abs/#1}}}

\bibitem[{Abadi {et~al.}(2015)Abadi, Agarwal, Barham, Brevdo, Chen, Citro,
  Corrado, Davis, Dean, Devin, Ghemawat, Goodfellow, Harp, Irving, Isard, Jia,
  Jozefowicz, Kaiser, Kudlur, Levenberg, Man\'{e}, Monga, Moore, Murray, Olah,
  Schuster, Shlens, Steiner, Sutskever, Talwar, Tucker, Vanhoucke, Vasudevan,
  Vi\'{e}gas, Vinyals, Warden, Wattenberg, Wicke, Yu, \& Zheng}]{tensorflow}
Abadi, M., Agarwal, A., Barham, P., {et~al.} 2015, {TensorFlow}: Large-Scale
  Machine Learning on Heterogeneous Systems.
\newblock \url{http://tensorflow.org/}

\bibitem[{{Ackermann} {et~al.}(2018){Ackermann}, {Schawinski}, {Zhang},
  {Weigel}, \& {Turp}}]{ackermann2018}
{Ackermann}, S., {Schawinski}, K., {Zhang}, C., {Weigel}, A.~K., \& {Turp},
  M.~D. 2018, \mnras, 479, 415, \dodoi{10.1093/mnras/sty1398}

\bibitem[{{Adamo} {et~al.}(2015){Adamo}, {Kruijssen}, {Bastian}, {Silva-Villa},
  \& {Ryon}}]{adamo2015}
{Adamo}, A., {Kruijssen}, J.~M.~D., {Bastian}, N., {Silva-Villa}, E., \&
  {Ryon}, J. 2015, \mnras, 452, 246, \dodoi{10.1093/mnras/stv1203}

\bibitem[{{Adamo} {et~al.}(2017){Adamo}, {Ryon}, {Messa}, {Kim}, {Grasha},
  {Cook}, {Calzetti}, {Lee}, {Whitmore}, {Elmegreen}, {Ubeda}, {Smith},
  {Bright}, {Runnholm}, {Andrews}, {Fumagalli}, {Gouliermis}, {Kahre}, {Nair},
  {Thilker}, {Walterbos}, {Wofford}, {Aloisi}, {Ashworth}, {Brown}, {Chandar},
  {Christian}, {Cignoni}, {Clayton}, {Dale}, {de Mink}, {Dobbs}, {Elmegreen},
  {Evans}, {Gallagher}, {Grebel}, {Herrero}, {Hunter}, {Johnson}, {Kennicutt},
  {Krumholz}, {Lennon}, {Levay}, {Martin}, {Nota}, {{\"O}stlin}, {Pellerin},
  {Prieto}, {Regan}, {Sabbi}, {Sacchi}, {Schaerer}, {Schiminovich}, {Shabani},
  {Tosi}, {Van Dyk}, \& {Zackrisson}}]{adamo2017}
{Adamo}, A., {Ryon}, J.~E., {Messa}, M., {et~al.} 2017, \apj, 841, 131,
  \dodoi{10.3847/1538-4357/aa7132}

\bibitem[{{Astropy Collaboration} {et~al.}(2013){Astropy Collaboration},
  {Robitaille}, {Tollerud}, {Greenfield}, {Droettboom}, {Bray}, {Aldcroft},
  {Davis}, {Ginsburg}, {Price-Whelan}, {Kerzendorf}, {Conley}, {Crighton},
  {Barbary}, {Muna}, {Ferguson}, {Grollier}, {Parikh}, {Nair}, {Unther},
  {Deil}, {Woillez}, {Conseil}, {Kramer}, {Turner}, {Singer}, {Fox}, {Weaver},
  {Zabalza}, {Edwards}, {Azalee Bostroem}, {Burke}, {Casey}, {Crawford},
  {Dencheva}, {Ely}, {Jenness}, {Labrie}, {Lim}, {Pierfederici}, {Pontzen},
  {Ptak}, {Refsdal}, {Servillat}, \& {Streicher}}]{2013A&A...558A..33A}
{Astropy Collaboration}, {Robitaille}, T.~P., {Tollerud}, E.~J., {et~al.} 2013,
  \aap, 558, A33, \dodoi{10.1051/0004-6361/201322068}

\bibitem[{Ba {et~al.}(2016)Ba, Kiros, \& Hinton}]{ln}
Ba, J.~L., Kiros, J.~R., \& Hinton, G.~E. 2016, Layer Normalization.
\newblock \doarXiv{1607.06450}

\bibitem[{{Barchi} {et~al.}(2020){Barchi}, {de Carvalho}, {Rosa}, {Sautter},
  {Soares-Santos}, {Marques}, {Clua}, {Gon{\c{c}}alves}, {de S{\'a}-Freitas},
  \& {Moura}}]{barchi2020}
{Barchi}, P.~H., {de Carvalho}, R.~R., {Rosa}, R.~R., {et~al.} 2020, Astronomy
  and Computing, 30, 100334, \dodoi{10.1016/j.ascom.2019.100334}

\bibitem[{{Bastian}(2008)}]{bastian2008}
{Bastian}, N. 2008, Monthly Notices of the Royal Astronomical Society, 390

\bibitem[{{Bastian} {et~al.}(2005){Bastian}, {Gieles}, {Lamers}, {Scheepmaker},
  \& {de Grijs}}]{bastian2005}
{Bastian}, N., {Gieles}, M., {Lamers}, H.~J.~G.~L.~M., {Scheepmaker}, R.~A., \&
  {de Grijs}, R. 2005, \aap, 431, 905, \dodoi{10.1051/0004-6361:20041078}

\bibitem[{{Bastian} {et~al.}(2012){Bastian}, {Adamo}, {Gieles}, {Silva-Villa},
  {Lamers}, {Larsen}, {Smith}, {Konstantopoulos}, \&
  {Zackrisson}}]{bastian2012}
{Bastian}, N., {Adamo}, A., {Gieles}, M., {et~al.} 2012, \mnras, 419, 2606,
  \dodoi{10.1111/j.1365-2966.2011.19909.x}

\bibitem[{{Bertin} \& {Arnouts}(1996)}]{SExtractor}
{Bertin}, E., \& {Arnouts}, S. 1996, \aaps, 117, 393,
  \dodoi{10.1051/aas:1996164}

\bibitem[{{Bik} {et~al.}(2003){Bik}, {Lamers}, {Bastian}, {Panagia}, \&
  {Romaniello}}]{bik2003}
{Bik}, A., {Lamers}, H.~J.~G.~L.~M., {Bastian}, N., {Panagia}, N., \&
  {Romaniello}, M. 2003, \aap, 397, 473, \dodoi{10.1051/0004-6361:20021384}

\bibitem[{{Calzetti} {et~al.}(2015){Calzetti}, {Lee}, {Sabbi}, \&
  {Adamo}}]{legus}
{Calzetti}, D., {Lee}, J., {Sabbi}, E., \& {Adamo}, A. 2015, The Astronomical
  Journal, 149, 25

\bibitem[{{Chandar} {et~al.}(2015){Chandar}, {Fall}, \&
  {Whitmore}}]{chandar2015}
{Chandar}, R., {Fall}, S.~M., \& {Whitmore}, B.~C. 2015, \apj, 810, 1,
  \dodoi{10.1088/0004-637X/810/1/1}

\bibitem[{{Chandar} {et~al.}(2017){Chandar}, {Fall}, {Whitmore}, \&
  {Mulia}}]{chandar2017}
{Chandar}, R., {Fall}, S.~M., {Whitmore}, B.~C., \& {Mulia}, A.~J. 2017, \apj,
  849, 128, \dodoi{10.3847/1538-4357/aa92ce}

\bibitem[{{Chandar} {et~al.}(2014){Chandar}, {Whitmore}, {Calzetti}, \&
  {O'Connell}}]{chandar2014}
{Chandar}, R., {Whitmore}, B.~C., {Calzetti}, D., \& {O'Connell}, R. 2014,
  \apj, 787, 17, \dodoi{10.1088/0004-637X/787/1/17}

\bibitem[{{Chandar} {et~al.}(2016){Chandar}, {Whitmore}, {Dinino}, {Kennicutt},
  {Chien}, {Schinnerer}, \& {Meidt}}]{chandar2016}
{Chandar}, R., {Whitmore}, B.~C., {Dinino}, D., {et~al.} 2016, \apj, 824, 71,
  \dodoi{10.3847/0004-637X/824/2/71}

\bibitem[{{Cook} {et~al.}(2019){Cook}, {Lee}, {Adamo}, {Kim}, {Chand ar},
  {Whitmore}, {Mok}, {Ryon}, {Dale}, {Calzetti}, {Andrews}, {Aloisi},
  {Ashworth}, {Bright}, {Brown}, {Christian}, {Cignoni}, {Clayton}, {da Silva},
  {de Mink}, {Dobbs}, {Elmegreen}, {Elmegreen}, {Evans}, {Fumagalli},
  {Gallagher}, {Gouliermis}, {Grasha}, {Grebel}, {Herrero}, {Hunter}, {Jensen},
  {Johnson}, {Kahre}, {Kennicutt}, {Krumholz}, {Lee}, {Lennon}, {Linden},
  {Martin}, {Messa}, {Nair}, {Nota}, {{\"O}stlin}, {Parziale}, {Pellerin},
  {Regan}, {Sabbi}, {Sacchi}, {Schaerer}, {Schiminovich}, {Shabani}, {Slane},
  {Small}, {Smith}, {Smith}, {Taibi}, {Thilker}, {de la Torre}, {Tosi},
  {Turner}, {Ubeda}, {Van Dyk}, {Walterbos}, \& {Wofford}}]{cook2019}
{Cook}, D.~O., {Lee}, J.~C., {Adamo}, A., {et~al.} 2019, \mnras, 484, 4897,
  \dodoi{10.1093/mnras/stz331}

\bibitem[{Deng {et~al.}(2009)Deng, Dong, Socher, Li, Li, \& Fei-Fei}]{imagenet}
Deng, J., Dong, W., Socher, R., {et~al.} 2009, in CVPR09

\bibitem[{Dieleman {et~al.}(2015)Dieleman, Willett, \& Dambre}]{Dieleman2015}
Dieleman, S., Willett, K.~W., \& Dambre, J. 2015, Monthly Notices of the Royal
  Astronomical Society, 450, 1441–1459, \dodoi{10.1093/mnras/stv632}

\bibitem[{{Dom{\'\i}nguez S{\'a}nchez} {et~al.}(2018){Dom{\'\i}nguez
  S{\'a}nchez}, {Huertas-Company}, {Bernardi}, {Tuccillo}, \&
  {Fischer}}]{dominguezsanchez2018}
{Dom{\'\i}nguez S{\'a}nchez}, H., {Huertas-Company}, M., {Bernardi}, M.,
  {Tuccillo}, D., \& {Fischer}, J.~L. 2018, \mnras, 476, 3661,
  \dodoi{10.1093/mnras/sty338}

\bibitem[{Duchi {et~al.}(2010)Duchi, Hazan, \& Singer}]{adagrad}
Duchi, J., Hazan, E., \& Singer, Y. 2010, Adaptive Subgradient Methods for
  Online Learning and Stochastic Optimization, Tech. Rep. UCB/EECS-2010-24,
  EECS Department, University of California, Berkeley

\bibitem[{{Gieles}(2009)}]{gieles2009}
{Gieles}, M. 2009, \mnras, 394, 2113, \dodoi{10.1111/j.1365-2966.2009.14473.x}

\bibitem[{{Gieles} {et~al.}(2006{\natexlab{a}}){Gieles}, {Larsen}, {Bastian},
  \& {Stein}}]{gieles2006b}
{Gieles}, M., {Larsen}, S.~S., {Bastian}, N., \& {Stein}, I.~T.
  2006{\natexlab{a}}, \aap, 450, 129, \dodoi{10.1051/0004-6361:20053589}

\bibitem[{{Gieles} {et~al.}(2006{\natexlab{b}}){Gieles}, {Larsen},
  {Scheepmaker}, {Bastian}, {Haas}, \& {Lamers}}]{gieles2006c}
{Gieles}, M., {Larsen}, S.~S., {Scheepmaker}, R.~A., {et~al.}
  2006{\natexlab{b}}, \aap, 446, L9, \dodoi{10.1051/0004-6361:200500224}

\bibitem[{{Gieles} {et~al.}(2006{\natexlab{c}}){Gieles}, {Portegies Zwart},
  {Baumgardt}, {Athanassoula}, {Lamers}, {Sipior}, \&
  {Leenaarts}}]{gieles2006a}
{Gieles}, M., {Portegies Zwart}, S.~F., {Baumgardt}, H., {et~al.}
  2006{\natexlab{c}}, \mnras, 371, 793,
  \dodoi{10.1111/j.1365-2966.2006.10711.x}

\bibitem[{{Glorot} \& {Bengio}(2010)}]{xavier}
{Glorot}, X., \& {Bengio}, Y. 2010, in JMLR W\&CP: Proceedings of the
  Thirteenth International Conference on Artificial Intelligence and Statistics
  (AISTATS 2010), Vol.~9, 249--256

\bibitem[{Goodfellow {et~al.}(2016)Goodfellow, Bengio, \&
  Courville}]{goodfellow}
Goodfellow, I., Bengio, Y., \& Courville, A. 2016, Deep Learning (MIT Press)

\bibitem[{{Grasha}(2018)}]{grasha2019b}
{Grasha}, K. 2018, PhD thesis, University of Massachusetts

\bibitem[{{Grasha} {et~al.}(2015){Grasha}, {Calzetti}, {Adamo}, {Kim},
  {Elmegreen}, {Gouliermis}, {Aloisi}, {Bright}, {Christian}, {Cignoni},
  {Dale}, {Dobbs}, {Elmegreen}, {Fumagalli}, {Gallagher}, {Grebel}, {Johnson},
  {Lee}, {Messa}, {Smith}, {Ryon}, {Thilker}, {Ubeda}, \&
  {Wofford}}]{grasha2015}
{Grasha}, K., {Calzetti}, D., {Adamo}, A., {et~al.} 2015, \apj, 815, 93,
  \dodoi{10.1088/0004-637X/815/2/93}

\bibitem[{{Grasha} {et~al.}(2017){Grasha}, {Calzetti}, {Adamo}, {Kim},
  {Elmegreen}, {Gouliermis}, {Dale}, {Fumagalli}, {Grebel}, {Johnson}, {Kahre},
  {Kennicutt}, {Messa}, {Pellerin}, {Ryon}, {Smith}, {Shabani}, {Thilker}, \&
  {Ubeda}}]{grasha2017a}
---. 2017, \apj, 840, 113, \dodoi{10.3847/1538-4357/aa6f15}

\bibitem[{{Grasha} {et~al.}(2019){Grasha}, {Calzetti}, {Adamo}, {Kennicutt},
  {Elmegreen}, {Messa}, {Dale}, {Fedorenko}, {Mahadevan}, {Grebel},
  {Fumagalli}, {Kim}, {Dobbs}, {Gouliermis}, {Ashworth}, {Gallagher}, {Smith},
  {Tosi}, {Whitmore}, {Schinnerer}, {Colombo}, {Hughes}, {Leroy}, \&
  {Meidt}}]{grasha2019}
---. 2019, \mnras, 483, 4707, \dodoi{10.1093/mnras/sty3424}

\bibitem[{He {et~al.}(2015)He, Zhang, Ren, \& Sun}]{resnet}
He, K., Zhang, X., Ren, S., \& Sun, J. 2015, Deep Residual Learning for Image
  Recognition.
\newblock \doarXiv{1512.03385}

\bibitem[{{Hollyhead} {et~al.}(2016){Hollyhead}, {Adamo}, {Bastian}, {Gieles},
  \& {Ryon}}]{hollyhead2016}
{Hollyhead}, K., {Adamo}, A., {Bastian}, N., {Gieles}, M., \& {Ryon}, J.~E.
  2016, \mnras, 460, 2087, \dodoi{10.1093/mnras/stw1142}

\bibitem[{Hu {et~al.}(2017)Hu, Shen, Albanie, Sun, \& Wu}]{squeeze-excite}
Hu, J., Shen, L., Albanie, S., Sun, G., \& Wu, E. 2017, Squeeze-and-Excitation
  Networks.
\newblock \doarXiv{1709.01507}

\bibitem[{Huang {et~al.}(2016)Huang, Liu, van~der Maaten, \&
  Weinberger}]{densenet}
Huang, G., Liu, Z., van~der Maaten, L., \& Weinberger, K.~Q. 2016, Densely
  Connected Convolutional Networks.
\newblock \doarXiv{1608.06993}

\bibitem[{Iandola {et~al.}(2016)Iandola, Han, Moskewicz, Ashraf, Dally, \&
  Keutzer}]{squeezenet}
Iandola, F.~N., Han, S., Moskewicz, M.~W., {et~al.} 2016, SqueezeNet:
  AlexNet-level accuracy with 50x fewer parameters and <0.5MB model size.
\newblock \doarXiv{1602.07360}

\bibitem[{{Ioffe} \& {Szegedy}(2015)}]{bn}
{Ioffe}, S., \& {Szegedy}, C. 2015, in Proceedings of the 32Nd International
  Conference on International Conference on Machine Learning - Volume 37,
  ICML'15 (JMLR.org), 448--456

\bibitem[{{Johnson} {et~al.}(2015){Johnson}, {Seth}, {Dalcanton}, {Wallace},
  {Simpson}, {Lintott}, {Kapadia}, {Skillman}, {Caldwell}, {Fouesneau},
  {Weisz}, {Williams}, {Beerman}, {Gouliermis}, \& {Sarajedini}}]{johnson2015}
{Johnson}, L.~C., {Seth}, A.~C., {Dalcanton}, J.~J., {et~al.} 2015, \apj, 802,
  127, \dodoi{10.1088/0004-637X/802/2/127}

\bibitem[{{Johnson} {et~al.}(2016){Johnson}, {Seth}, {Dalcanton}, {Beerman},
  {Fouesneau}, {Lewis}, {Weisz}, {Williams}, {Bell}, {Dolphin}, {Larsen},
  {Sandstrom}, \& {Skillman}}]{johnson2016}
---. 2016, \apj, 827, 33, \dodoi{10.3847/0004-637X/827/1/33}

\bibitem[{{Johnson} {et~al.}(2017){Johnson}, {Seth}, {Dalcanton}, {Beerman},
  {Fouesneau}, {Weisz}, {Bell}, {Dolphin}, {Sandstrom}, \&
  {Williams}}]{johnson2017}
---. 2017, \apj, 839, 78, \dodoi{10.3847/1538-4357/aa6a1f}

\bibitem[{{Khan} {et~al.}(2019){Khan}, {Huerta}, {Wang}, {Gruendl}, {Jennings},
  \& {Zheng}}]{khan2019}
{Khan}, A., {Huerta}, E.~A., {Wang}, S., {et~al.} 2019, Physics Letters B, 795,
  248, \dodoi{10.1016/j.physletb.2019.06.009}

\bibitem[{{Kingma} \& {Ba}(2015)}]{Adam}
{Kingma}, D., \& {Ba}, J. 2015, in 3rd International Conference on Learning
  Representations, {ICLR} 2015, San Diego, CA, USA, May 7-9, 2015, Conference
  Track Proceedings

\bibitem[{Krizhevsky {et~al.}(2012)Krizhevsky, Sutskever, \&
  Hinton}]{Krizhevsky2012}
Krizhevsky, A., Sutskever, I., \& Hinton, G.~E. 2012, in Advances in Neural
  Information Processing Systems 25, ed. F.~Pereira, C.~J.~C. Burges,
  L.~Bottou, \& K.~Q. Weinberger (Curran Associates, Inc.), 1097--1105

\bibitem[{{Lada} \& {Lada}(2003)}]{lada2003}
{Lada}, C., \& {Lada}, E. 2003, ARAA, 41

\bibitem[{{Larsen}(2002)}]{larsen2002}
{Larsen}, S.~S. 2002, \aj, 124, 1393, \dodoi{10.1086/342381}

\bibitem[{Lin {et~al.}(2015)Lin, RoyChowdhury, \& Maji}]{bilinear}
Lin, T.-Y., RoyChowdhury, A., \& Maji, S. 2015, in The IEEE International
  Conference on Computer Vision (ICCV)

\bibitem[{{Longmore} {et~al.}(2014){Longmore}, {Kruijssen}, {Bastian}, {Bally},
  {Rathborne}, {Testi}, {Stolte}, {Dale}, {Bressert}, \&
  {Alves}}]{Longmore2014}
{Longmore}, S.~N., {Kruijssen}, J.~M.~D., {Bastian}, N., {et~al.} 2014, in
  Protostars and Planets VI, ed. H.~{Beuther}, R.~S. {Klessen}, C.~P.
  {Dullemond}, \& T.~{Henning}, 291,
  \dodoi{10.2458/azu_uapress_9780816531240-ch013}

\bibitem[{Maas {et~al.}(2013)Maas, Hannun, \& Ng}]{leakyrelu}
Maas, A.~L., Hannun, A.~Y., \& Ng, A.~Y. 2013, in in ICML Workshop on Deep
  Learning for Audio, Speech and Language Processing

\bibitem[{{Messa} {et~al.}(2018{\natexlab{a}}){Messa}, {Adamo}, {{\"O}stlin},
  {Calzetti}, {Grasha}, {Grebel}, {Shabani}, {Chandar}, {Dale}, {Dobbs},
  {Elmegreen}, {Fumagalli}, {Gouliermis}, {Kim}, {Smith}, {Thilker}, {Tosi},
  {Ubeda}, {Walterbos}, {Whitmore}, {Fedorenko}, {Mahadevan}, {Andrews},
  {Bright}, {Cook}, {Kahre}, {Nair}, {Pellerin}, {Ryon}, {Ahmad}, {Beale},
  {Brown}, {Clarkson}, {Guidarelli}, {Parziale}, {Turner}, \&
  {Weber}}]{messa2018a}
{Messa}, M., {Adamo}, A., {{\"O}stlin}, G., {et~al.} 2018{\natexlab{a}},
  \mnras, 473, 996, \dodoi{10.1093/mnras/stx2403}

\bibitem[{{Messa} {et~al.}(2018{\natexlab{b}}){Messa}, {Adamo}, {Calzetti},
  {Reina-Campos}, {Colombo}, {Schinnerer}, {Chand ar}, {Dale}, {Gouliermis},
  {Grasha}, {Grebel}, {Elmegreen}, {Fumagalli}, {Johnson}, {Kruijssen},
  {{\"O}stlin}, {Shabani}, {Smith}, \& {Whitmore}}]{messa2018b}
{Messa}, M., {Adamo}, A., {Calzetti}, D., {et~al.} 2018{\natexlab{b}}, \mnras,
  477, 1683, \dodoi{10.1093/mnras/sty577}

\bibitem[{{Mok} {et~al.}(2019){Mok}, {Chandar}, \& {Fall}}]{mok2019}
{Mok}, A., {Chandar}, R., \& {Fall}, S.~M. 2019, \apj, 872, 93,
  \dodoi{10.3847/1538-4357/aaf6ea}

\bibitem[{{Mok} {et~al.}(2020){Mok}, {Chandar}, \& {Fall}}]{mok2020}
---. 2020, \apj, 893, 135, \dodoi{10.3847/1538-4357/ab7a14}

\bibitem[{{Mora} {et~al.}(2009){Mora}, {Larsen}, {Kissler-Patig}, {Brodie}, \&
  {Richtler}}]{mora2009}
{Mora}, M.~D., {Larsen}, S.~S., {Kissler-Patig}, M., {Brodie}, J.~P., \&
  {Richtler}, T. 2009, \aap, 501, 949, \dodoi{10.1051/0004-6361/200810614}

\bibitem[{Nair \& Hinton(2010)}]{relu2}
Nair, V., \& Hinton, G.~E. 2010, in ICML, ed. J.~Fürnkranz \& T.~Joachims
  (Omnipress), 807--814

\bibitem[{Paszke {et~al.}(2019)Paszke, Gross, Massa, Lerer, Bradbury, Chanan,
  Killeen, Lin, Gimelshein, Antiga, Desmaison, Kopf, Yang, DeVito, Raison,
  Tejani, Chilamkurthy, Steiner, Fang, Bai, \& Chintala}]{pytorch}
Paszke, A., Gross, S., Massa, F., {et~al.} 2019, in Advances in Neural
  Information Processing Systems 32, ed. H.~Wallach, H.~Larochelle,
  A.~Beygelzimer, F.~d~Alch\'{e}-Buc, E.~Fox, \& R.~Garnett (Curran Associates,
  Inc.), 8024--8035

\bibitem[{Pedregosa {et~al.}(2011)Pedregosa, Varoquaux, Gramfort, Michel,
  Thirion, Grisel, Blondel, Prettenhofer, Weiss, Dubourg, Vanderplas, Passos,
  Cournapeau, Brucher, Perrot, \& Duchesnay}]{scikit-learn}
Pedregosa, F., Varoquaux, G., Gramfort, A., {et~al.} 2011, Journal of Machine
  Learning Research, 12, 2825

\bibitem[{{Ryon} {et~al.}(2015){Ryon}, {Bastian}, {Adamo}, {Konstantopoulos},
  {Gallagher}, {Larsen}, {Hollyhead}, {Silva-Villa}, \& {Smith}}]{ryon2015}
{Ryon}, J.~E., {Bastian}, N., {Adamo}, A., {et~al.} 2015, \mnras, 452, 525,
  \dodoi{10.1093/mnras/stv1282}

\bibitem[{{Ryon} {et~al.}(2017){Ryon}, {Gallagher}, {Smith}, {Adamo},
  {Calzetti}, {Bright}, {Cignoni}, {Cook}, {Dale}, {Elmegreen}, {Fumagalli},
  {Gouliermis}, {Grasha}, {Grebel}, {Kim}, {Messa}, {Thilker}, \&
  {Ubeda}}]{ryon2017}
{Ryon}, J.~E., {Gallagher}, J.~S., {Smith}, L.~J., {et~al.} 2017, \apj, 841,
  92, \dodoi{10.3847/1538-4357/aa719e}

\bibitem[{{Sabbi} {et~al.}(2018){Sabbi}, {Calzetti}, {Ubeda}, {Adamo},
  {Cignoni}, {Thilker}, {Aloisi}, {Elmegreen}, {Elmegreen}, {Gouliermis},
  {Grebel}, {Messa}, {Smith}, {Tosi}, {Dolphin}, {Andrews}, {Ashworth},
  {Bright}, {Brown}, {Chandar}, {Christian}, {Clayton}, {Cook}, {Dale}, {de
  Mink}, {Dobbs}, {Evans}, {Fumagalli}, {Gallagher}, {Grasha}, {Herrero},
  {Hunter}, {Johnson}, {Kahre}, {Kennicutt}, {Kim}, {Krumholz}, {Lee},
  {Lennon}, {Martin}, {Nair}, {Nota}, {{\"O}stlin}, {Pellerin}, {Prieto},
  {Regan}, {Ryon}, {Sacchi}, {Schaerer}, {Schiminovich}, {Shabani}, {Van Dyk},
  {Walterbos}, {Whitmore}, \& {Wofford}}]{sabbi2018}
{Sabbi}, E., {Calzetti}, D., {Ubeda}, L., {et~al.} 2018, \apjs, 235, 23,
  \dodoi{10.3847/1538-4365/aaa8e5}

\bibitem[{{Silva-Villa} {et~al.}(2014){Silva-Villa}, {Adamo}, {Bastian},
  {Fouesneau}, \& {Zackrisson}}]{silva-villa2014}
{Silva-Villa}, E., {Adamo}, A., {Bastian}, N., {Fouesneau}, M., \&
  {Zackrisson}, E. 2014, \mnras, 440, L116, \dodoi{10.1093/mnrasl/slu028}

\bibitem[{{Simonyan} \& {Zisserman}(2014)}]{VGG}
{Simonyan}, K., \& {Zisserman}, A. 2014, CoRR, abs/1409.1556

\bibitem[{Srivastava {et~al.}(2014)Srivastava, Hinton, Krizhevsky, Sutskever,
  \& Salakhutdinov}]{dropout}
Srivastava, N., Hinton, G., Krizhevsky, A., Sutskever, I., \& Salakhutdinov, R.
  2014, J. Mach. Learn. Res., 15, 1929–1958

\bibitem[{Srivastava {et~al.}(2015)Srivastava, Greff, \& Schmidhuber}]{highway}
Srivastava, R.~K., Greff, K., \& Schmidhuber, J. 2015, Highway Networks

\bibitem[{Szegedy {et~al.}(2015)Szegedy, Liu, Jia, Sermanet, Reed, Anguelov,
  Erhan, Vanhoucke, \& Rabinovich}]{googlenet}
Szegedy, C., Liu, W., Jia, Y., {et~al.} 2015, in Computer Vision and Pattern
  Recognition (CVPR).
\newblock \url{http://arxiv.org/abs/1409.4842}

\bibitem[{Van~der Walt {et~al.}(2014)Van~der Walt, Sch{\"o}nberger,
  Nunez-Iglesias, Boulogne, Warner, Yager, Gouillart, \& Yu}]{scikit-image}
Van~der Walt, S., Sch{\"o}nberger, J.~L., Nunez-Iglesias, J., {et~al.} 2014,
  PeerJ, 2, e453

\bibitem[{{Virtanen} {et~al.}(2020){Virtanen}, {Gommers}, {Oliphant},
  {Haberland}, {Reddy}, {Cournapeau}, {Burovski}, {Peterson}, {Weckesser},
  {Bright}, {van der Walt}, {Brett}, {Wilson}, {Jarrod Millman}, {Mayorov},
  {Nelson}, {Jones}, {Kern}, {Larson}, {Carey}, {Polat}, {Feng}, {Moore}, {Vand
  erPlas}, {Laxalde}, {Perktold}, {Cimrman}, {Henriksen}, {Quintero}, {Harris},
  {Archibald}, {Ribeiro}, {Pedregosa}, {van Mulbregt}, \&
  {Contributors}}]{scipy}
{Virtanen}, P., {Gommers}, R., {Oliphant}, T.~E., {et~al.} 2020, Nature
  Methods, \dodoi{https://doi.org/10.1038/s41592-019-0686-2}

\bibitem[{Walmsley {et~al.}(2018)Walmsley, Ferguson, Mann, \&
  Lintott}]{Walmsley2018}
Walmsley, M., Ferguson, A. M.~N., Mann, R.~G., \& Lintott, C.~J. 2018, Monthly
  Notices of the Royal Astronomical Society, 483, 2968–2982,
  \dodoi{10.1093/mnras/sty3232}

\bibitem[{{Wei} {et~al.}(2020){Wei}, {Huerta}, {Whitmore}, {Lee}, {Hannon},
  {Chandar}, {Dale}, {Larson}, {Thilker}, {Ubeda}, {Boquien}, {Chevance},
  {Kruijssen}, {Schruba}, {Blanc}, \& {Congiu}}]{wei2020}
{Wei}, W., {Huerta}, E.~A., {Whitmore}, B.~C., {et~al.} 2020, \mnras, 493,
  3178, \dodoi{10.1093/mnras/staa325}

\bibitem[{{Whitmore} {et~al.}(2014){Whitmore}, {Chandar}, {Bowers}, {Larsen},
  {Lindsay}, {Ansari}, \& {Evans}}]{whitmore2014}
{Whitmore}, B.~C., {Chandar}, R., {Bowers}, A.~S., {et~al.} 2014, \aj, 147, 78,
  \dodoi{10.1088/0004-6256/147/4/78}

\bibitem[{{Whitmore} {et~al.}(1999){Whitmore}, {Zhang}, {Leitherer}, {Fall},
  {Schweizer}, \& {Miller}}]{whitmore1999}
{Whitmore}, B.~C., {Zhang}, Q., {Leitherer}, C., {et~al.} 1999, \aj, 118, 1551,
  \dodoi{10.1086/301041}

\bibitem[{{Whitmore} {et~al.}(2010){Whitmore}, {Chandar}, {Schweizer},
  {Rothberg}, {Leitherer}, {Rieke}, {Rieke}, {Blair}, {Mengel}, \&
  {Alonso-Herrero}}]{whitmore2010}
{Whitmore}, B.~C., {Chandar}, R., {Schweizer}, F., {et~al.} 2010, \aj, 140, 75,
  \dodoi{10.1088/0004-6256/140/1/75}

\bibitem[{{Wu} \& {He}(2018)}]{gn}
{Wu}, Y., \& {He}, K. 2018, in Computer Vision - {ECCV} 2018 - 15th European
  Conference, Munich, Germany, September 8-14, 2018, Proceedings, Part {XIII},
  3--19, \dodoi{10.1007/978-3-030-01261-8\_1}

\bibitem[{Zhang {et~al.}(2017)Zhang, Zhou, Lin, \& Sun}]{shufflenet}
Zhang, X., Zhou, X., Lin, M., \& Sun, J. 2017, ShuffleNet: An Extremely
  Efficient Convolutional Neural Network for Mobile Devices.
\newblock \doarXiv{1707.01083}

\bibitem[{Ćiprijanović {et~al.}(2020)Ćiprijanović, Snyder, Nord, \&
  Peek}]{iprijanovi2020}
Ćiprijanović, A., Snyder, G.~F., Nord, B., \& Peek, J. E.~G. 2020, DeepMerge:
  Classifying High-redshift Merging Galaxies with Deep Neural Networks.
\newblock \doarXiv{2004.11981}

\end{thebibliography}
\bibliographystyle{aasjournal}



\end{document}